\documentclass[a4paper,11pt]{article}
\usepackage{jinstpub} 
\usepackage{lineno}
\usepackage{hyperref}
\usepackage{comment}
\usepackage{xcolor}
\usepackage{float}
\usepackage[numbers,sort&compress]{natbib}

\newcommand{\epem}{$e^{+}e^{-}$}

\title{\boldmath Design and beam-test characterization of the CRILIN semi-homogeneous crystal calorimeter}
\author[a]{S.~Ceravolo}
\author[b]{M.~Cavallina}
\author[a]{V.~Ciccarella}
\author[a,1]{E.~Di Meco\note{Corresponding authors.}}
\author[a]{E.~Diociaiuti}
\author[c,1]{R.~Gargiulo}
\author[d]{Q.~Han}
\author[c]{E.~Leonardi}
\author[d]{D.~Lucchesi}
\author[a]{M.~Moulson}
\author[d]{L.~Palombini}
\author[e]{N.~Pastrone}
\author[f]{I.~Raghib}
\author[a]{A.~Russo}
\author[b]{A.~Saputi}
\author[a,1]{I.~Sarra}
\author[a]{S.~Salamino}
\author[f]{L.~Sestini}
\author[b]{S.~Squerzanti}
\author[d]{D.~Zuliani}
\affiliation[a]{Laboratori Nazionali di Frascati dell'INFN, Frascati, Italy}
\affiliation[b]{INFN Sezione di Ferrara, Ferrara, Italy}
\affiliation[c]{INFN Sezione di Roma1, Rome, Italy}
\affiliation[d]{INFN Sezione di Padova, Padua, Italy}
\affiliation[e]{INFN Sezione di Torino, Turin, Italy}
\affiliation[f]{INFN Sezione di Firenze, Florence, Italy}
\emailAdd{elisa.dimeco@lnf.infn.it,\\ruben.gargiulo@roma1.infn.it,\\ivano.sarra@lnf.infn.it}

\abstract{
CRILIN is a high-granularity semi-homogeneous electromagnetic calorimeter based on longitudinally segmented PbF$_2$ crystal matrices read out by UV-extended silicon photomultipliers. The concept combines fast Cherenkov response, fine transverse granularity, longitudinal shower information, and radiation tolerance for future lepton-collider experiments. This paper reports the construction of a large-area prototype and its performance measured in beam tests at the CERN SPS. The detector comprises five $7\times7$ PbF$_2$ crystal matrices, has a depth of about $22X_0$, and is read out by four $3\times3~\mathrm{mm}^2$ SiPMs per crystal integrated in a single electronic channel.

Electron data between 10 and 120~GeV and dedicated 150~GeV muon data were used to characterize the detector response. A time resolution below 50~ps is achieved for electron energies above 10~GeV, reaching values below 20~ps above 60~GeV. 

The energy resolution is described by a stochastic term of $(6.58\pm0.04)\%/\sqrt{E/\mathrm{GeV}}$ and a constant term of $(0.23\pm0.02)\%$, with an additional noise contribution fixed from pedestal data. The longitudinal segmentation enables event-by-event corrections based on the reconstructed shower development, resulting in a significant improvement of the energy resolution. A light yield of approximately 0.54~photoelectrons/MeV is measured consistently using both electron showers and minimum-ionizing particles. A Geant4-based simulation incorporating the relevant experimental effects reproduces the measured energy resolution. These results validate the CRILIN architecture as a compact, fast, and longitudinally segmented electromagnetic calorimeter for future collider experiments.
}

\begin{document}
\maketitle
\flushbottom

\section{Introduction and physics motivations}
 \label{sec:introduction}
 The physics program of future lepton colliders, i.e., $e^+e^-$ or Muon Colliders, places stringent requirements on electromagnetic calorimetry. 
At a Muon Collider, the decay of beam muons along the machine lattice generates an intense flux of secondary and tertiary particles, producing a substantial out-of-time background in the detector~\cite{Accettura:2023MuonCollider}. In the electromagnetic calorimeter region, simulations predict particle fluxes dominated by low-energy photons, with an average photon energy of about 1.7 MeV, together with a non-negligible neutron component. In this environment, precise time-of-arrival measurements and longitudinal information are essential to reject background contributions, correctly associate hits with the primary interaction, and preserve the performance of event reconstruction algorithms. In addition to good energy resolution, electromagnetic calorimeters are expected to provide fine transverse granularity, longitudinal segmentation, and precise timing to support particle-flow reconstruction, improve the separation of nearby deposits, and enhance the identification of electromagnetic shower topologies in dense environments~\cite{ParticleFlow}.
These requirements are relevant not only for Muon Collider detector concepts \cite{MUSIC}, where beam-induced backgrounds represent a major challenge, but also for future \epem machines such as FCC-ee, where high-granularity calorimetry is expected to play a key role in jet and photon reconstruction performance for precision measurements and for searches for rare or beyond-the-Standard-Model processes~\cite{FCC:FeasibilityStudy}.

A conventional high-granularity silicon-tungsten calorimeter is a natural benchmark for these applications, thanks to its mature technology and excellent segmentation~\cite{Linssen:2012CLIC}. However, such a solution implies a very large number of readout channels, significant integration complexity, and a substantial cost, especially when extended to detector volumes relevant for future collider experiments. These considerations motivate the study of alternative calorimeter concepts that can retain the advantages of high granularity and longitudinal sampling while reducing the overall system complexity.

The calorimeter concept developed in this work is therefore driven by a set of requirements emerging from both general future-collider needs and the more demanding Muon Collider environment. The detector must provide a time resolution at the level of O(100~ps) or better, fine enough granularity to reduce occupancy and cluster confusion, sufficient longitudinal segmentation to identify the shower profile, and adequate energy resolution for precision electromagnetic measurements. At the same time, it should remain compact, scalable, and economically sustainable when extrapolated to a full detector system. These considerations motivated the development of CRILIN(CRystal calorImeter with Longitudinal InformatioN), a crystal-based semi-homogeneous calorimeter optimized to combine fast response, compactness, and longitudinal information within a reduced channel count. This concept combines the strengths of homogeneous and sampling calorimeters~\cite{Ceravolo:2022CRILIN}. The detector is based on matrices of dense, fast crystals, each individually read out by silicon photomultipliers, and arranged in multiple longitudinal layers. In this way, CRILIN combines compactness, high transverse granularity, intrinsic longitudinal segmentation, and fast signal formation from Cherenkov light, making it an attractive solution for future collider calorimetry. The use of PbF$_2$ crystals, together with UV-extended SiPMs, allows the detector to exploit a prompt optical response while maintaining a mechanically simple and scalable architecture.

Over the past few years, the CRILIN concept has been progressively validated through simulation studies, component characterization, irradiation campaigns, and beam tests on increasingly complex prototypes~\cite{Ceravolo:2022CRILIN},\cite{Cantone:2023Frontiers},\cite{Cantone:2024TNS},\cite{Cemmi:2022JINST},\cite{Cemmi:2024JINST}. The prototypes demonstrated excellent timing capabilities, reaching sub-50~ps performance for electromagnetic energy deposits at the GeV scale. In parallel, dedicated studies on crystals, wrapping configurations, front-end electronics, and SiPM irradiation behavior have provided the information needed to optimize the detector design for a larger large-area module.~\cite{VittoriaTipp},\cite{Cantone:2024Biodola}.
Recent R\&D activities led to the adoption of 15~$\mu$m pixel-size SiPMs in order to improve photostatistics for applications beyond the original Muon Collider-driven optimization, with measured values of about 0.54~photoelectrons/MeV under the selected operating conditions.

The latest step in this development program is the construction of a new large-area CRILIN prototype, consisting of five layers of 7$\times$7 PbF$_2$ crystal matrices. The detector was designed to provide sufficient transverse and longitudinal shower containment for precise calorimetric measurements, while implementing the mechanical and readout solutions developed in the previous prototype iterations. In particular, the new module integrates an aluminum alveolar structure with thin inter-crystal septa, a compact external envelope, four SiPMs per crystal, and a custom readout architecture based on flexible Kapton strips for remote signal and bias routing. The detector was exposed to the electron beam at the CERN SPS from 24 to 29 June 2026 to perform a complete characterization of its calorimetric response and to validate the full detector chain from mechanics and electronics to calibration and reconstruction.

The first part of the paper presents the detector layout, mechanics, front-end electronics, and the test beam setup. Later, the paper discusses the reconstruction procedure, the calibration based on minimum-ionizing particles, the Monte Carlo simulation, and the studies on energy and time resolutions.


\section{Detector concept and design}
\label{sec:Detector concept and design}

\subsection{Semi-homogeneous calorimeter concept}
CRILIN was conceived as an electromagnetic calorimeter architecture positioned between the traditional homogeneous and sampling approaches~\cite{Ceravolo:2022CRILIN}. The detector is based on a stack of crystal matrices interleaved with thin readout layers, so that the active material also provides most of the shower development volume, while the segmented structure preserves longitudinal information. In this sense, the detector can be regarded as semi-homogeneous: unlike a sampling calorimeter, it minimizes passive absorber layers, while unlike a fully homogeneous calorimeter, it retains a layered architecture that can be exploited for shower imaging and timing.

Each crystal acts as an independent calorimetric cell and is read out by silicon photomultipliers, allowing the detector to achieve fine transverse granularity together with fast signal formation. The layered geometry provides a natural handle on the longitudinal shower development, which is important both for background rejection and for advanced reconstruction strategies based on timing and topology. In the original Muon Collider-oriented configuration, the concept was optimized with crystal cells of about 10\(\times\)10~mm$^2$ transverse size and layer thicknesses of about 4~$X_0$, obtaining a calorimeter design able to suppress BIB fluctuations while preserving competitive electromagnetic performance.

A key advantage of this approach is its design flexibility. The crystal transverse size, longitudinal segmentation, and total depth can be adjusted according to the target application, allowing the detector to be optimized for different collider environments and physics goals. Compared with highly segmented silicon-tungsten calorimeters, the CRILIN concept can reduce the total number of layers and readout channels, with a corresponding simplification of mechanics, electronics, and integration. This feature is particularly attractive for future large-scale calorimeter systems, where channel count, cooling, services, and cost all become central design constraints. The semi-homogeneous concept not only allows particle flow techniques for jet reconstructions, but also grants a large degree of software compensation, via the usage of hit-based machine learning techniques ~\cite{RubenPaper}, to avoid spoiling HCAL performance on jet energy resolutions due to the natural extreme non-compensation of the crystal matrices.

\subsection{Crystal choice and SiPM selection}
The choice of the active material and of the photodetectors was primarily guided by the need for fast response, compactness, radiation tolerance, and compatibility with a highly segmented geometry. Among the crystal candidates considered within the CRILIN R\&D program, lead fluoride (PbF$_2$) emerged as the reference option because of its high density, short radiation length, compact Moli\`ere radius, and prompt Cherenkov light emission. These properties make PbF$_2$ particularly attractive for precise timing applications, since the signal formation is intrinsically fast and the calorimeter can remain compact while preserving longitudinal segmentation.

Alternative crystal options, including PbWO$_4$-UF, were also investigated during the early R\&D phase~\cite{Cantone:2023Frontiers}. Prototype studies showed that PbF$_2$ provides superior timing performance in the adopted readout configuration, despite its lower light yield compared with ultrafast scintillating crystals, confirming the benefit of a purely Cherenkov-based response for this application. In addition, irradiation campaigns performed on PbF$_2$ crystals demonstrated a behavior compatible with the expected operating conditions of the Muon Collider electromagnetic calorimeter, although the response depends on crystal production quality and optical treatment~\cite{Cemmi:2022JINST},\cite{Cemmi:2024JINST}. These results supported the consolidation of PbF$_2$ as the baseline crystal choice for the new large-area prototype.

The photodetector choice followed a similar optimization logic. Since each crystal is read out individually, the sensor must combine compactness, high photon detection efficiency in the near-UV, stable operation, and acceptable behavior under irradiation. UV-extended surface-mount SiPMs from the Hamamatsu S14160~\cite{Hamamatsu} family were selected for this purpose and tested in different pixel-size configurations during the project development. Earlier prototype stages were mainly optimized around 10~$\mu$m pixel devices, which offered better radiation hardness and lower dark-current increase under neutron exposure~\cite{Cemmi:2022JINST}. More recently, the detector development moved toward the adoption of 15~$\mu$m pixel SiPMs in order to improve photostatistics and increase the collected light yield in the larger prototype geometry.

Within the present detector design, the selected 15~$\mu$m pixel SiPM configuration provided a measured light yield of about 0.54~photoelectrons/MeV under the adopted operating conditions. The larger pixel size also improves the single-channel signal amplitude, which is advantageous for low-energy deposits and for stable reconstruction after thresholding, although at the price of a reduced margin in dynamic range and radiation hardness with respect to smaller-pixel sensors. As demonstrated in previous beam tests and discussed in~\cite{Cantone:2023Frontiers}, CRILIN adopts a back-incidence configuration, in which the incident particles enter the crystal from the SiPM side. Since the Cherenkov light is emitted predominantly in the forward direction, it must first be reflected at the far end of the crystal before reaching the photosensors. This reflection randomizes the photon trajectories and suppresses the correlation between the light-collection response and the particle impact position, thereby improving response uniformity and mitigating position-dependent timing effects. In addition, the four lateral crystal surfaces are ground to promote diffuse reflections, while the two small end faces are polished to preserve the optical coupling to the SiPMs and the reflective properties of the opposite face.


\section{Prototype construction and readout}
\label{sec:Prototype Construction and Readout}

\subsection{Mechanical layout and longitudinal segmentation}

The large-area prototype consists of five independent layers, each based on a $7\times7$ mechanical matrix of PbF$_2$ crystal positions. In each layer, the four corner positions are occupied by passive mechanical placeholders and are not instrumented; the remaining 45 positions host active crystals coupled to the readout. This results in $45\times5=225$ calorimetric cells in total, providing a longitudinal depth of approximately $22\,X_0$, sufficient to contain electromagnetic showers in the energy range addressed by the test-beam program (10--120~GeV). The transverse dimensions of each crystal are $1.3\times1.3~\mathrm{cm}^2$, with a nominal length of 40~mm. The active transverse size of the module corresponds to approximately two Moli\`ere radii, offering an adequate balance between containment, channel count, and mechanical complexity.

\begin{figure}[htbp!]
    \centering
    \includegraphics[width=0.35\textwidth]{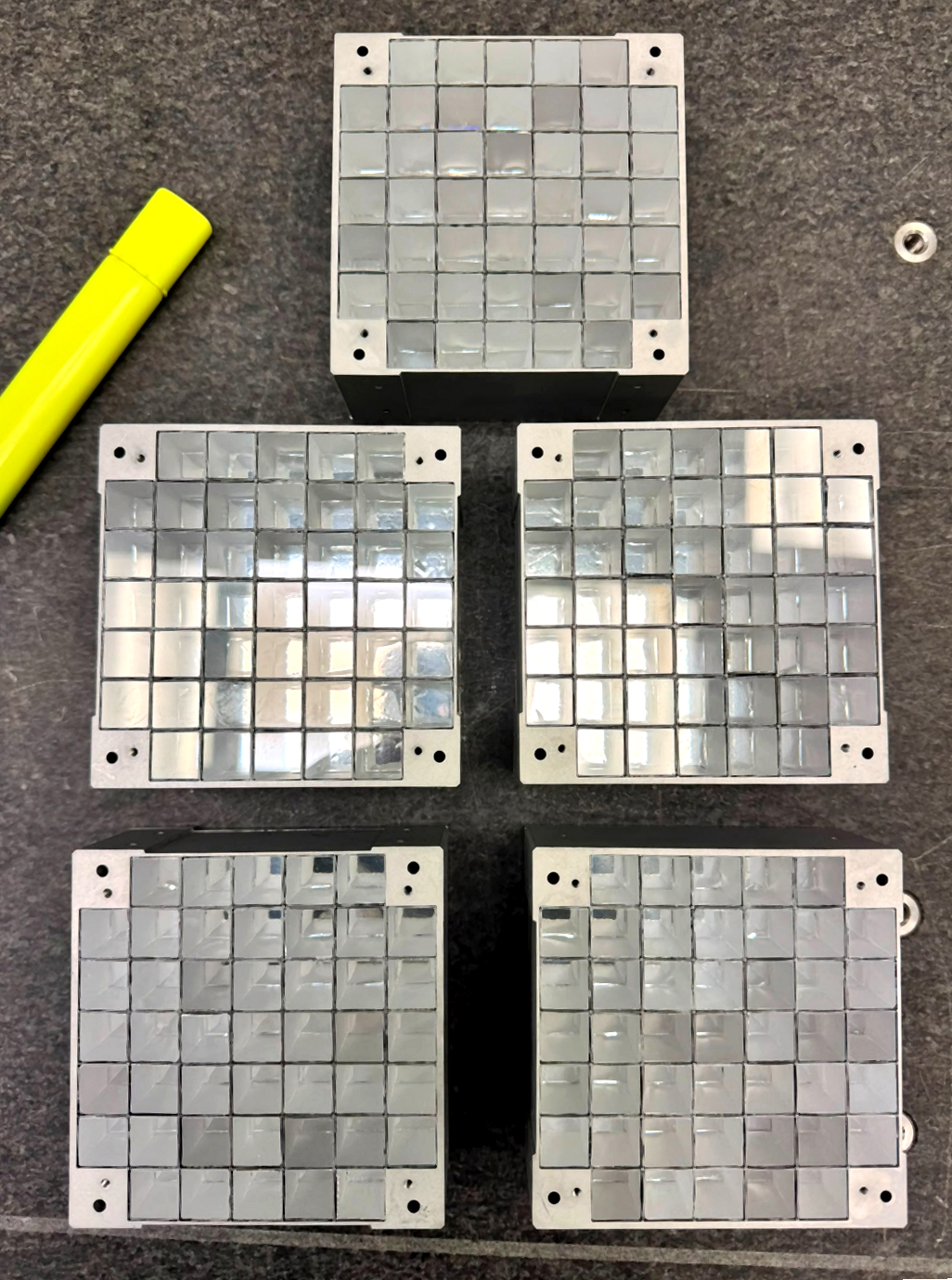}
    \includegraphics[width=0.35\textwidth]{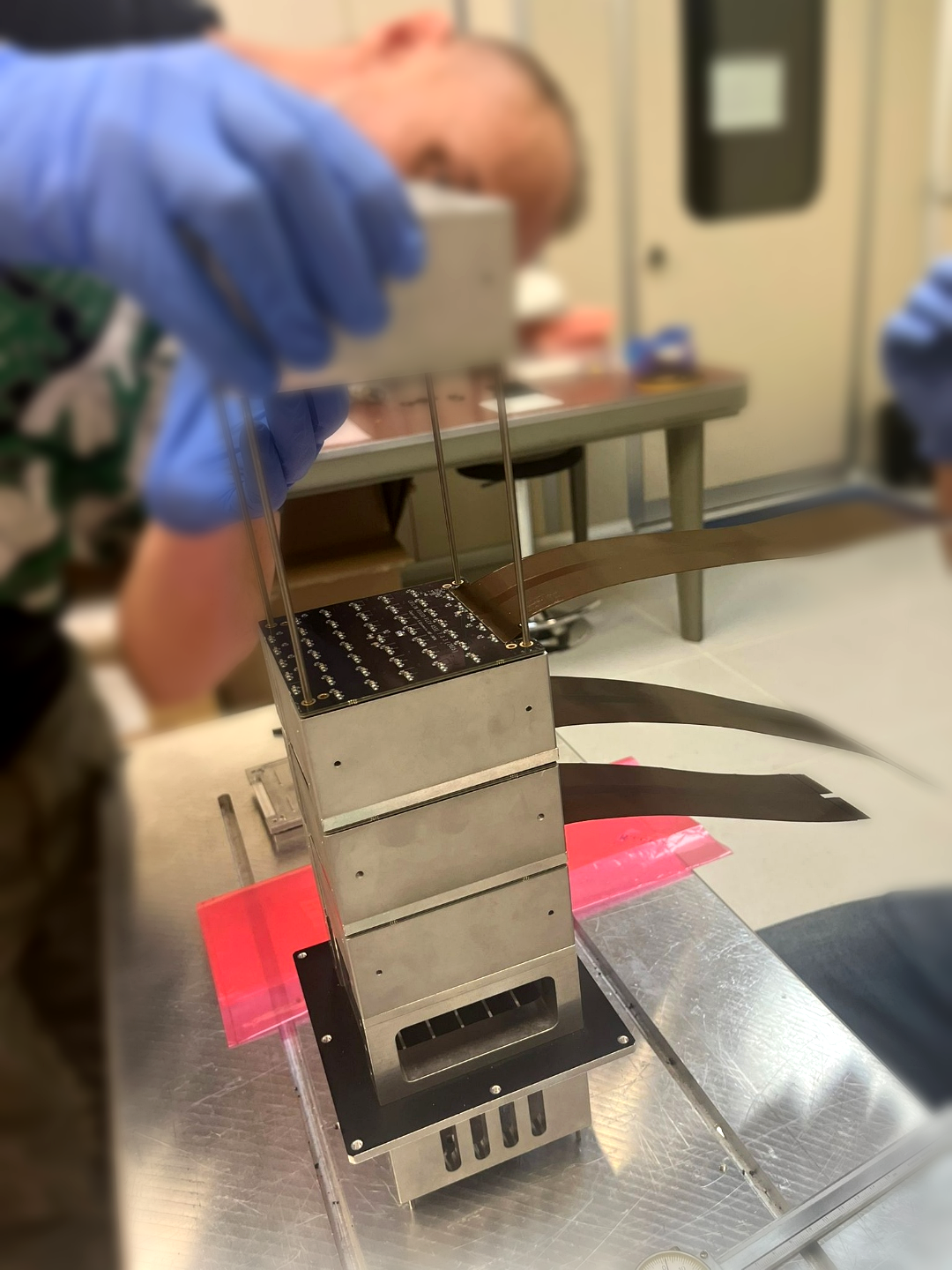}
    \caption{Left: front view of the five $\mathrm{7\times7}$ PbF$_2$ crystal matrices housed in the aluminum alveolar structure. Right: side view of the five-layer module, showing the longitudinally segmented arrangement of the crystal matrices. }
    \label{fig:crilin_7x7_stack}
\end{figure}

The five-layer structure, shown in Fig.~\ref{fig:crilin_7x7_stack}, preserves the main feature of the CRILIN concept: each layer provides independent information on the longitudinal development of the shower.  Each layer is hosted in an identical monolithic alveolar support manufactured from a single block of 7000-series aluminum alloy by Wire Electrical Discharge Machining (Wire EDM). This material was selected for its mechanical strength and dimensional stability during machining, allowing thin structural elements to be produced while maintaining the required rigidity and manufacturing accuracy. Each support comprises 49 cells designed to house the crystals, forming a monolithic honeycomb structure directly machined from the aluminum block. The inter-crystal aluminum septa are approximately 200~$\mu$m thick, while the external envelope is constrained to remain below approximately 2~mm. The Wire EDM manufacturing process ensures the geometrical tolerances required for the crystal positioning while preserving the continuity of the aluminum matrix. This configuration provides a mechanically stiff structure with a limited material budget, minimizes assembly tolerances, and ensures mechanical and optical isolation between neighboring crystals. In addition, the high thermal conductivity of aluminum allows the support structure to act as an efficient heat spreader, enabling direct cooling of the crystal matrices and improving temperature uniformity throughout the calorimeter.

The five alveolar modules are assembled into a self-supporting structure by means of four high-strength steel tie rods. The preload applied by the tie rods ensures the mechanical integrity and rigidity of the calorimeter during handling, installation, and operation, while preserving the relative alignment of the individual layers.
The wrapping configuration was investigated during the previous prototype campaigns, including dedicated measurements with different reflective materials ~\cite{VittoriaTipp}. The choice fell on 4 layers of 50~$\mu$m Aluminized Mylar foil, on the four lateral faces and on the face opposite to the readout.

\begin{figure}[htbp!]
    \centering
    \includegraphics[height=5cm]{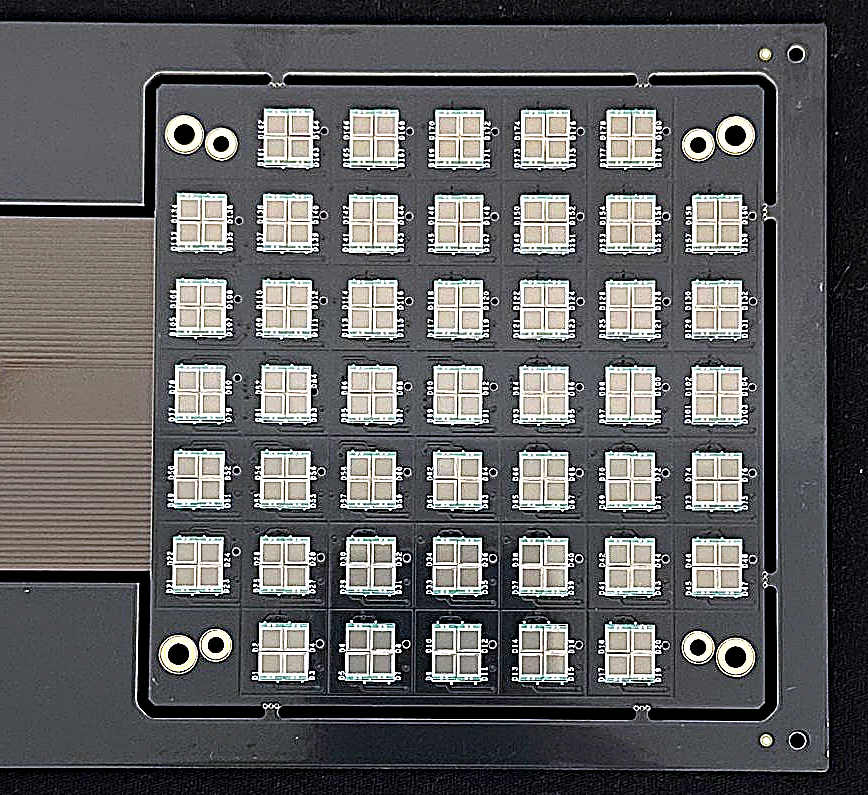}
    \includegraphics[height=5cm]{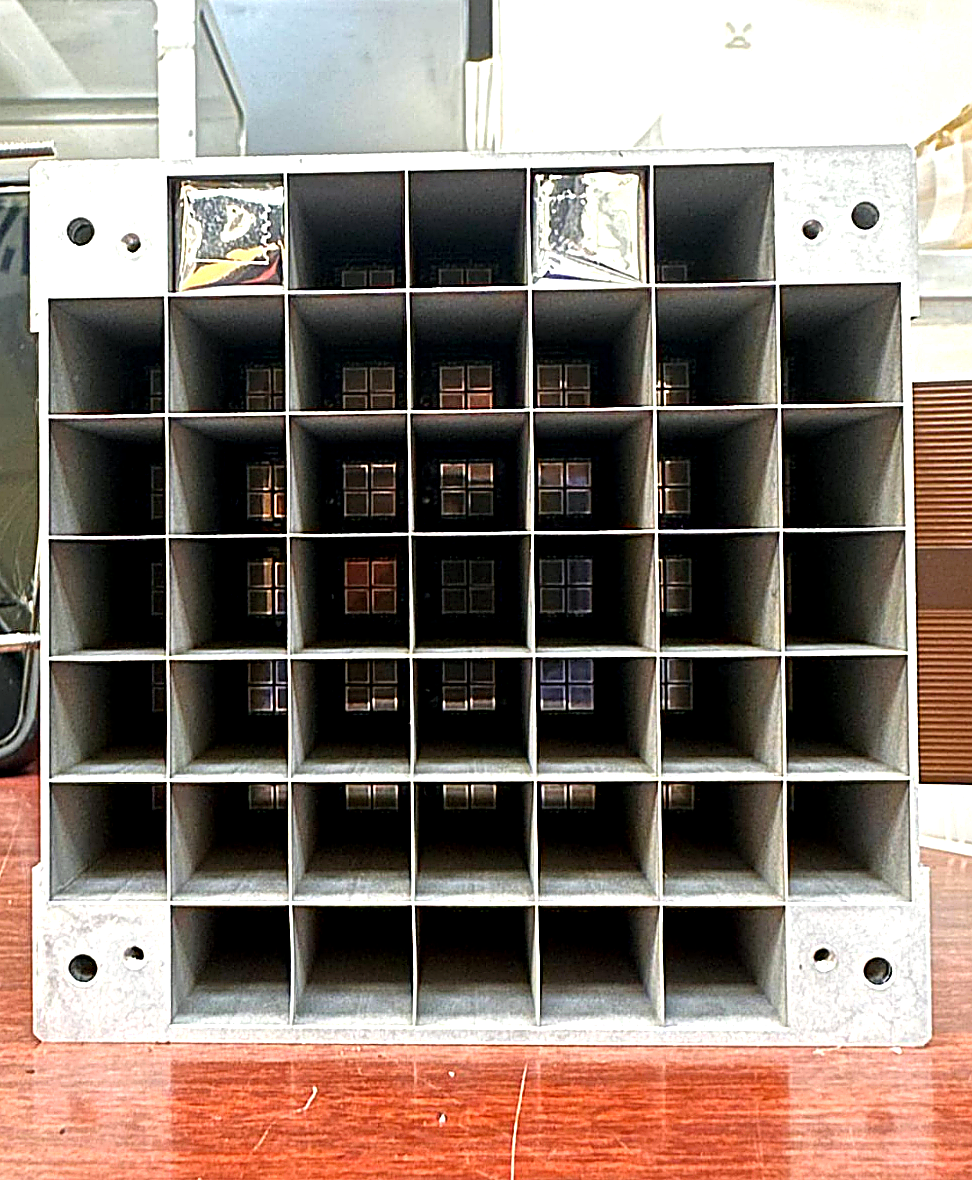}
    \caption{Readout integration of the $\mathrm{7\times7}$ module: (left) SiPM PCB instrumented with four photosensors per calorimetric cell, (right) CRILIN module prototype with its mechanical frame, partially populated crystal matrix, and the SiPM readout board.}
    \label{fig:crilin_readout_integration}
\end{figure}

Each PbF$_2$ crystal is read out by four surface-mount SiPMs with $3\times3~\mathrm{mm^2}$ active area. The photosensors are mounted on a dedicated PCB positioned directly behind the corresponding crystal matrix. The four SiPMs of each cell are arranged as two parallel branches, each branch consisting of two SiPMs connected in series, as shown in Figure~\ref{fig:crilin_readout_integration}-right. This readout scheme was chosen to maximize the collected light yield while maintaining a fast pulse shape and a sufficiently low effective sensor capacitance, thereby ensuring an adequate signal-to-noise ratio for low-energy deposits and preserving the timing performance required by the CRILIN specification.

\begin{figure}[htbp!]
    \centering
    \includegraphics[width=0.8\textwidth]{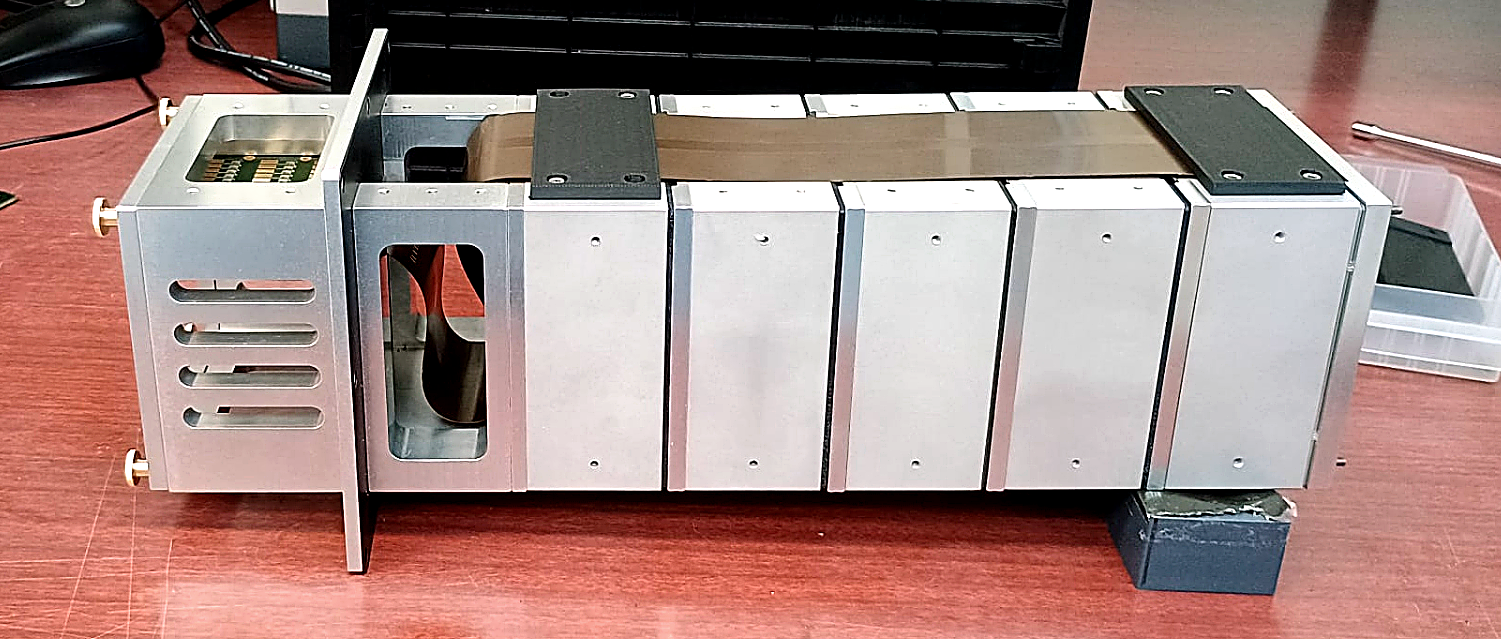}
    \caption{CRILIN module prototype with the mechanical support structure, crystal matrix, SiPM board, and rear transition region for the readout routing.}
    \label{fig:Al_cups}
\end{figure}

As shown in Fig.~\ref{fig:Al_cups} and~\ref{fig:crilin_readout_integration}-left , the crystals are kept against the front face by five 4-mm-thick aluminum cups, combined with a 2-mm-thick compression foam insert for each crystal. The cups are machined to reduce the radiation length contributed by the support structure. On the readout side, a rear basket, also manufactured by Wire EDM, houses the transition boards. These boards route the signals from the Kapton flex cables to the front-end electronics, located approximately 3.2~m from the detector. Only the SiPM bias distribution is integrated on the photosensor board. The remaining front-end electronics are moved away from the calorimeter volume in order to improve the radiation tolerance of the system, minimize passive material between longitudinal layers, and avoid local power dissipation in the SiPM region. An expansion region between the detector modules and the electronics accommodates the slack of the flexible flat cables, preventing excessive mechanical stress on the electrical connections and simplifying assembly and maintenance operations.

\begin{figure}[htbp!]
    \centering
    \includegraphics[width=0.45\textwidth]{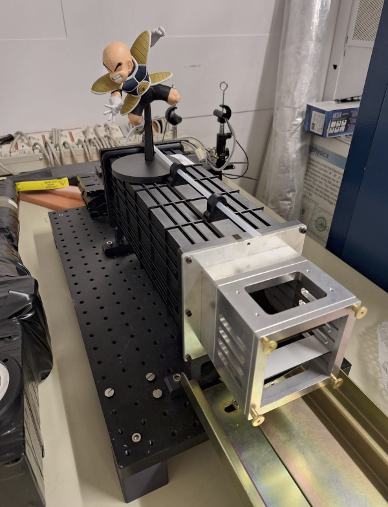}
    \caption{Black glass-fiber-reinforced polymer housing enclosing the CRILIN calorimeter prototype and providing a light-tight volume for the photosensors.}
    \label{fig:crilin_housing}
\end{figure}

The complete calorimeter is enclosed in a black glass-fiber-reinforced polymer housing, shown in Fig.~\ref{fig:crilin_housing}, which provides a light-tight environment for the photosensors. The volume between the housing and the calorimeter modules can be continuously flushed with dry gas to suppress moisture condensation during low-temperature operation. Alternatively, this volume can be used as a cooling channel by circulating temperature-controlled dry gas, providing an additional means of stabilizing the detector operating temperature.

\subsection{Front-end electronics}
The readout architecture is designed to keep the active electronics outside the calorimeter volume, thereby reducing the radiation exposure and power dissipation in the vicinity of the photosensors. Each SiPM layer is connected to a rear transition board through a rigid-flex printed circuit, in which the SiPMs are mounted on the rigid section, and the analog signals and bias voltages are routed through a Kapton-based flexible section. The latter provides individually shielded $50~\Omega$ stripline paths for the analog signals, preserving signal integrity and limiting electromagnetic interference and inter-channel cross-talk. The transition boards interface the five detector layers with the remote front-end system, located approximately $3.2~\mathrm{m}$ from the calorimeter and connected through coaxial cables. This arrangement avoids discrete cables within the active volume, accommodates the relative displacement of the longitudinal modules, and minimizes the passive material between calorimeter layers. Further details on the rigid-flex interconnect and its routing are provided in Section~\ref{app:kapton_routing} of the Appendix.

\begin{figure}[htbp!]
    \centering
    \includegraphics[width=0.7\textwidth]{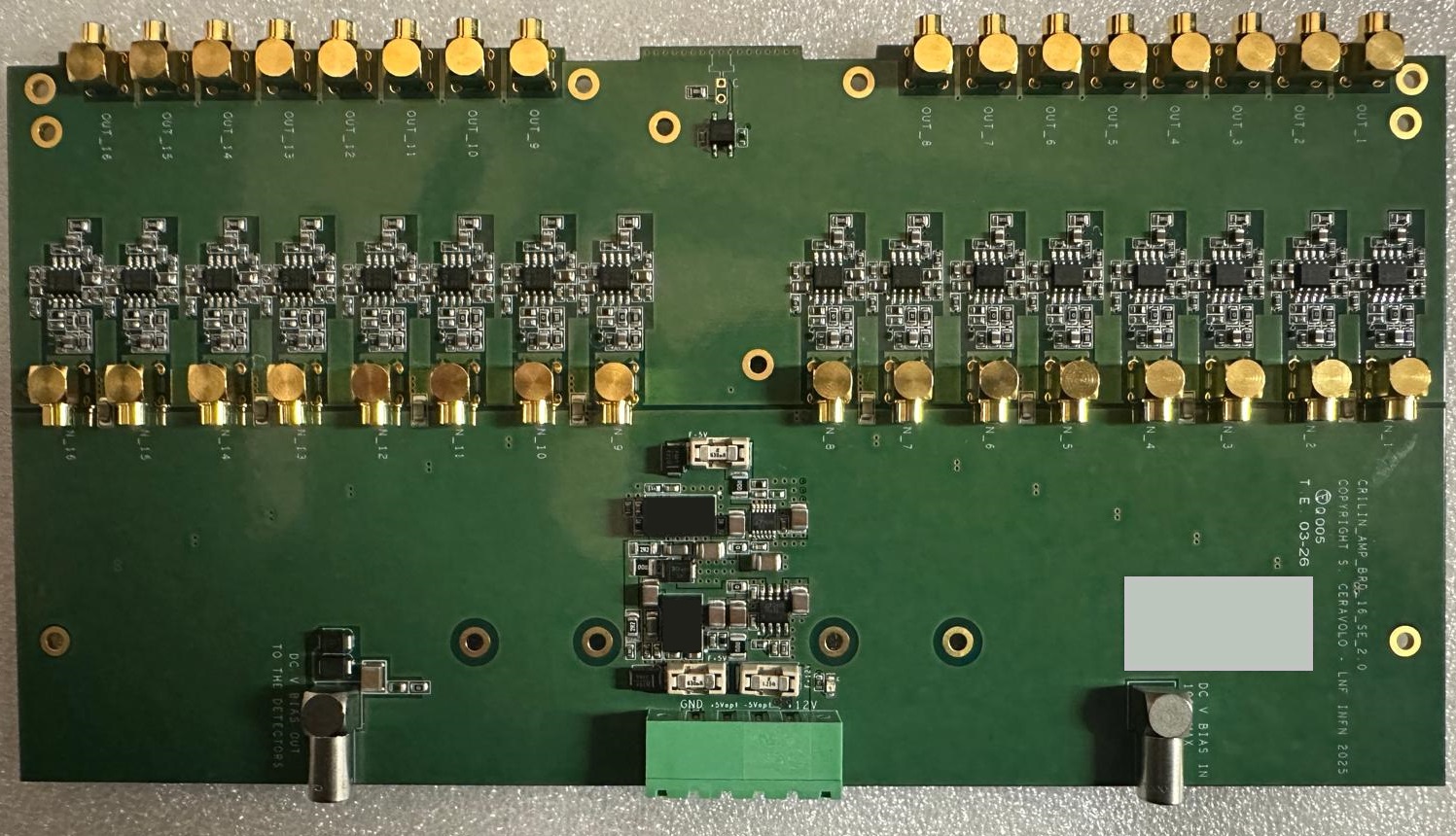}
    \caption{Sixteen-channel CRILIN front-end board. Fifteen channels are used for the calorimeter readout, while one channel is kept as a spare.}
    \label{fig:fee_board}
\end{figure}

The analog front-end consists of 16-channel amplification boards. At the input stage, each channel holds the SiPM anode at ground potential and couples the signal to the amplifier through a capacitive input network. Both the input and output stages of the amplification chain are impedance-matched to ensure signal transmission over the coaxial connections. The output stage saturates at approximately $1.6~\mathrm{V}$ when terminated with a $50~\Omega$ load. A 16-channel amplification board is shown in Fig.~\ref{fig:fee_board}. For each board, the first 15 channels amplify one-third of the channels of a single SiPM layer, while the 16th channel is left unconnected and reserved as a spare. Therefore, three amplification boards are associated with each calorimeter layer, which comprises 45 readout channels, for a total of 15 amplification boards for the five-layer calorimeter.

\begin{figure}[htbp!]
    \centering
    \includegraphics[width=0.8\textwidth]{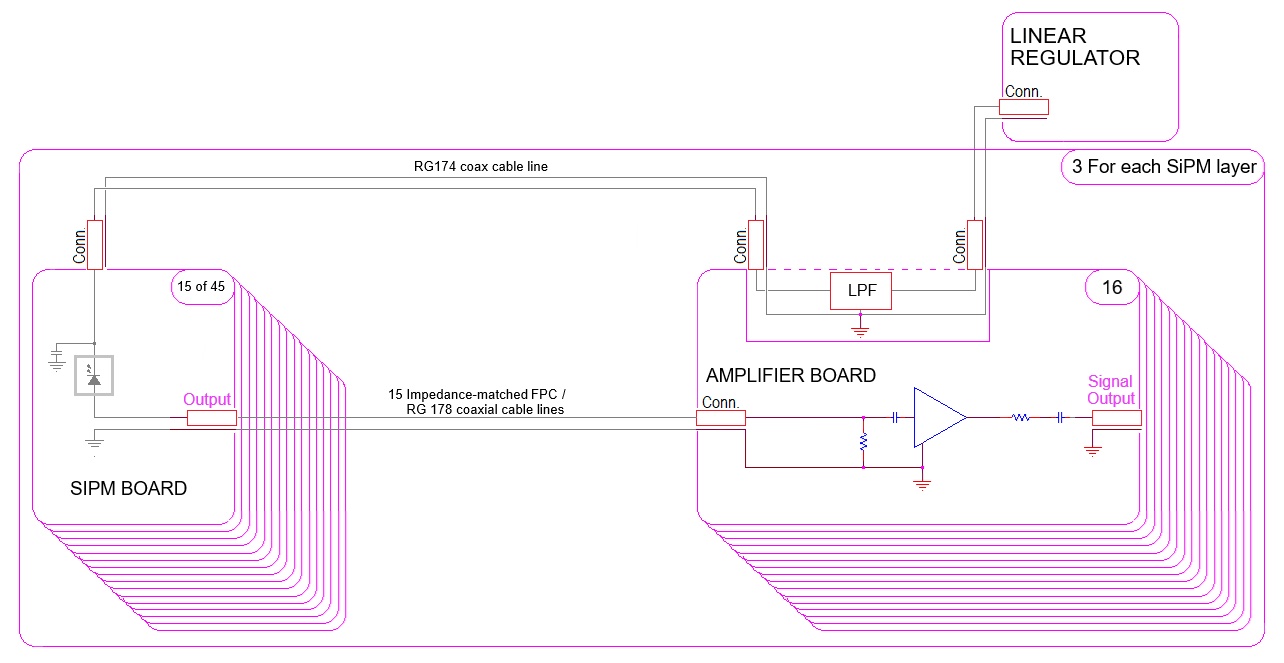}
    \caption{Equivalent circuit of the CRILIN front-end electronics, showing the capacitive input stage, the impedance-matched amplification chain, and the bias distribution network for the 15 connected SiPMs.}
    \label{fig:schematic_all}
\end{figure}

Through an RG174 coaxial cable, each amplification board receives, filters, and distributes the bias voltage to the 15 connected SiPMs. Each group of 15 SiPMs shares a common bias voltage within a given calorimeter layer. The return currents of the 15 channels, for both the signal paths and the bias distribution, converge on a common ground plane on the amplification board. The equivalent circuit of the CRILIN front-end electronics is shown in Fig.~\ref{fig:schematic_all}.

\begin{figure}[htbp!]
    \centering
    \includegraphics[width=0.6\textwidth]{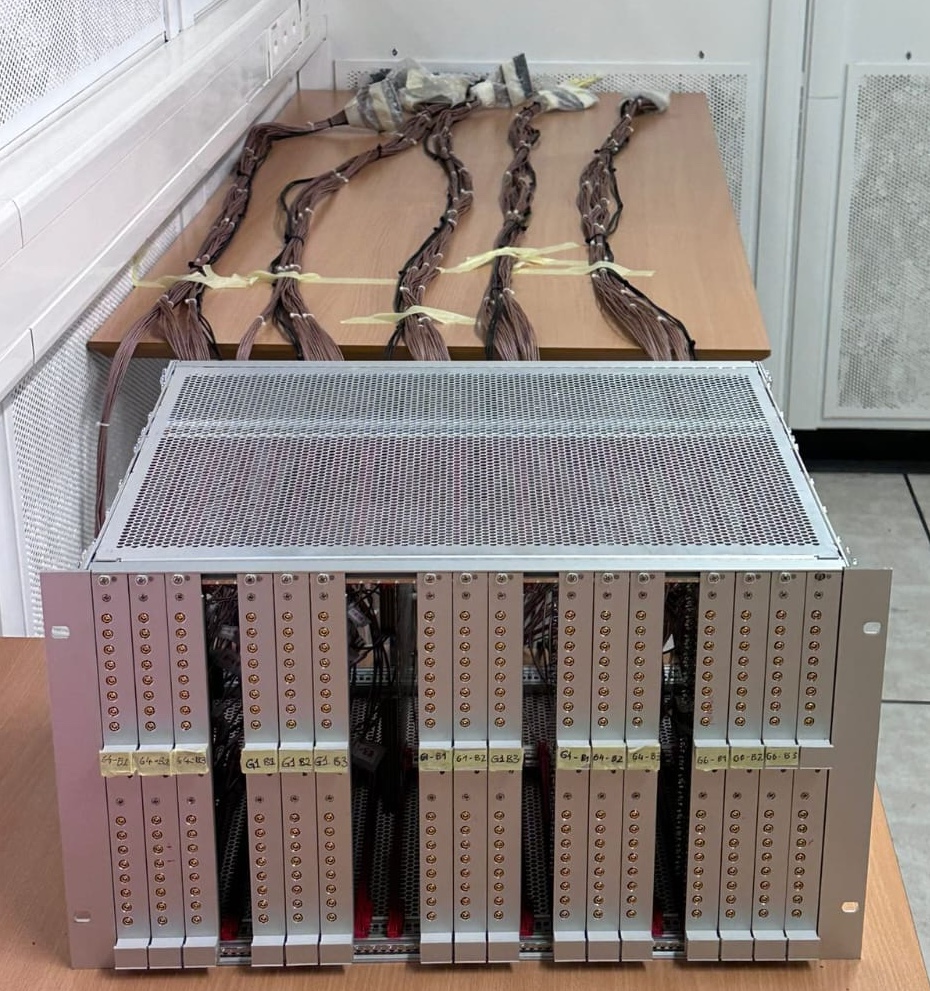}
    \caption{Custom crate hosting the 15 front-end boards used for the five-layer CRILIN calorimeter, with three boards assigned to each layer.}
    \label{fig:electronics_all}
\end{figure}

The 15 amplification boards, together with one spare board, are housed in a 19-inch, 6U, 84-HP, 355-mm-deep subrack, as shown in Fig.~\ref{fig:electronics_all}. Locating the amplification and bias-distribution electronics outside the calorimeter stack reduces their radiation exposure, minimizes the passive material within the active detector volume, and lowers power dissipation in the vicinity of the SiPMs.

\subsection{Monte Carlo simulation}
\label{sec:monte_carlo}
In order to provide a detailed understanding of the detector response observed during the beam-test campaign, a dedicated Monte Carlo framework has been developed. In this framework, the detector response is simulated using Geant4 with the QGSP\_BERT physics list for hadronic interactions and the PENELOPE physics list for the electromagnetic interactions \cite{geant4_1, geant4_2}. The range parameter of the electron and positron is set to 100 $\mu$m to accurately model the energy that is converted into Cherenkov radiation. 
The choice of Geant4 physics list affects the quantitative results, as detailed in Section \ref{app:physics_list} of the Appendix.

The detector geometry implemented in the framework reproduces the active medium consisting of PbF$_2$ crystals, the SiPMs, and the mechanical structure of the CRILIN calorimeter. The simulation also includes the front-end electronics, aluminum support elements, shielding structures, and Kapton services, providing a realistic description of the detector material budget.
Several parameters employed in the simulation are obtained from the test-beam data analysis. These are: the light yield, used to model the photo-statistics \ref{sec:photo_model}; the electronic noise and the thresholds \ref{sec:noise_model}. Simulation samples were produced using primary electron beams matching the energies explored during the test-beam campaign. A total of 10000 events were generated for each energy point, corresponding to incident electron energies of 10, 20, 34, 52, 74, 99, and 120 GeV. This configuration allows a direct comparison between simulated and experimental data over the full energy range investigated during the beam test.

\subsubsection{Cherenkov modeling and Cherenkov-energy scale}

\label{sec:cherenkov_mc_scale}

A full optical-photon simulation was not performed because of its high computational cost. Instead, the number of photoelectrons produced in each cell was estimated through a dedicated parametrized procedure. 

The Cherenkov response was modeled analytically using the Frank--Tamm relation: 
\begin{equation} \frac{\mathrm{d}^2 N_C}{\mathrm{d}x\,\mathrm{d}\lambda} = \frac{2\pi\alpha}{\lambda^2} \left( 1-\frac{1}{\beta^2 n^2(\lambda)} \right), \end{equation} 
where \(N_C\) is the number of emitted Cherenkov photons, \(x\) is the particle path length at each Geant4 step, \(\lambda\) is the photon wavelength, \(\alpha\) is the fine-structure constant, \(\beta\) is the particle velocity in units of the speed of light, and \(n(\lambda)\) is the wavelength-dependent refractive index of the radiator. Only charged-particle track segments satisfying the Cherenkov emission condition $\beta > \frac{1}{n(\lambda)}$
were considered in the calculation. Contributions from particles below the Cherenkov threshold were therefore discarded. 

The Cherenkov photon yield was evaluated in the wavelength range from 350 to 550~nm, corresponding to the region of highest effective sensitivity of the SiPM-based readout. In this wavelength range, the refractive index $n(\lambda)$ of the PbF$_2$ crystals is considered constant and equal to 1.82 \cite{pbf2}. The number of Cherenkov photons in the detectable wavelength range ($N_C$) was estimated from the charged-particle track lengths, velocities, and charges recorded by Geant4 by integrating the Frank--Tamm spectrum over the selected wavelength interval. 

The scatter plot of the Cherenkov photon yield and the deposited energy in the active material obtained from the electron simulation samples is shown in Fig. \ref{fig:energy_scale_MC}. It is evident that the relation between these two quantities is linear, and a linear fit is performed to describe the trend.

The Cherenkov-energy conversion factor obtained from the fit is 
$C/E_{\mathrm{dep}} = 30.31 \, \gamma/\mathrm{MeV}$.

\subsubsection{Photo-statistics modeling}
\label{sec:photo_model}

At this stage, the finite photoelectron statistics of the SiPM readout is modeled. The number of Cherenkov photons produced in each crystal is converted into an equivalent energy using the calibration constant \(C/E_{\mathrm{dep}}\) and subsequently into photoelectrons using the measured light yield \(LY\) discussed in Section~\ref{sec:light_yield}. Statistical fluctuations associated with the finite number of detected photoelectrons are then applied. The resulting photoelectron yield is converted back into an equivalent cell energy through the inverse light-yield calibration and used as the detector response in the subsequent stages of the simulation.

\begin{figure}[H]
    \centering
    \includegraphics[height=5.3cm]{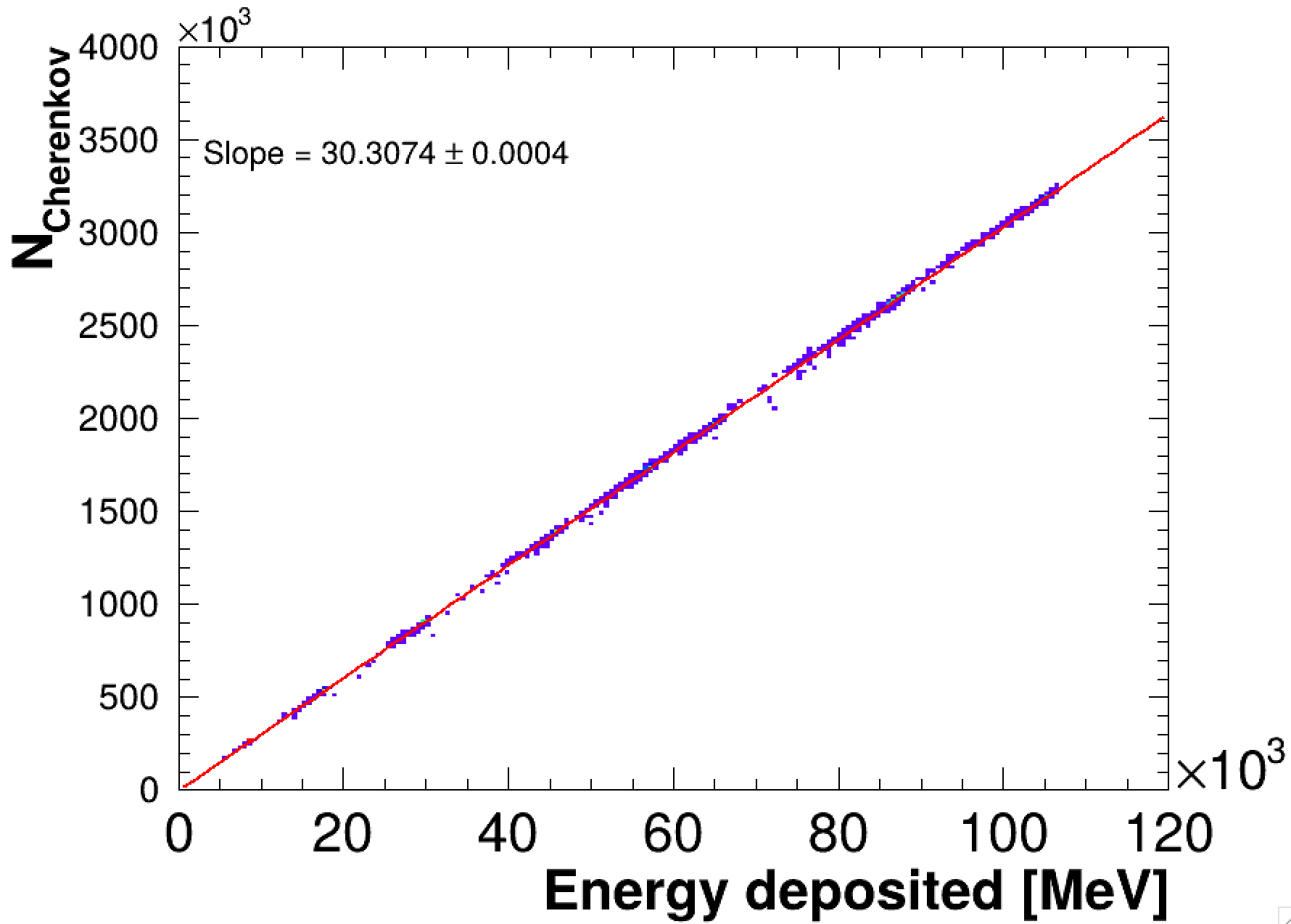}
        \includegraphics[height=5.6cm]{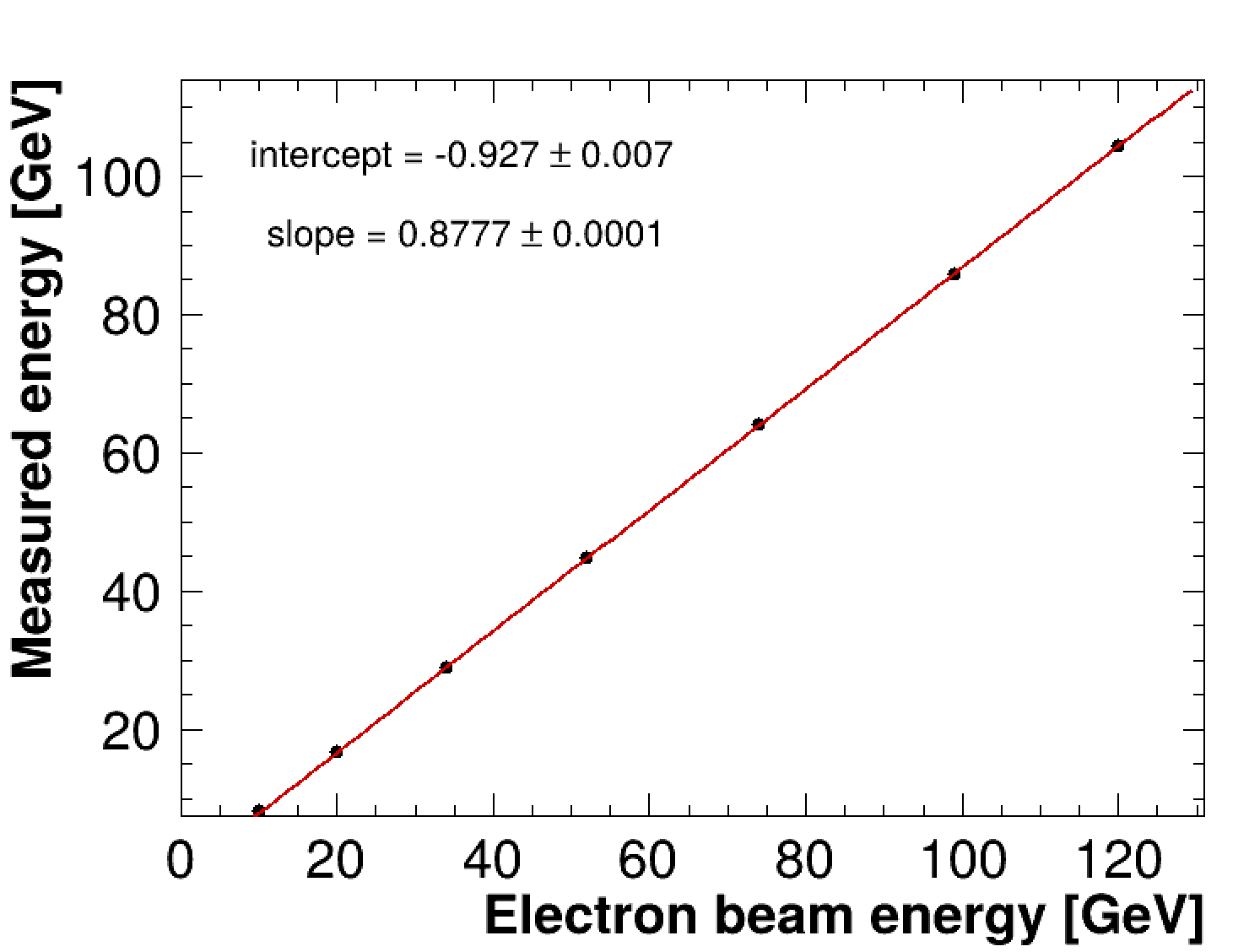}
    \caption{Left: Cherenkov photon yield in the wavelength range from 350 to 550~nm versus deposited energy in the active material obtained from the electron simulation samples, showing also the result of the linear fit. Right: fitted mean of the measured energy distribution as a function of the electron beam energy, obtained in the simulation.}
    \label{fig:energy_scale_MC}
\end{figure}

In order to verify the linearity of the calorimeter response in the simulation, the total measured energy distribution is determined for each electron energy value. In this check, a light yield of 0.6 is used. These distributions are fitted with Crystal Ball functions, and the fitted mean as a function of the electron beam energy is obtained and reported in Fig. \ref{fig:energy_scale_MC}. It can be seen that the relation between these two quantities is linear, as expected.

\subsubsection{Noise and thresholds implementation in simulation}
\label{sec:noise_model}

Instrumental effects related to electronic noise and channel thresholds were implemented in the simulation through a parametrized model. Both noise and thresholds were expressed in units of energy, ensuring consistency with the calorimeter energy reconstruction.

Electronic noise was added independently to the reconstructed energy of each cell according to

\begin{equation}
E^{\mathrm{cell}}_{\mathrm{noise}}
=
E^{\mathrm{cell}}
+
n_{cell},
\end{equation}
where $n_{cell}$ is taken from a random noise vector representing all channels $\vec{n} \in \mathbb{R}^{225}$, drawn for each event, following a multivariate Gaussian distribution with the noise covariance matrix derived in Section \ref{sec:noise} from pedestal events using data.
A channel-dependent threshold was then applied to emulate the offline threshold application on hits applied in data. The threshold was defined according to the configuration described in Section~\ref{sec:thresholds}.
Only cells with reconstructed energies above threshold were retained in the event reconstruction. The total reconstructed calorimetric energy was therefore computed as
\begin{equation}
E^{\mathrm{reco}}
=
\sum_i
E^{\mathrm{cell}}_{i,\mathrm{noise}}
\,
\Theta\!\left(
E^{\mathrm{cell}}_{i,\mathrm{noise}}
-
E^{\mathrm{cell}}_{i,\mathrm{thr}}
\right),
\end{equation}
where \(\Theta\) denotes the Heaviside step function. Equivalently, the sum extends only over cells whose noise-corrected energy exceeds the threshold assigned to that channel.


\section{Beam-test setup}
\label{sec:Beam-Test Setup}

\subsection{SPS beam line and run conditions}
The CRILIN prototype was tested at the CERN SPS H2 beam line, a secondary beam line in the SPS North Area providing mixed beams with selectable momentum and composition. The H2 line provides particle momenta ranging from approximately 10 GeV/c up to 400 GeV/c. The beam is delivered in slow-extracted spills of about 4.8~s duration, allowing stable data acquisition conditions. 
The data discussed in this work were collected with nominal electron beam momenta of 10, 20, 34, 52, 74, 99, and 120~GeV/$c$, as well as a data sample of muons at 150 GeV used for calibration purposes. During the test-beam campaign, no active temperature control or cooling system was installed on the CRILIN prototype, and the ambient temperature of the SPS H2 beam line was not monitored. The detector response is therefore sensitive to temperature variations of the facility, which may induce time-dependent drifts in the SiPM gain and in the overall calorimeter response.
For the beam settings up to 99~GeV/$c$, the beam composition included about 30\% pion contamination, according to private communication from the SPS H2 beam-line experts~\cite{SPSH2privateComm}, which is compatible with the one seen in data, increasing for higher energy settings.
No event cuts are applied in the analysis to remove this pion contamination, since the simple total energy sum, employed to measure energy linearity and resolution, already reduces the contamination to a negligible level.

The openings of the C6 and C12 collimators defined the beam momentum selection. For beam momenta below 150~GeV/$c$, where synchrotron-radiation effects along the line are still small, the relative momentum spread (RMS) can be approximated as
\begin{equation}
\frac{\Delta p}{p} = \frac{\sqrt{\left(\frac{C6}{2}\right)^2 + \left(\frac{C12}{2}\right)^2}}{27\sqrt{3}} \,[\%],
\label{eq:beam_momentum_spread}
\end{equation}
where $C6$ and $C12$ are the full openings of the corresponding collimators in mm. 
This parametrization does not include the contribution from scattering in the last section of the beam line before the detector, nor the additional degradation induced by synchrotron-radiation effects, which becomes more relevant (after 0.1\%) only above 150~GeV/$c$, where the maximum energy employed in the test is 120 GeV.

For the beam energies used in this analysis, the collimator settings and the corresponding estimated momentum spreads are summarized in Table~\ref{tab:beam_settings}.
\begin{table}[htbp]
    \centering
    \begin{tabular}{cccc}
        \hline
        $p_{\mathrm{nominal}}$ [GeV/$c$] & $C6$ [mm] & $C12$ [mm] & $\Delta p/p$ [\%] \\
        \hline
        10  & 16 & 16 & 0.24 \\
        20  & 16 & 16 & 0.24 \\
        34  & 16 & 16 & 0.24 \\
        52  & 20 & 16 & 0.27 \\
        74  & 30 & 16 & 0.36 \\
        99  & 50 & 16 & 0.56 \\
        120 & 70 & 16 & 0.77 \\
        \hline
    \end{tabular}
    \caption{H2 beam-line collimator settings used during the CRILIN beam test and the corresponding estimated relative momentum spreads.}
    \label{tab:beam_settings}
\end{table}

\subsection{Trigger and reference detectors}
\label{subsec:trigger_reference_detectors}

The trigger logic for the CRILIN beam test was based on the coincidence of two independent scintillation detectors placed upstream of the calorimeter. Two trigger configurations were used during data-taking and could be selected via software through the acquisition system, described in Sec.~\ref{subsec:DAQ}. In both cases, the trigger rates were individually limited at 150 Hz via a hold-off system, to avoid saturating the data-acquisition chain.

\begin{figure}[htbp!]
    \centering
    \includegraphics[height=5.3cm]{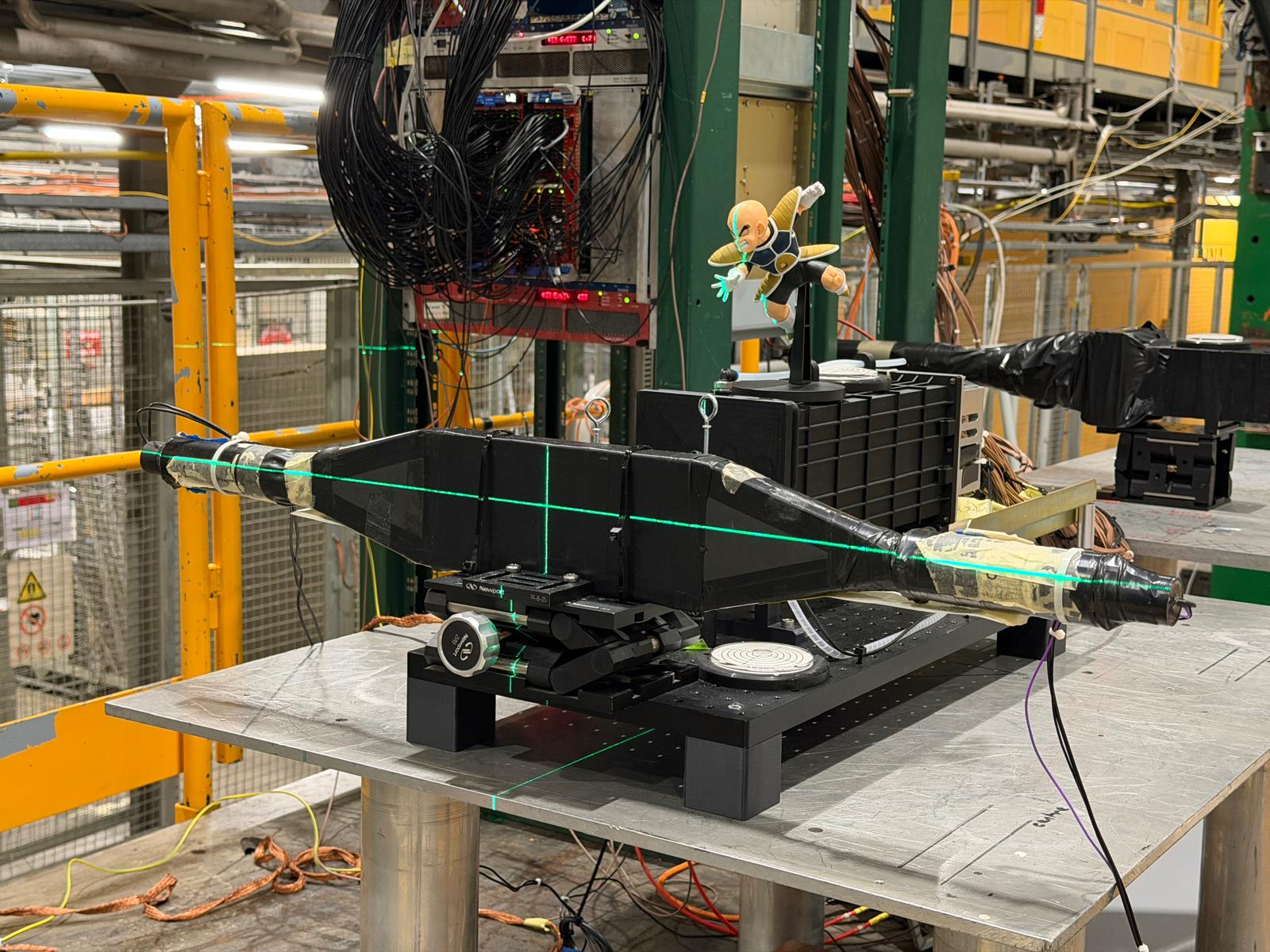}
        \includegraphics[height=5.3cm]{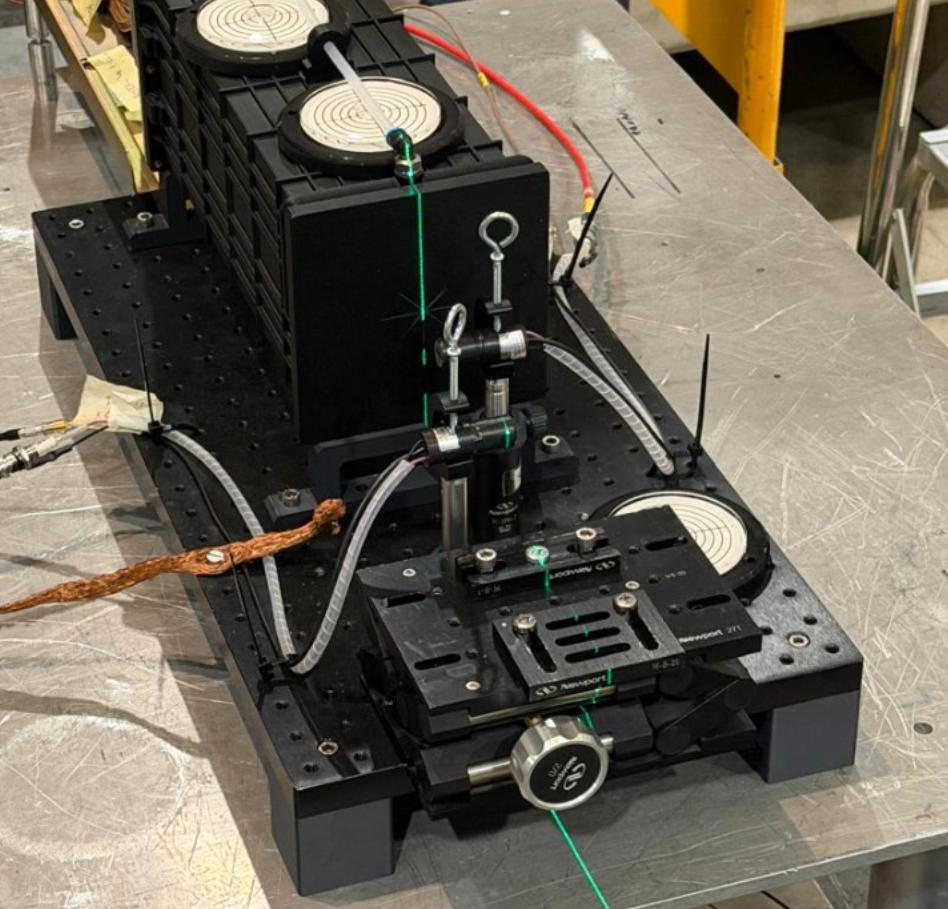}
    \caption{Large plastic scintillator used as one of the trigger detectors (left), pair of fast trigger scintillators mounted upstream of CRILIN on the optical alignment system (right)}
    \label{fig:trigger_paddle_cindy}
\end{figure}

The first trigger element was a large plastic scintillator, shown in Fig.~\ref{fig:trigger_paddle_cindy}, with an active area of approximately $20\times10~\mathrm{cm^2}$ and a thickness of 2~cm. The scintillation light was collected through light guides and read out by two photomultiplier tubes, one on each side. Due to its large acceptance, this detector was primarily used for fast monitoring of the beam position, using the charge centroid reconstructed in the calorimeter itself. As visible in Fig.~\ref{fig:trigger_paddle_cindy}, the scintillator was mounted upstream of the detector on a mechanical support aligned with the beam axis and integrated in the same optical-bench setup used for the rest of the beam instrumentation.

A second trigger configuration was based on two small fast plastic scintillators, each with dimensions of $5\times5\times5~\mathrm{mm^3}$, optically aligned with the geometrical center of the CRILIN module. This configuration, shown in Fig.~\ref{fig:trigger_paddle_cindy}, provided a more localized trigger acceptance and was mainly used to increase the trigger efficiency for particles crossing the central region of the calorimeter. The figure shows the compact mounting of the fast scintillators on a Thorlabs optical plate, along with the alignment system used to position them at the detector center.

\subsection{DAQ system and data quality monitoring}
\label{subsec:DAQ}
All 225 channels of CRILIN, as well as the trigger system and beam catcher, are connected to a set of 8 CAEN~V1742~\cite{CAENV1742, testV1742} switched-capacitor ADC boards based on the DRS4 chip, each featuring a 1024-cell switched-capacitor array (SCA) acting as a circular buffer. The usual dynamic range allowed for this digitizer is V$_{pp}$= 1~V, while larger signals, such as the central crystals per layer and the trigger system ones, were instead acquired using a modified version of the CAEN~V1742 that provides a V$ _ {pp} $ = 2~V instead.
 
\begin{figure}[htbp!]
    \centering
    \includegraphics[width=0.95\textwidth]{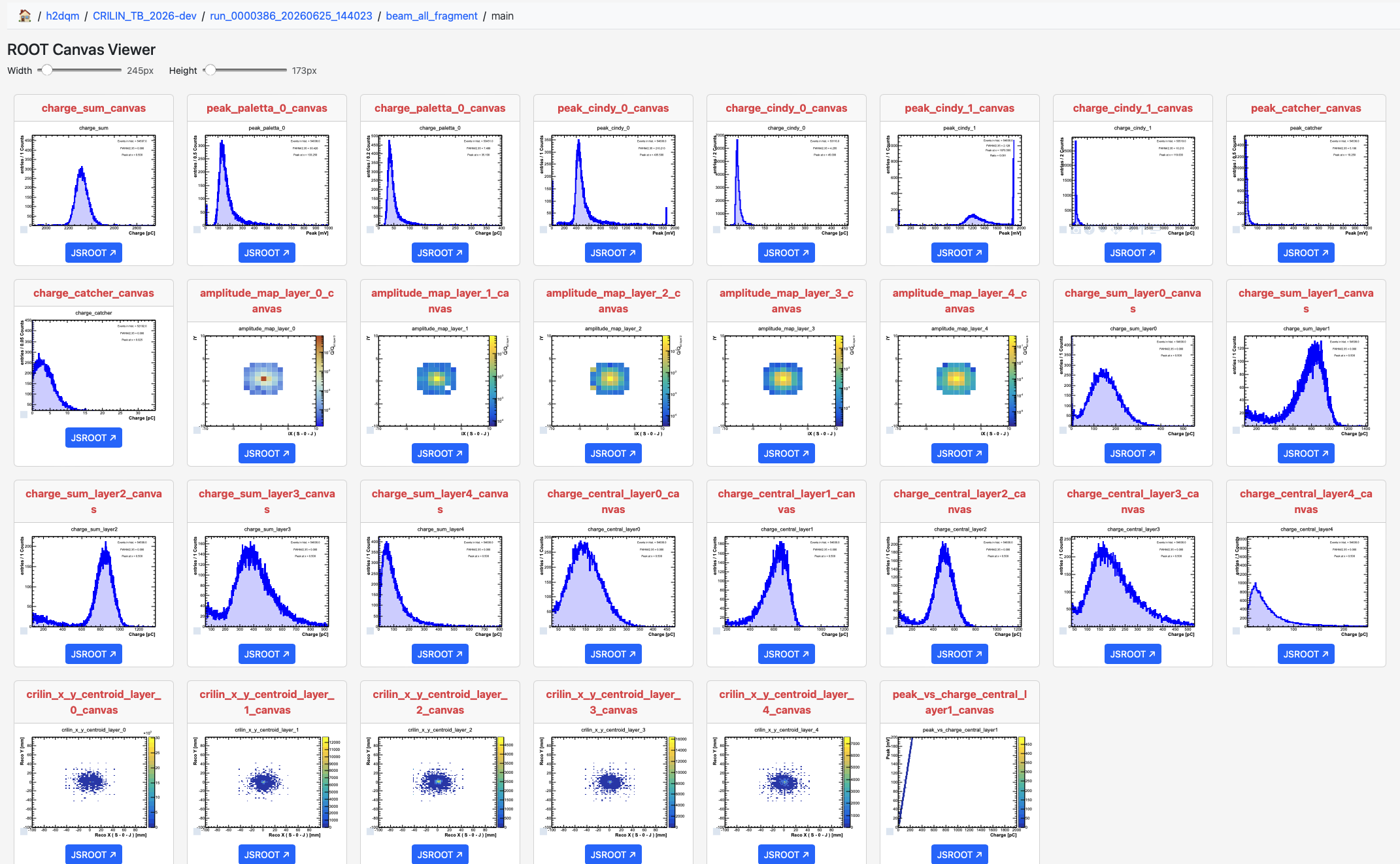}
    \caption{Web-based online monitoring page for a typical run with a set of plots of interest needed to have prompt feedback on the quality of the data collected.}
    \label{fig:online_monitor}
\end{figure}

Since all the detectors used are characterized by very fast pulses, the sampling frequency was set to 2.5~GS/s. 
A CAEN~A5818 PCIe board collects data from all ADC modules via optical links, capable of a maximum throughput of 80~MB/s.
All ADC boards interface with the Trigger and Timing System (TTS), which consists of a main trigger board and two trigger-distribution units providing NIM-standard signals to the digitizers \cite{testV1742}.
A random trigger, with a frequency set to 0.5~Hz, for pedestal evaluations and monitoring purposes active simultaneously with the trigger on physics events. It should be noted that the DRS4 chip needs data corrections due to the unavoidable differences in the chip construction process. One of interest for this application is the cell index offset correction. Indeed, the analog capacitors of the DRS4 chip might have small differences between each other due to the construction processes. According to the cell index where the start acquisition arrives, the same input signal can be reconstructed in different ways. This requires calibration to correct non-uniformity and offsets introduced by variations in capacitor properties and charge injection effects.
The digitizer has pre-loaded calibration done by CAEN in its flash memory, but a more fine tuning is needed to properly calibrate the data; hence, it is essential to keep track of the index of the first read cell to perform offline corrections that will be shown later on (~\ref{subsec:digi-corr}).
An online monitoring system was developed to allow for prompt feedback on the data acquired. On a beam spill basis, the DAQ provides a typical 300~MB ROOT file collecting all the waveforms from around 700 events and ready to be analyzed with its full statistics by a fast reconstruction tool running on a GPU with 12 GB of VRAM. This provided a first version of the events reconstruction output as well as quality control plots such as trigger and beam catcher signals, calorimeter centroids, channel status, and a first raw estimate of the energy response. An example of the web interface online monitor for a typical run is presented in Figure~\ref{fig:online_monitor}.


\section{Reconstruction, calibration and noise studies}
\label{sec:Calibration and Reconstruction}

\subsection{Signal reconstruction peak and template fit}
\label{sec:template}
As mentioned above, the CAEN V1742 switched-capacitor arrays used in this test beam provide full waveform digitization for all 225 channels of the calorimeter module, allowing for different and complementary approaches to energy reconstruction. As a first step, the signals are processed with a Gaussian low-pass filter with a cutoff frequency of 200~MHz. The corresponding transfer function is centered at zero frequency and exhibits an attenuation of approximately 6~dB at 200~MHz.
As a second step, depending on the channel, all waveforms are divided by the FEE gain and by the ADC/voltage conversion, based on the digitizer employed (1 or 2 V dynamic range), to provide sample values in units of mV measured at the input stage of the FEE boards, and, consequently, at the SiPMs output.
The quantities derived from these converted waveforms, evaluated in terms of $V_{iFEE}$, i.e. the voltage and the input (i) of the front-end (FEE), are therefore proportional to the light collected by the SiPMs and to the energy deposited in the crystals.
For the digitizers at 1V, the conversion factors from ADC counts to $V_{iFEE}$ are about, respectively, 0.0597 mV/ADC, 0.244 mV/ADC, 0.244 mV/ADC, 0.0597 mV/ADC, 0.0402 mV/ADC for the first, second, third, fourth, and fifth layers, at gains 4, 1, 1, 4, 6. For the digitizer at 2V, these conversion factors are doubled.

The filtered waveforms are used in the reconstruction phase to obtain a first estimate of the pulse peak amplitude and related time. 
These quantities are subsequently employed to define the signal and baseline integration windows, defined using fixed offsets relative to the peak positions, from which a raw charge estimate is derived.
For each event and channel, the peak positions correspond to the index of the maximum sample, with the constraint that all peak positions should lie in a 10-ns-wide window centered around the peak position of the channel with the highest peak.
For pedestal events, acquired using the dedicated random trigger events, the signal and baseline windows are fixed to the region where a physics pulse would be expected, since no genuine signal is present and a peak-search procedure would yield an arbitrary result.

\begin{figure}[htbp!]
    \centering
    \includegraphics[width=0.6\linewidth]{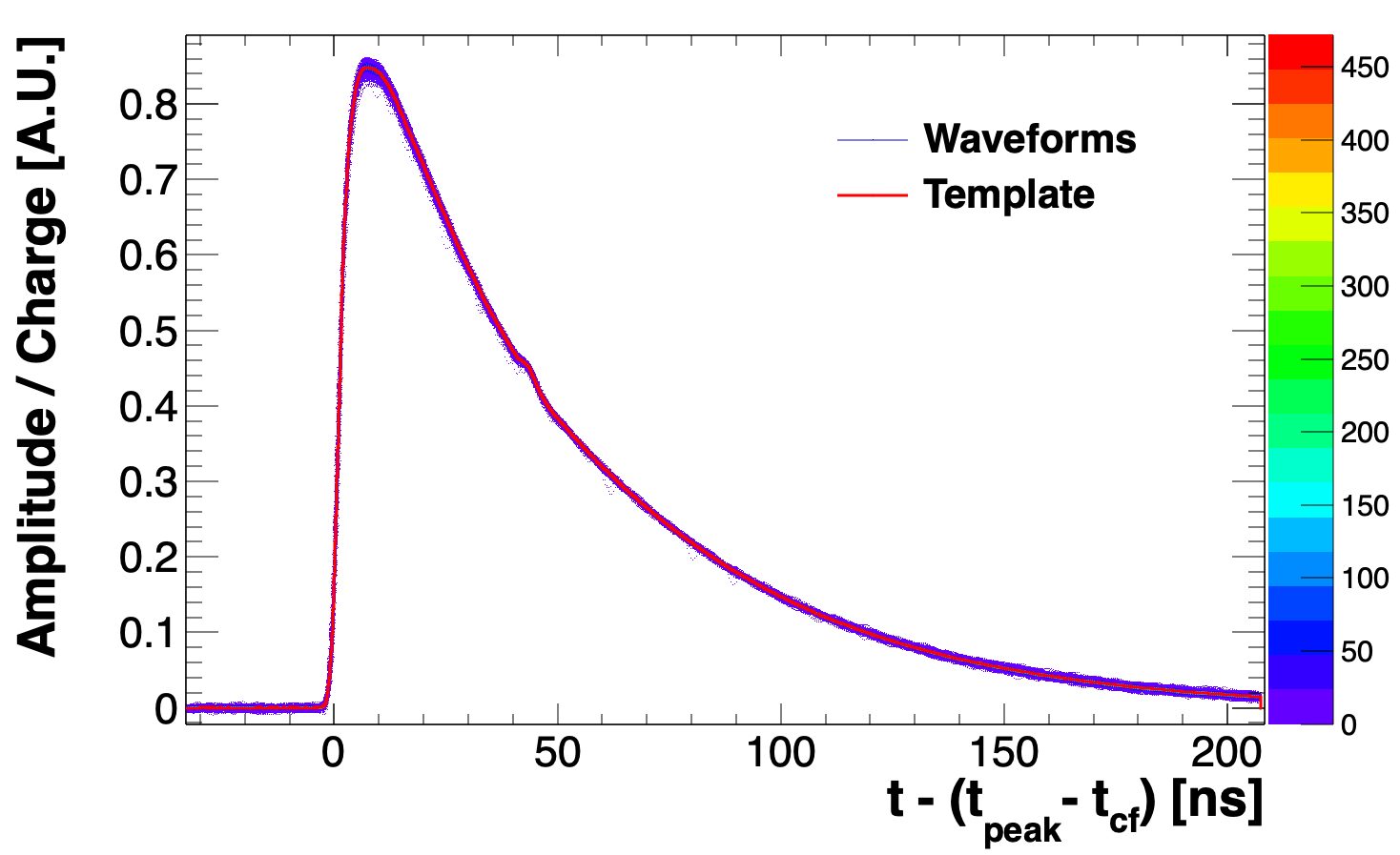}
    \caption{Example of a set of waveforms normalized to their charge and shifted according to the difference between the channel peak position in time and its constant fraction time, with the template shape superimposed in red.}
    \label{fig:template_over_wf}
\end{figure}

No threshold is applied at any stage of the reconstruction.
An estimate of the constant-fraction time is also computed by performing a polynomial interpolation of the waveform rising edge and determining the time at which the interpolated pulse reaches a threshold corresponding to 11\% of the peak amplitude. This threshold was optimized to minimize the effect on the time resolution.
Pedestal quantities are evaluated using the dedicated random-trigger events recorded through the reserved trigger mask. In this case, the signal and baseline windows are fixed to the region where a physics pulse would be expected, since no genuine signal is present and a peak-search procedure would yield an arbitrary result.
For the actual observables used in this analysis, an alternative waveform reconstruction method was applied based on a linearized template fit, exploiting the information collected up to now.
The measured signal $w(t)$ is modeled as a scaled and time-shifted version of a reference pulse shape $P(t)$,
\begin{equation}
w(t)=A\,P(t-\Delta t),
\end{equation}
where $A$ and $\Delta t$ represent the pulse amplitude and time offset, respectively. To obtain a computationally efficient solution, the template is expanded to first order around the current estimate of the time offset,
\begin{equation}
P(t-\Delta t)\simeq P(t)-\Delta t\,P'(t),
\end{equation}
thereby transforming the problem into a linear least-squares fit involving the template and its derivative, with amplitude and time as parameters of interest, solved by matrix inversion. Since the first-order approximation is valid only for small time shifts, the procedure is iterated four times, by updating the template position at each step. The method is equivalent to a Gauss--Newton minimization of the original nonlinear template-fitting problem and provides simultaneous and precise estimates of the pulse amplitude and arrival time, and is applied to all the calorimeter channels on an event-by-event basis, in a vectorized fashion, where three-dimensional tensors (event, channel, sample) are processed in a single step. 
The template shape used in the fit is constructed from data by first normalizing each waveform to its integrated charge and aligning them in time using the constant-fraction time. The resulting ensemble of waveforms is then examined in the two-dimensional space of normalized amplitude versus time offset. For each time-offset slice, a Gaussian fit is performed along the amplitude axis, and the fitted mean is taken as the template value at that time offset, as illustrated in Figure~\ref{fig:template_over_wf}. Finally, the template is re-normalized such that its peak value is unity.
A single template shape, derived from 99~GeV electron data using the seed channel of the second layer, is employed for the reconstruction of all channels and beam energies.

\subsection{Digitizers starting cell index correction}
\label{subsec:digi-corr}
To correct the DRS4 capacitor offsets, a two-step procedure is applied. The first step uses pedestal events, namely events acquired with a random trigger, to derive a channel-by-channel correction. The second step removes a residual run-dependent offset by using the peak of the total reconstructed energy as a standard candle.

Both corrections are applied as a function of the index of the first read cell in a given DRS4 chip of one digitizer, hereafter referred to as \textit{StartCell}. It was verified that the \textit{StartCell} values of all DRS4 chips in the acquisition are fully linearly correlated. In each run, the possible \textit{StartCell} values are separated by approximately 20 counts; therefore, \textit{StartCell} is binned into 50 slices over the range 0--1000.

\begin{figure}[htbp!]
    \centering
    \includegraphics[height=4.5cm]{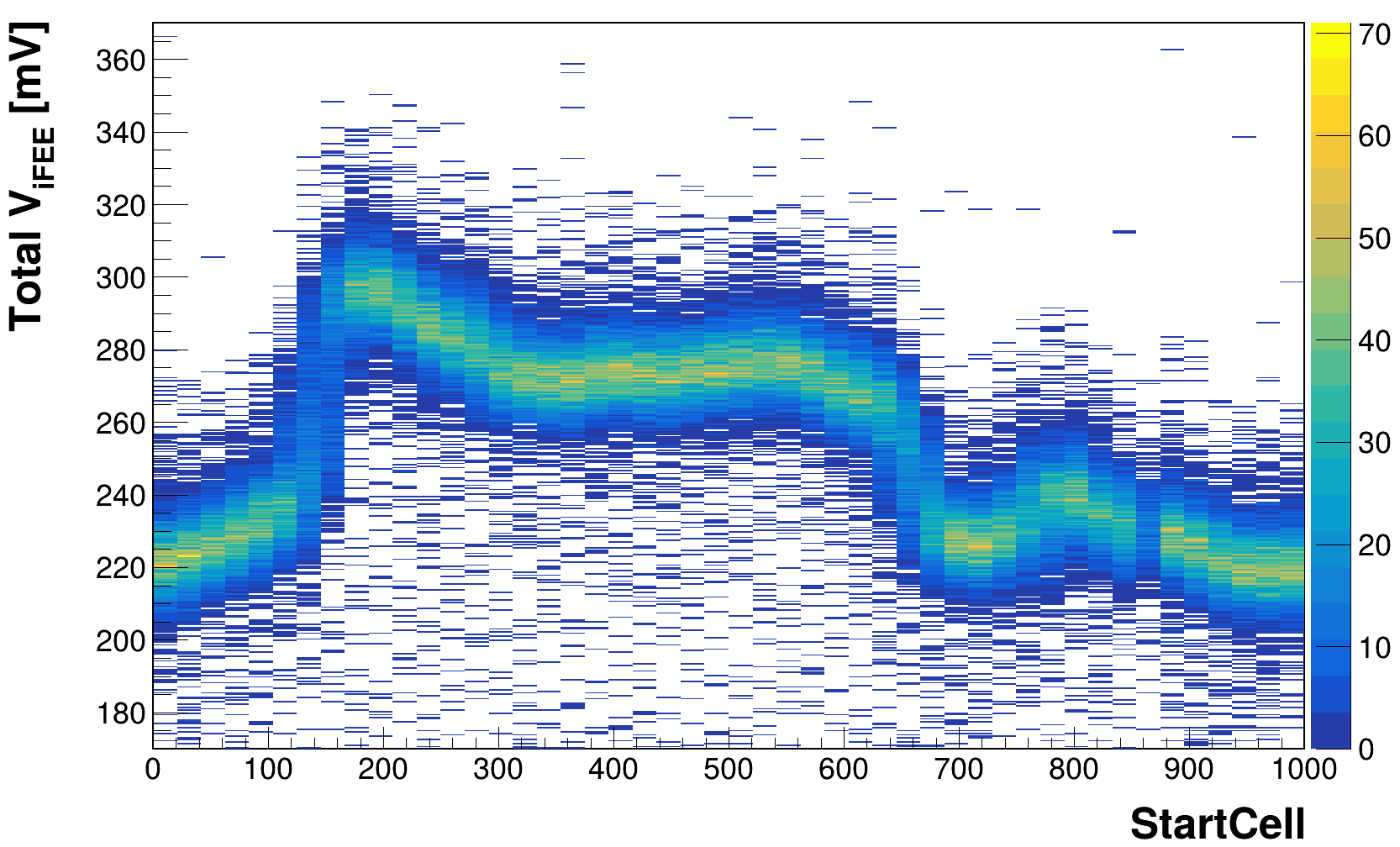}
    \includegraphics[height=4.5cm]{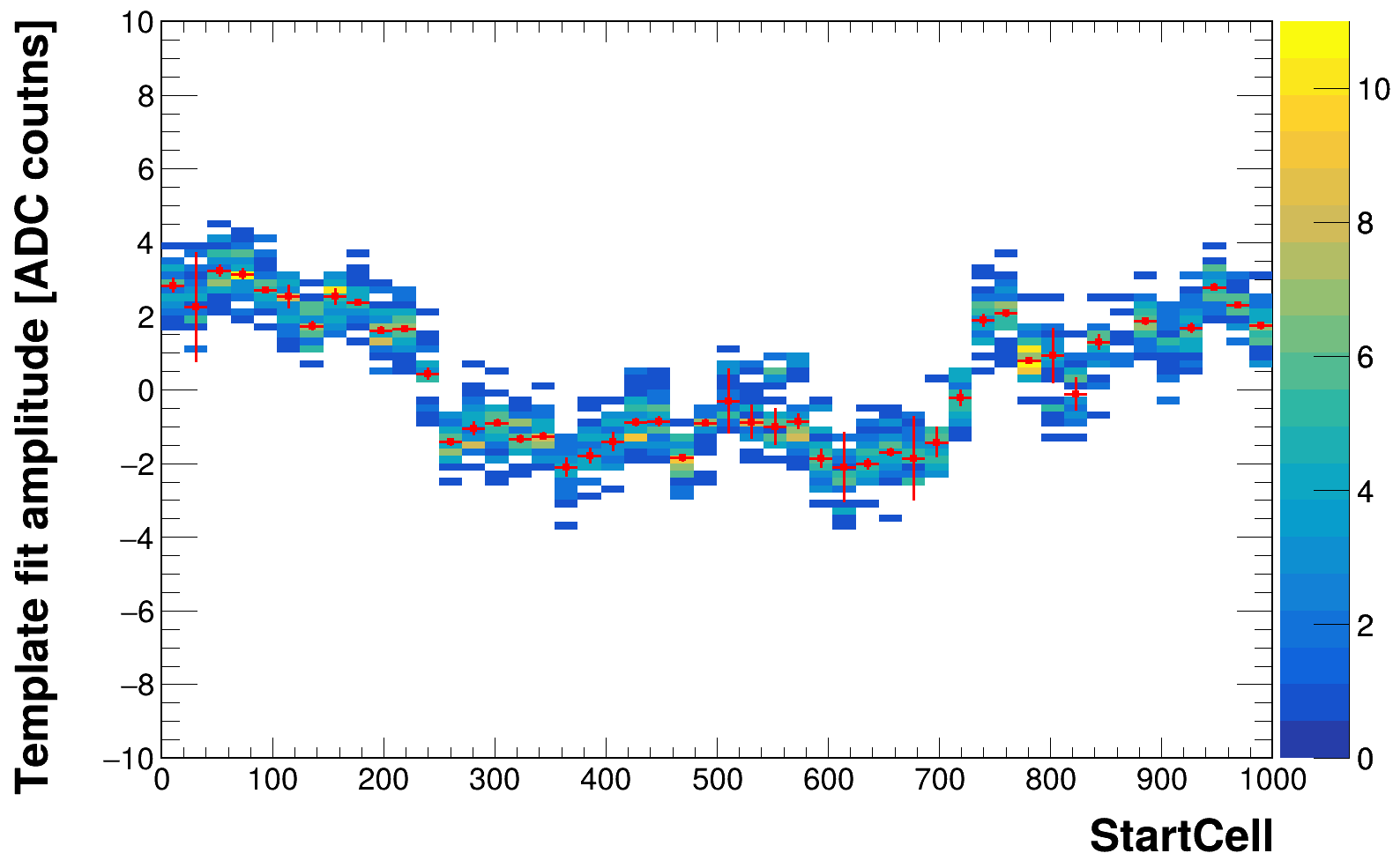}
    \caption{Left: two-dimensional distribution of the uncalibrated total amplitude of the 10~GeV run, obtained by summing the template-fit amplitudes $V_{iFEE}$, as a function of \textit{StartCell}.
    Right: two-dimensional distribution of the template-fit amplitude, in ADC counts, versus \textit{StartCell}, for pedestal events in a representative channel and run. The red points indicate the peak positions of Gaussian fits to the projected amplitude distributions in each \textit{StartCell} bin.}
    \label{fig:ped_corr}
\end{figure}

To illustrate the need for these corrections, Fig.~\ref{fig:ped_corr}-left shows the total energy of the 10~GeV run, obtained by summing the template-fit amplitudes $V_{iFEE}$ and dividing each channel contribution by its MIP calibration coefficient, as a function of \textit{StartCell}. The reconstructed energy exhibits a large variation, of the order of 25\%, as a function of \textit{StartCell}, demonstrating the necessity of the correction procedure. Two regions around \textit{StartCell} $\sim$150 and $\sim$270 show a particularly steep dependence and an especially large spread; events in these regions are excluded from the energy-resolution analysis.

In the first correction step, the two-dimensional distribution of the template-fit amplitude, expressed in ADC counts, versus \textit{StartCell} is used to derive a channel-by-channel correction from pedestal events. As shown in Fig.~\ref{fig:ped_corr}-right for a representative channel and run, the pedestal correction corresponds to the sequence of Gaussian peak positions extracted from the projected histograms in each \textit{StartCell} bin. These corrections span a range of about 10 ADC counts per channel.

Since the template-fit procedure is linear and additive, the signal amplitude can be decomposed into a noise contribution and a physical-signal contribution. The \textit{StartCell} dependence affects only the noise term and can therefore be corrected using the pedestal-based offset measured in the random-trigger sample. The same correction is then applied directly to signal events.

\subsection{MIP calibration}
\label{subsec:mip_calibration}

The relative calibration of the 225 calorimetric channels was obtained using minimum-ionizing particles (MIPs). Two dedicated overnight runs, each with a duration of approximately 6 hours, were collected with a 150~GeV/$c$ muon beam. The large plastic-scintillator paddle described in Sec.~\ref{subsec:trigger_reference_detectors} was used in the trigger logic. Before data taking, the beam profile was verified to ensure a sufficiently uniform illumination of the entire calorimeter surface.

Through-going MIP candidates were selected by requiring energy deposits compatible with a single traversing particle in the five crystals sharing the same transverse $(x,y)$ position across the five longitudinal layers. This selection suppresses events with shower development and identifies muons crossing the calorimeter approximately parallel to its longitudinal axis.

For each channel, the spectrum of the selected events was fitted with a Langauss function, defined as the convolution of a Landau distribution with a Gaussian function,
\begin{equation}
f(x) = \mathrm{Landau}(x) \otimes \mathrm{Gauss}(x).
\label{eq:langau_function}
\end{equation}
The Landau component describes the intrinsic fluctuations of the energy loss of a minimum-ionizing particle in the detector material, whereas the Gaussian component accounts primarily for detector resolution and electronic-noise effects. The most probable value (MPV) extracted from the Langauss fit was used as the characteristic MIP response of each calorimeter channel.

\begin{figure}[htbp!]
    \centering
    \includegraphics[width=0.65\textwidth]{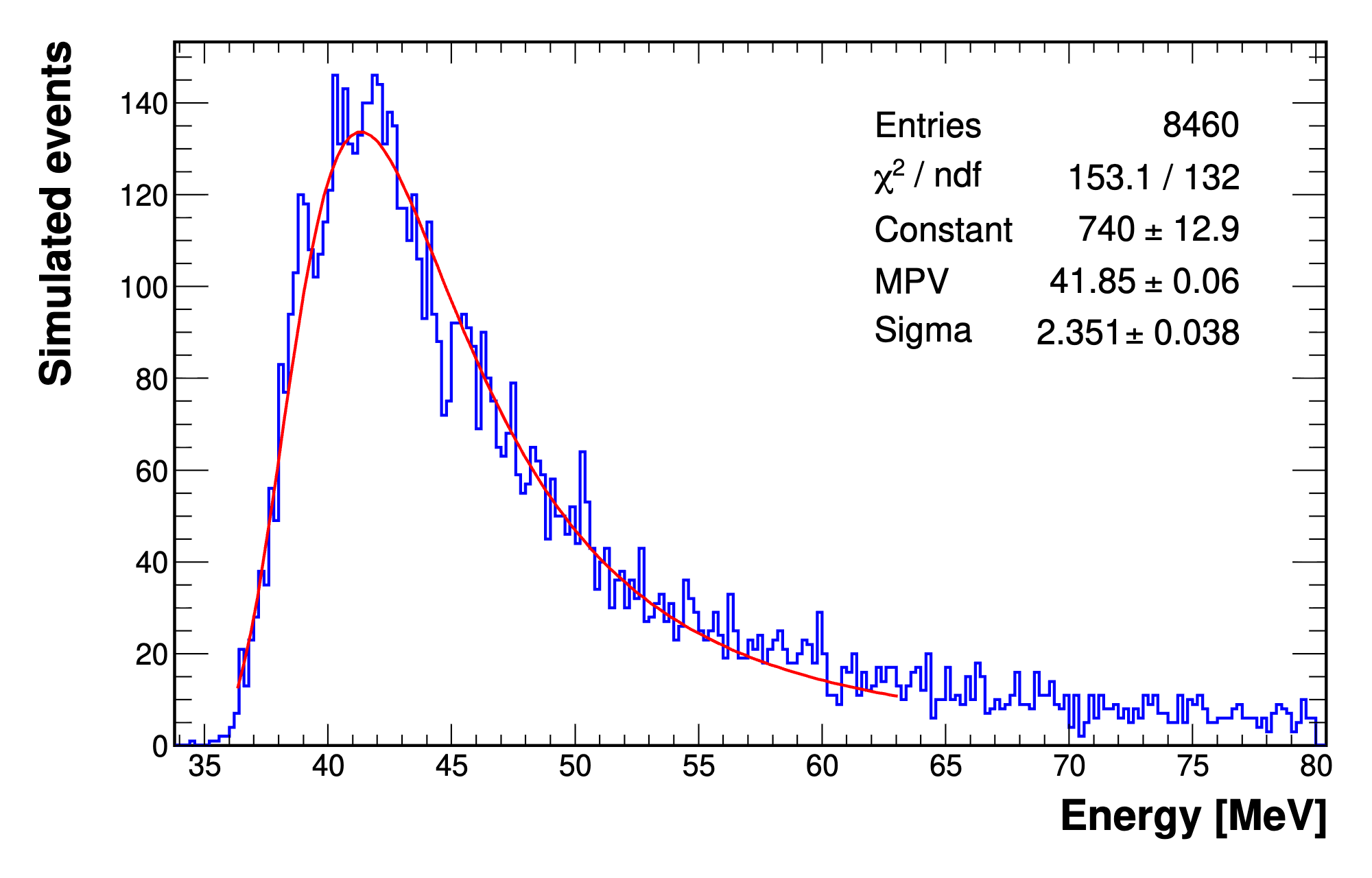}
    \caption{Monte Carlo distribution from through-going 150~GeV/$c$ muons after applying the same selection criteria used for the data. The response is rescaled to the electromagnetic energy scale according to the procedure described in Sec.~\ref{sec:cherenkov_calibration}. No experimental effects (noise, light yield, thresholds) are applied. The Langauss-fit MPV defines the reference MIP energy deposit used for the channel calibration.}
    \label{fig:mip_mpv_mc}
\end{figure}

The same through-going MIP selection applied to the data was implemented in the simulation. The simulated distribution of the rescaled energy deposits is shown in Fig.~\ref{fig:mip_mpv_mc}; its MPV $(E_{dep})^{\mu,G4} \approx 41.8$~MeV, obtained from a Langauss fit, defines the reference energy deposit associated with a traversing MIP and is used to convert the measured channel responses into a common energy scale.

For each channel, the calibration constant was determined from the ratio between the simulated reference MPV and the corresponding measured MPV. To maximize the signal-to-noise ratio while preserving the dynamic range of the CAEN flash ADCs, different analog gains were applied to the five longitudinal layers. The SiPMs were operated at a common nominal bias voltage, corresponding to a gain of $3.6\times10^{5}$, while the electronic gains of layers 1 to 5 were set to 4, 1, 1, 4, and 6, respectively. This approach allows the response of each layer to be optimized without changing the SiPM operating point.

Before the extraction of the calibration constants, the acquired waveforms were rescaled by the corresponding layer-dependent electronic gain. The MPVs obtained from the Langauss fits are evaluated in terms of $V_{iFEE}$. 

\begin{figure}[htbp!]
    \centering
    \includegraphics[width=0.45\textwidth]{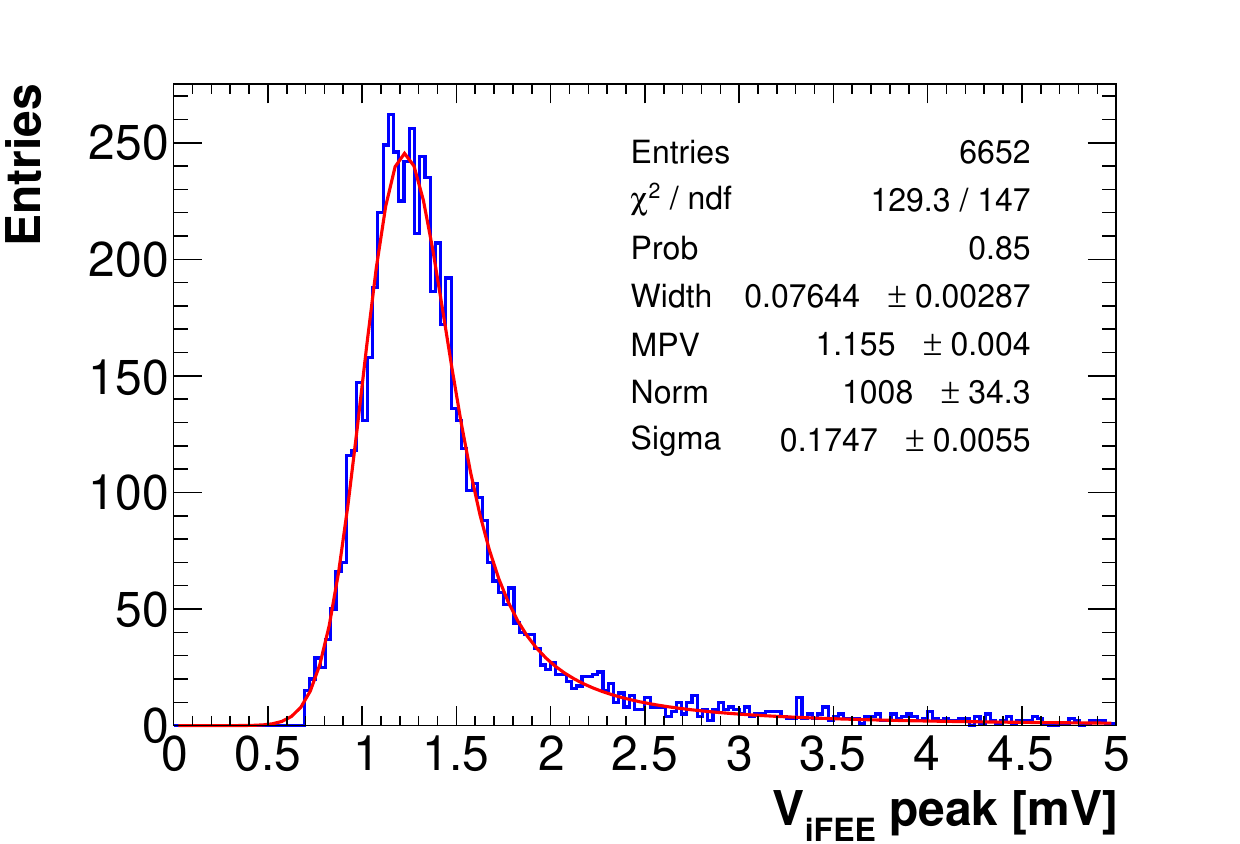}
    \includegraphics[width=0.45\textwidth]{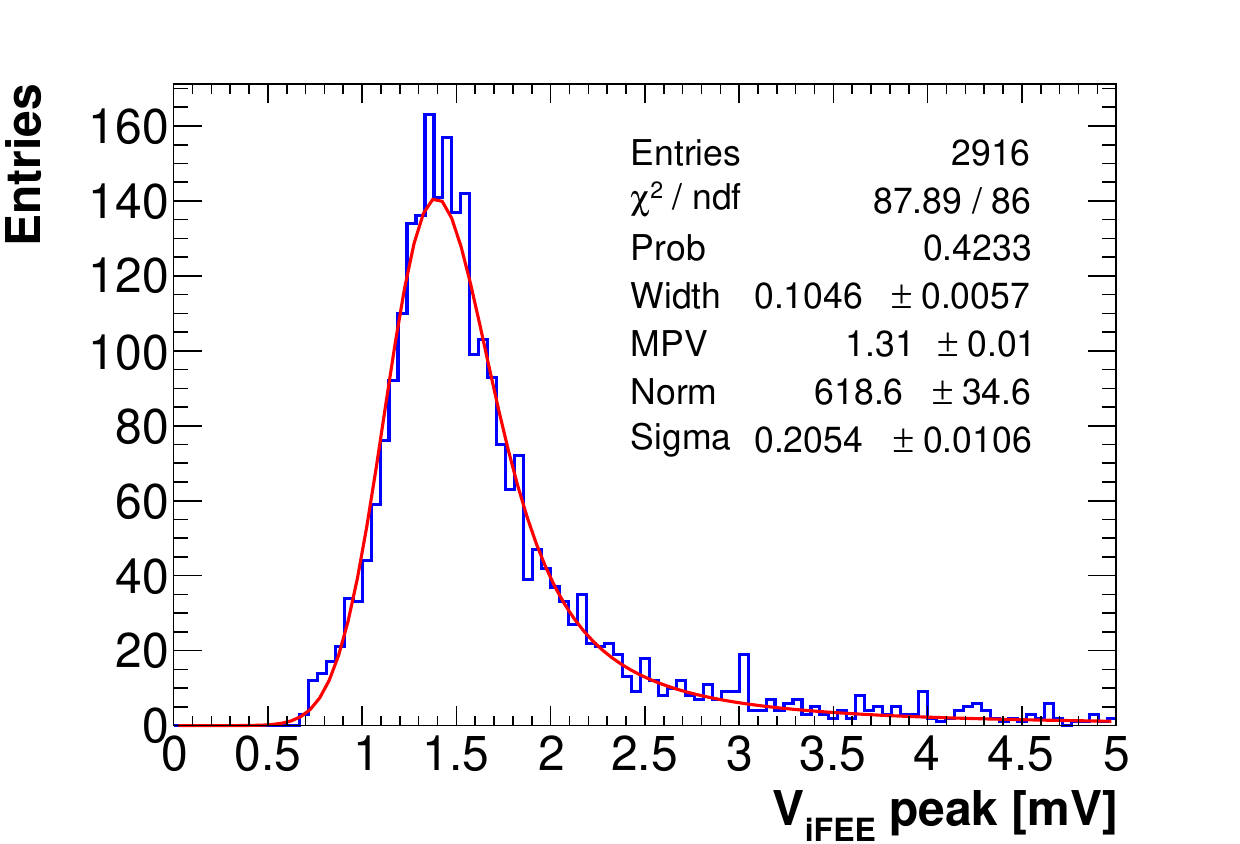}
    \caption{Representative Langauss fits to the MIP spectra measured in different calorimeter channels (left: channel 48, right: channel 203) after the through-going muon selection.}
    \label{fig:mip_langau_examples}
\end{figure}

To ensure a stable and unbiased extraction of the MPV for all channels, the fit was performed through an automated iterative procedure. A preliminary Gaussian fit was first applied near the spectrum maximum to estimate the peak position and its RMS. The full width at half maximum (FWHM) was then determined and used to initialize the Langauss fit parameters: the Landau width and the Gaussian resolution were initialized to $\mathrm{FWHM}/2$, the Gaussian peak position was used as the initial MPV estimate, and the histogram integral was used as the initial normalization factor. If the resulting fit did not describe the distribution adequately, the procedure was repeated with updated initial conditions, up to a maximum of ten iterations. This strategy ensured stable convergence for all calibrated channels while minimizing sensitivity to the initial parameter choice. Examples of the resulting Langauss fits for representative calorimeter channels are shown in Fig.~\ref{fig:mip_langau_examples}.

\begin{figure}[H]
    \centering
    \includegraphics[width=0.6\textwidth]{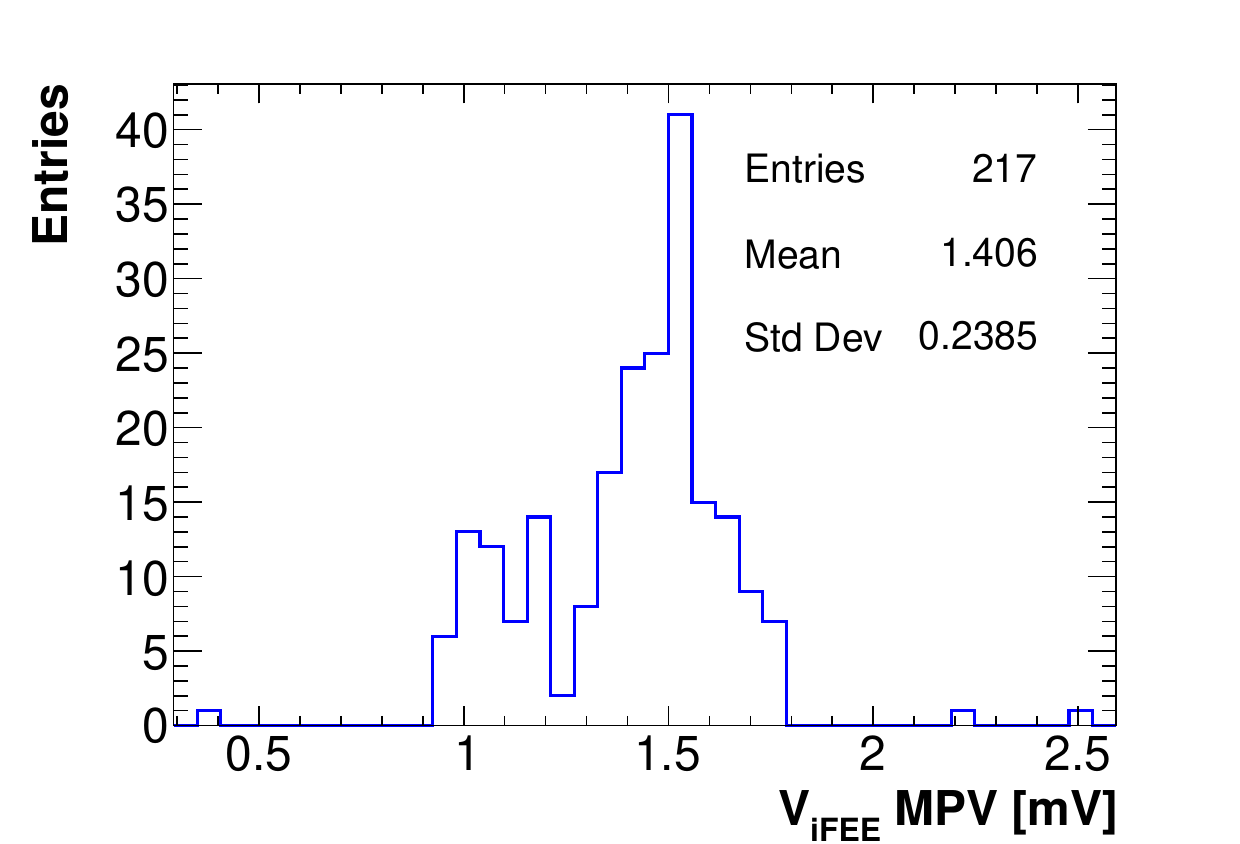}
\caption{Distribution of the  MPVs values of $V_{iFEE}$ extracted from Langauss fits to the 217 calorimeter channels with valid fits. The eight peripheral channels with insufficient MIP statistics were assigned the mean calibration value.}
    \label{fig:mip_mpv_distribution}
\end{figure}

The distribution of the rescaled MPVs for the 217 channels with valid fits is shown in Fig.~\ref{fig:mip_mpv_distribution}. 
It has a mean value of 1.406~mV and an RMS of 0.239~mV; the residual spread reflects channel-to-channel variations in the uncalibrated response, including differences in crystal light yield, optical coupling, SiPM response, and front-end electronic response. The remaining eight channels could not be calibrated individually because of the incomplete geometrical coverage of the large trigger paddle, resulting in insufficient statistics after the through-going MIP selection. These channels are all located at the calorimeter periphery and therefore do not affect the energy-resolution measurement. Their calibration constants were assigned using the mean MPV value of the calibrated channels.


\subsection{Muon and electron energy scale} 
\label{sec:cherenkov_calibration}
\begin{figure}[htbp!]
    \centering
    \includegraphics[width=0.45\linewidth]{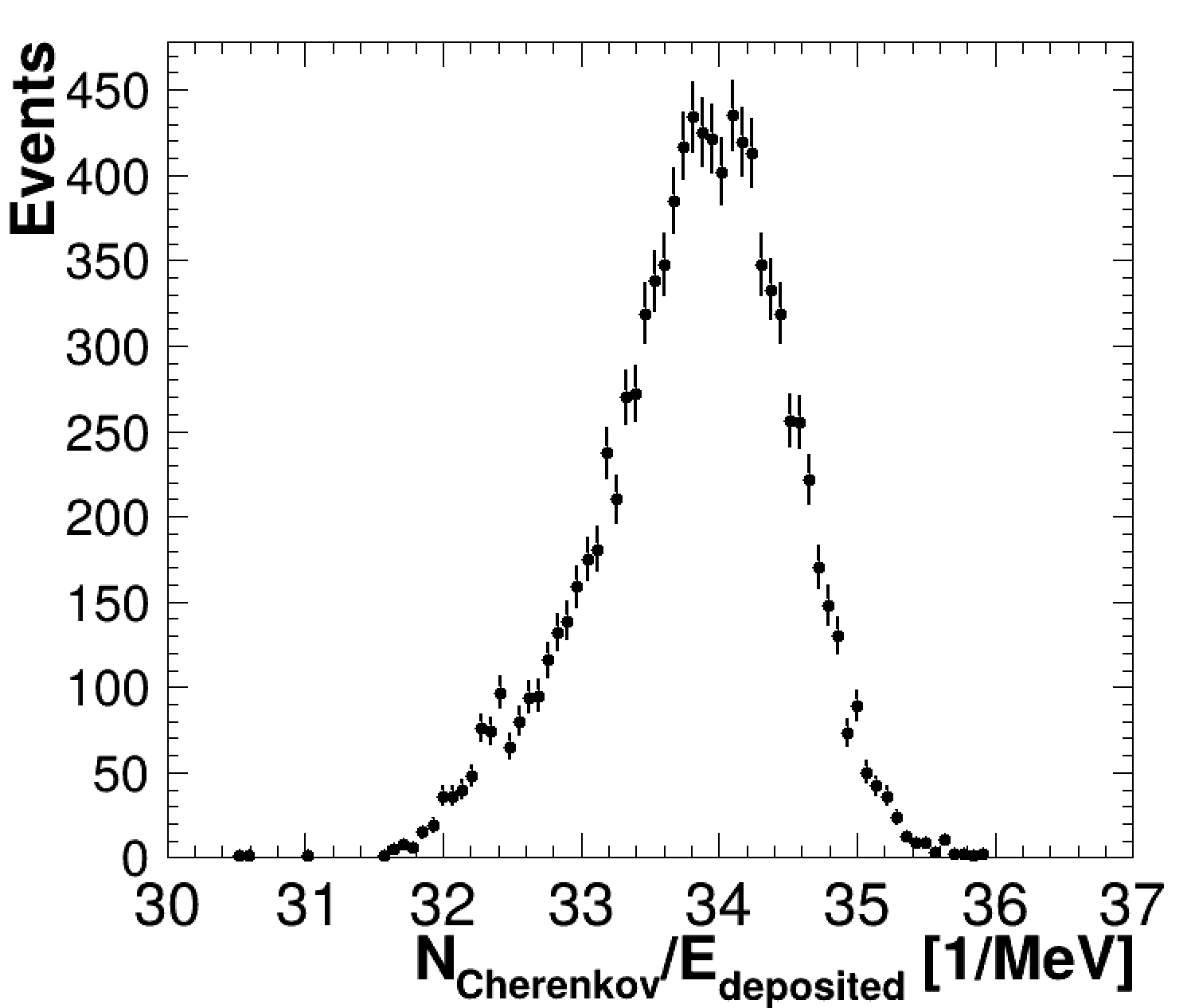}
    \caption{Distribution of the Cherenkov response, expressed as the number of photons produced per unit deposited energy, for through-going 150~GeV muons in the simulated sample.}
    \label{fig:muons_cherenkov}
\end{figure}

The conversion between the Cherenkov response expressed as the number of photons produced per 1~MeV of deposited energy for muons and for multi-GeV electromagnetic showers was evaluated with Monte Carlo simulations. A difference between the two responses is expected, as discussed in
\cite{RubenPaper}, due to the very different fractions of relativistic charged particles produced in the two cases; the present study updates that result using a more accurate physics list and detector description. For electron showers, the value derived in Sec.~\ref{sec:cherenkov_mc_scale} is used, namely $(C/E_{\mathrm{dep}})_e = 30.3~\gamma/\mathrm{MeV}$, while for muons the procedure described in \cite{RubenPaper} was repeated using a simulated sample of 150~GeV primary muons, corresponding to the energy of the beam used in the test-beam campaign for the MIP calibration runs. The conversion factor for muons is obtained as the maximum of the distribution shown in Fig.~\ref{fig:muons_cherenkov}, $(C/E_{\mathrm{dep}})^\mu = 33.9~\gamma/\mathrm{MeV}$. The resulting ratio between the muon and electron responses is therefore
\[
\frac{(C/E_{\mathrm{dep}})^\mu}{(C/E_{\mathrm{dep}})^e} = 1.12.
\]

To convert the calibrated values $V_{iFEE}^e/V_{iFEE}^\mu$ for electron events into deposited energy, the electron scale can be written as
\begin{equation}
    \frac{E_{\mathrm{dep}}^e}{V_{iFEE}^e}
    =
    \left(\frac{E_{\mathrm{dep}}}{C}\right)^e
    \cdot
    \frac{C}{V_{iFEE}} \, .
\end{equation}

Here, $C$ denotes the number of Cherenkov photons produced, and $V_{iFEE}^\mu$ is the MPV measured with muon events during the calibration for each channel. The factor $C/V_{iFEE}$ is independent of the particle, and can be expressed as
\begin{equation}
    \frac{C}{V_{iFEE}}
    =
    \left(\frac{C}{E_{\mathrm{dep}}}\right)^{\mu,\mathrm{G4}}
    \cdot
    \frac{E_{\mathrm{dep}}^{\mu,\mathrm{G4}}}{V_{iFEE}^\mu} \, ,
\end{equation}
therefore:
\begin{equation}
    \left(\frac{E_{\mathrm{dep}}}{V_{iFEE}}\right)^e
    =
    \frac{(C/E_{\mathrm{dep}})^\mu}{(C/E_{\mathrm{dep}})^e}
    \cdot
    \frac{E_{\mathrm{dep}}^{\mu,\mathrm{G4}}}{V_{iFEE}^\mu} \, .
\end{equation}
Using $(E_{\mathrm{dep}})^{\mu,\mathrm{G4}} \approx 41.8 $~MeV and
$(C/E_{\mathrm{dep}})^\mu/(C/E_{\mathrm{dep}})^e \approx 1.12$, the conversion factor for electron events becomes
\begin{equation}
   \frac{E_{\mathrm{dep}^e}}{V_{iFEE}^e/V_{iFEE}^\mu}
    \approx
    1.12 \times 41.8~\mathrm{MeV}
    \approx
    46.8~\mathrm{MeV}.
\end{equation}

Combining the conversion factors from mV to ADC counts for the different layers and digitizers in Sec.~\ref{sec:template}, and using the average value $\langle V_{iFEE} \rangle^\mu \approx 1.4$~mV, the corresponding MeV/ADC conversion factors are obtained, as reported in Table~\ref{tab:MeVADCconv}.

\begin{table}[htbp!]
    \centering
    \begin{tabular}{lcc}
        \hline
        Layer & Conversion (MeV/ADC), 1~V digitizers & Conversion (MeV/ADC), 2~V digitizer \\
        \hline
        Layer 1 (Gain 4) & 1.98  & 3.96  \\
        Layer 2 (Gain 1) & 8.12 & 16.24 \\
        Layer 3 (Gain 1) &  8.12 & 16.24 \\
        Layer 4 (Gain 4) & 1.98  & 3.96 \\
        Layer 5 (Gain 6) & 1.33  & 2.66  \\
        \hline 
    \end{tabular}
    \caption{Conversion factors from ADC counts to deposited energy for the different layers and digitizer operating voltages.}
    \label{tab:MeVADCconv}
\end{table}

\label{sec:Performance Results}

\subsection{Noise study}
\begin{figure}[htbp!]
    \centering
    \includegraphics[height=4.5cm]{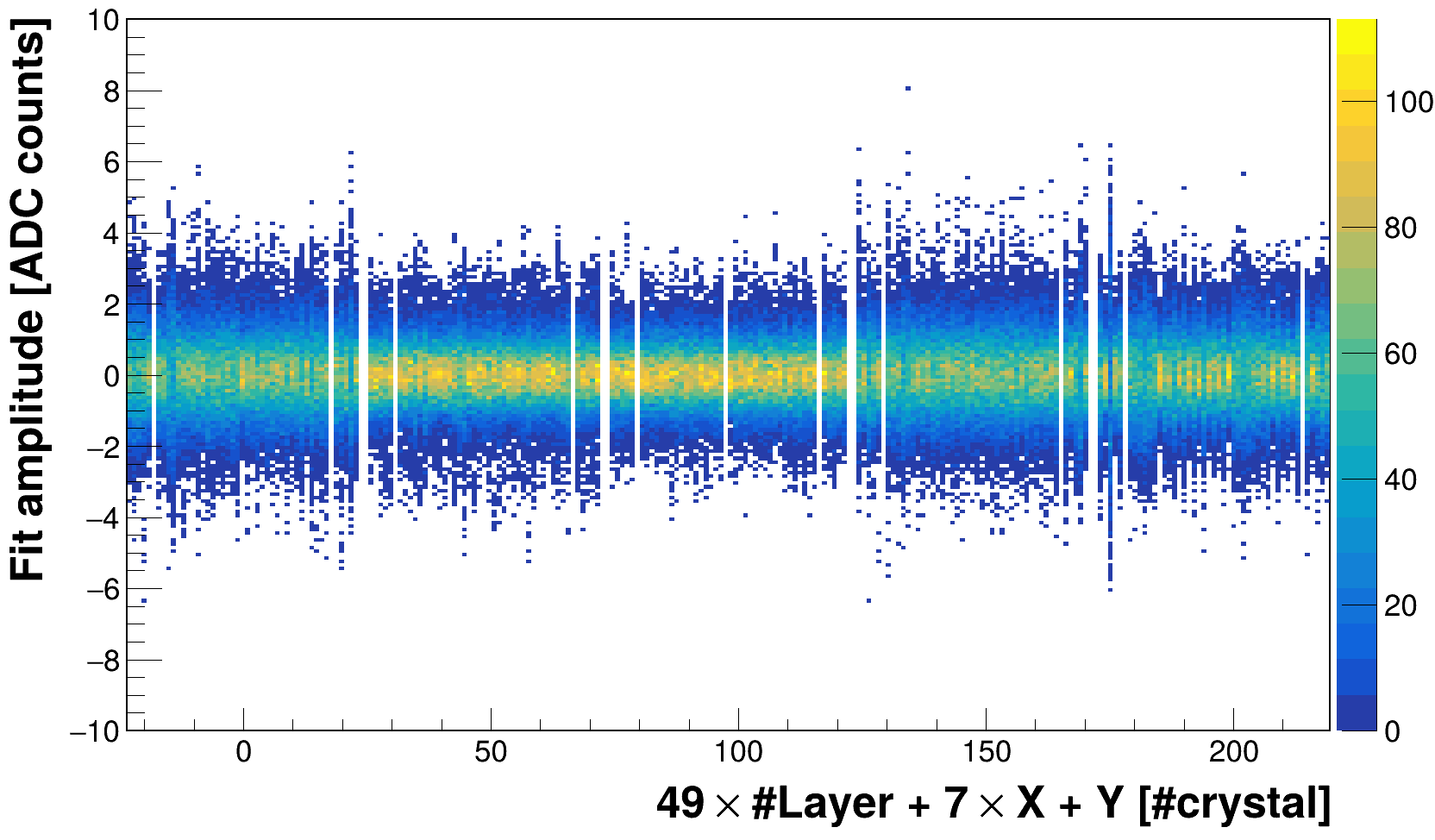}
    \includegraphics[height=4.5cm]{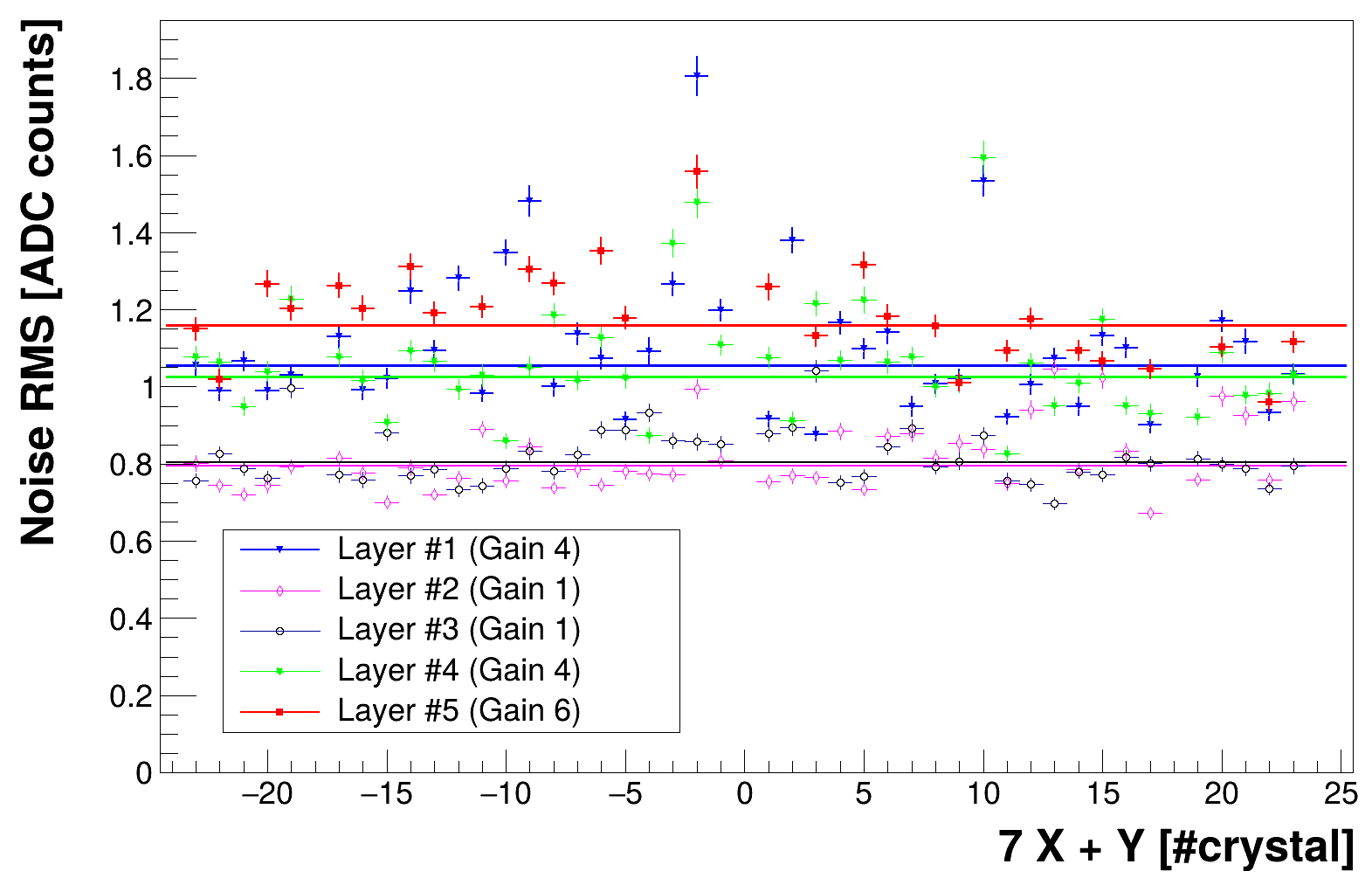}
    \caption{Left: two-dimensional distribution of the template-fit amplitude, in ADC counts, for pedestal events after applying the pedestal-based \textit{StartCell} corrections, as a function of the readout-channel index. Right: channel-by-channel noise RMS extracted from Gaussian fits to the pedestal distributions, restricted to the digitizers operated at 1~V.}
    \label{fig:ped_noise}
\end{figure}

The system noise is evaluated channel by channel by studying the distribution of the template-fit amplitude, expressed in ADC counts, for pedestal events after applying the pedestal-based \textit{StartCell} corrections, as shown in Fig.~\ref{fig:ped_noise}. The left panel shows the two-dimensional distribution of the template-fit amplitude as a function of the readout-channel index, while the right panel reports the Gaussian noise RMS extracted for each channel.

For the channels read out by the seven digitizers operated at 1~V, the noise RMS is shown separately for each layer. As expected, no dependence on the transverse crystal position is observed, while a clear positive correlation with the front-end gain is present. The average noise RMS values for the layers operated at gains of 1, 4, and 6 are approximately 0.68, 0.89, and 0.98 ADC counts, respectively, when considering only the 1~V digitizers.

The channels read out by the digitizer operated at 2~V show noise RMS values that are consistently lower by about 20\% with respect to the other digitizers. Since this concerns a limited number of channels, the average values obtained from the seven 1~V digitizers are used throughout the analysis.

The noise RMS can also be converted channel by channel into MeV by using Table \ref{tab:MeVADCconv}, obtaining, respectively, 6.6, 1.6, and 1.1 MeV, for the layers operated at gains of 1, 4, and 6, when considering only the 1~V digitizers. The values are doubled for the channels on the 2V digitizer.

The total noise term in the calorimeter energy sum, in the case of completely uncorrelated noise, obtained by summing in quadrature the energy-equivalent noise for all channels, is around $\sigma_{uncorr} = 70$ MeV.

\begin{figure}
    \centering
    \includegraphics[height=5cm]{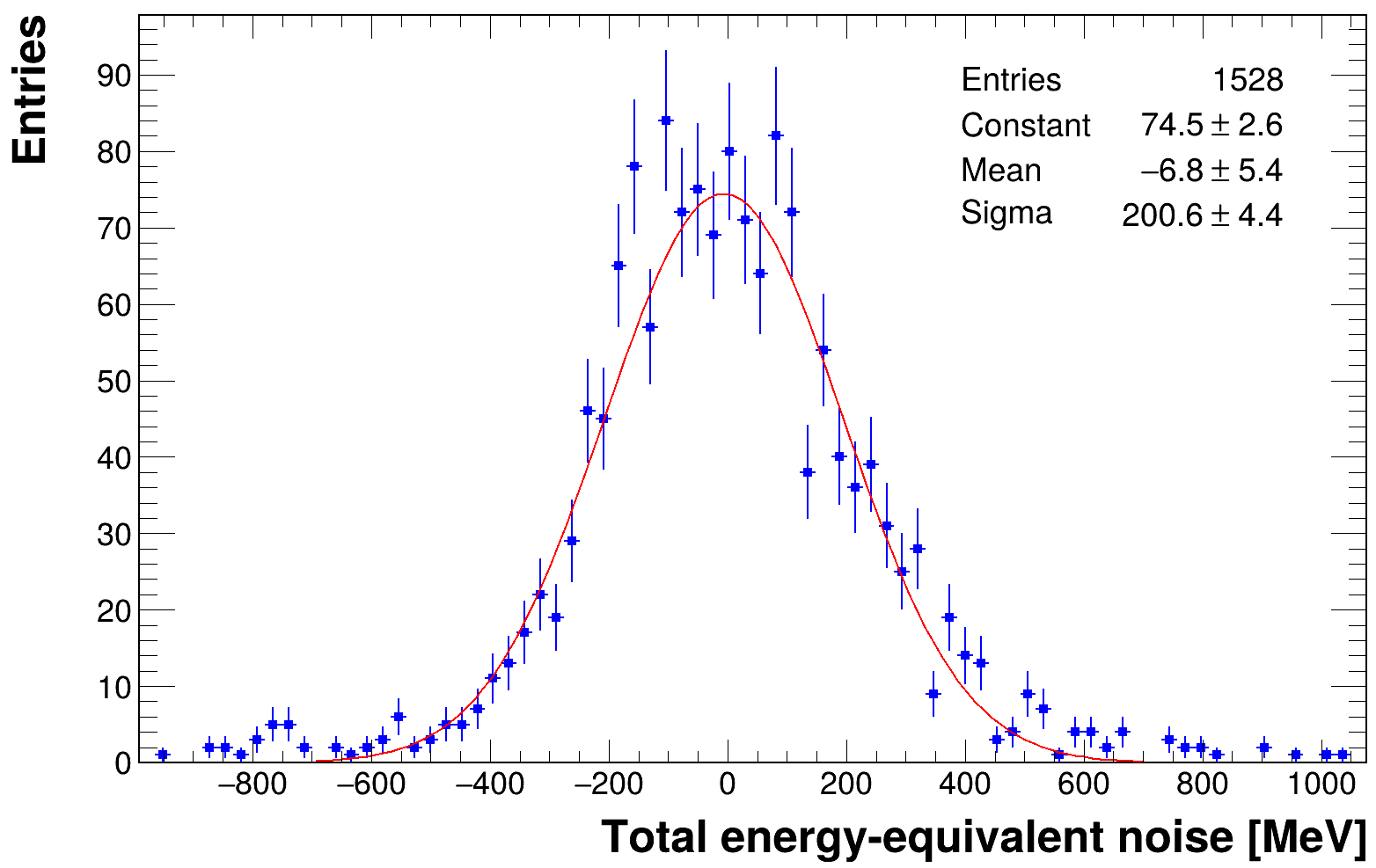}
     \includegraphics[height=5cm]{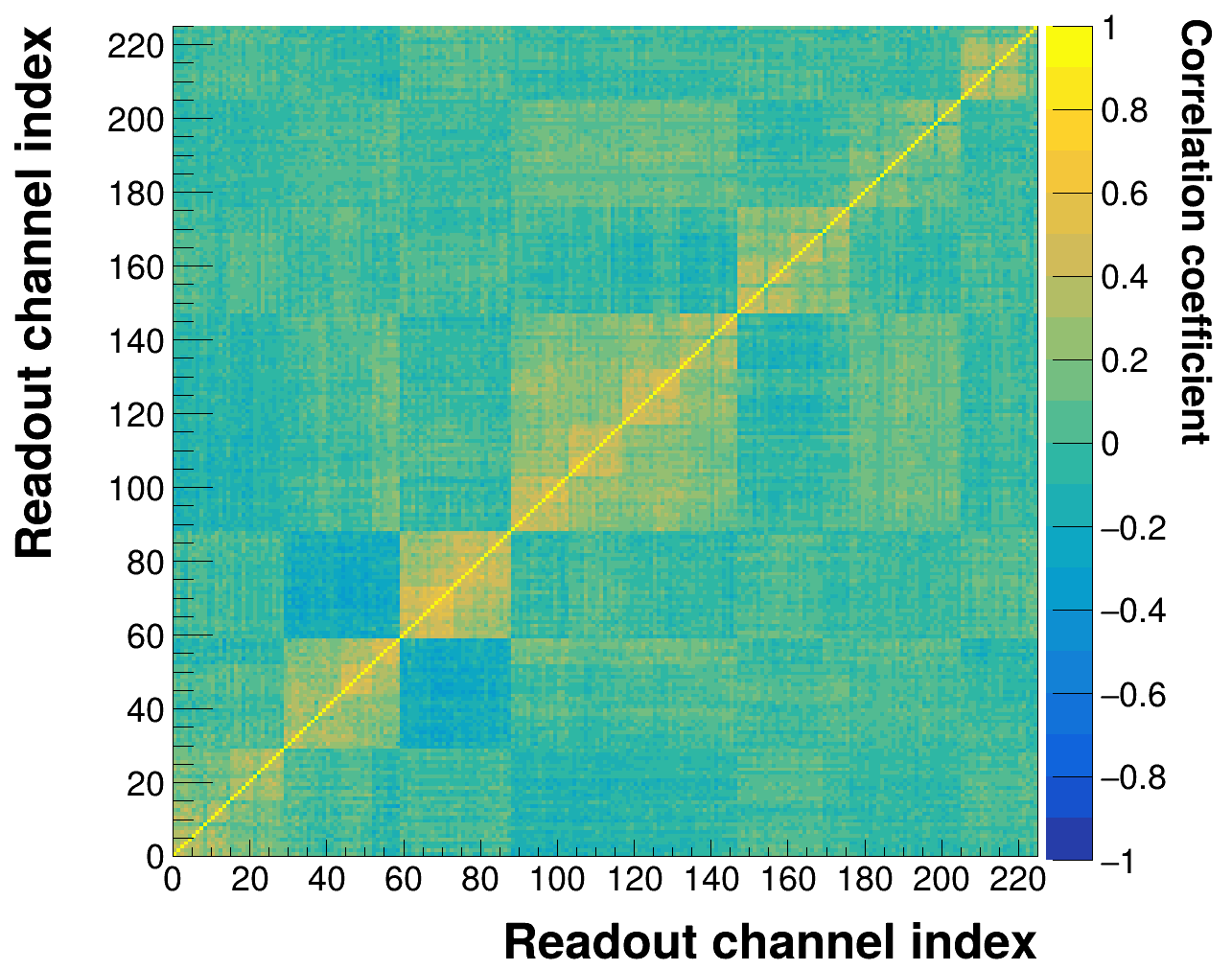}
    \caption{Left: distribution of the total sum of the template fit amplitude converted to energy, for pedestal events. Right: noise correlation matrix, built on the pedestal dataset, between all 225 channels.}
    \label{fig:ped_sum}
\end{figure}

The distribution of the total sum of the template fit amplitude converted in energy, for pedestal events, reported in Figure \ref{fig:ped_sum}, shows an RMS of about 208 MeV for the core of the distribution using a Gaussian fit, suggesting a large impact from noise correlations between the channels.

The correlation (reported in Figure \ref{fig:ped_sum}) matrix and the covariance matrix are evaluated from the same pedestal dataset, and the total variance of the energy sum can be evaluated as the sum of all the elements of the covariance matrix, obtaining a value of around $\sigma_{corr} = 213$ MeV, similar to the 208 MeV RMS of the energy sum distribution obtained before.

\label{sec:noise}

\section{Position reconstruction and time resolution}

\subsection{Transverse position reconstruction} 
The transverse position of electromagnetic showers within the calorimeter matrix was reconstructed using the energy-weighted centroid of the crystal matrix. To reduce the contribution of low-energy deposits originating from shower fluctuations and electronic noise, a logarithmic weighting scheme was adopted. The centroid coordinates were computed as
\begin{equation}
x_c=\frac{\sum_i x_i\,w_i}{\sum_i w_i},
\qquad
y_c=\frac{\sum_i y_i\,w_i}{\sum_i w_i},
\end{equation}
where the weights are defined as
\begin{equation}
w_i=\max\left[0,\;w_0+\ln\left(\frac{E_i}{\sum_j E_j}\right)\right].
\label{eq:logweight}
\end{equation}
Here, $E_i$ is the reconstructed energy deposited in crystal $i$, and $w_0$ was set to 7. This value was determined from Monte Carlo simulations by optimizing the linearity between the reconstructed centroid and the true impact position.

The centroid reconstruction was first validated using Monte Carlo simulations. A sample of 99~GeV electrons was generated with a Gaussian transverse beam profile of 2.5~mm RMS. The true impact position was binned along the $x$ and $y$ directions, and the distributions of the reconstructed centroid coordinates $x_c$ and $y_c$ were fitted with Gaussian functions in each bin. The mean value of each Gaussian fit was taken as the reconstructed centroid associated with the corresponding true-position interval.

\begin{figure}[htbp!]
    \centering
    \includegraphics[width=0.49\textwidth]{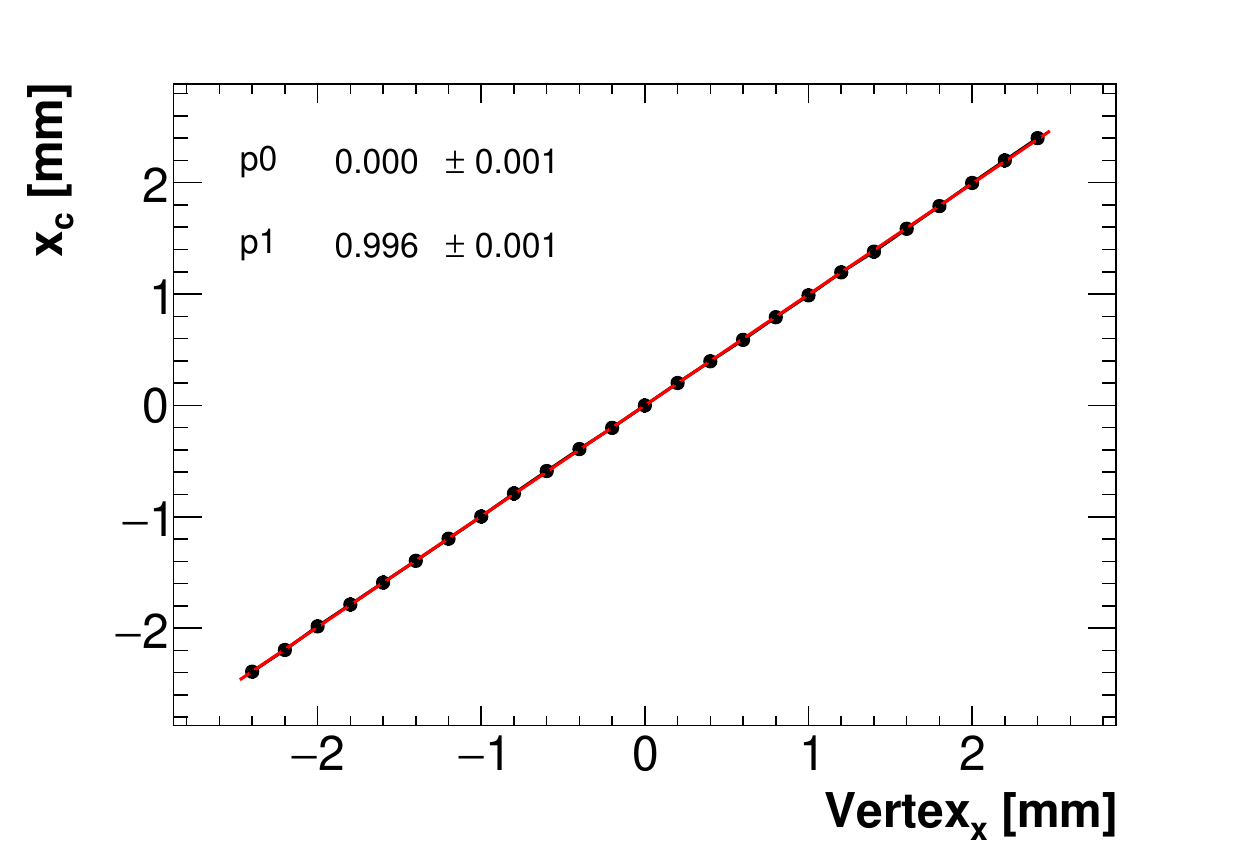}
    \includegraphics[width=0.49\textwidth]{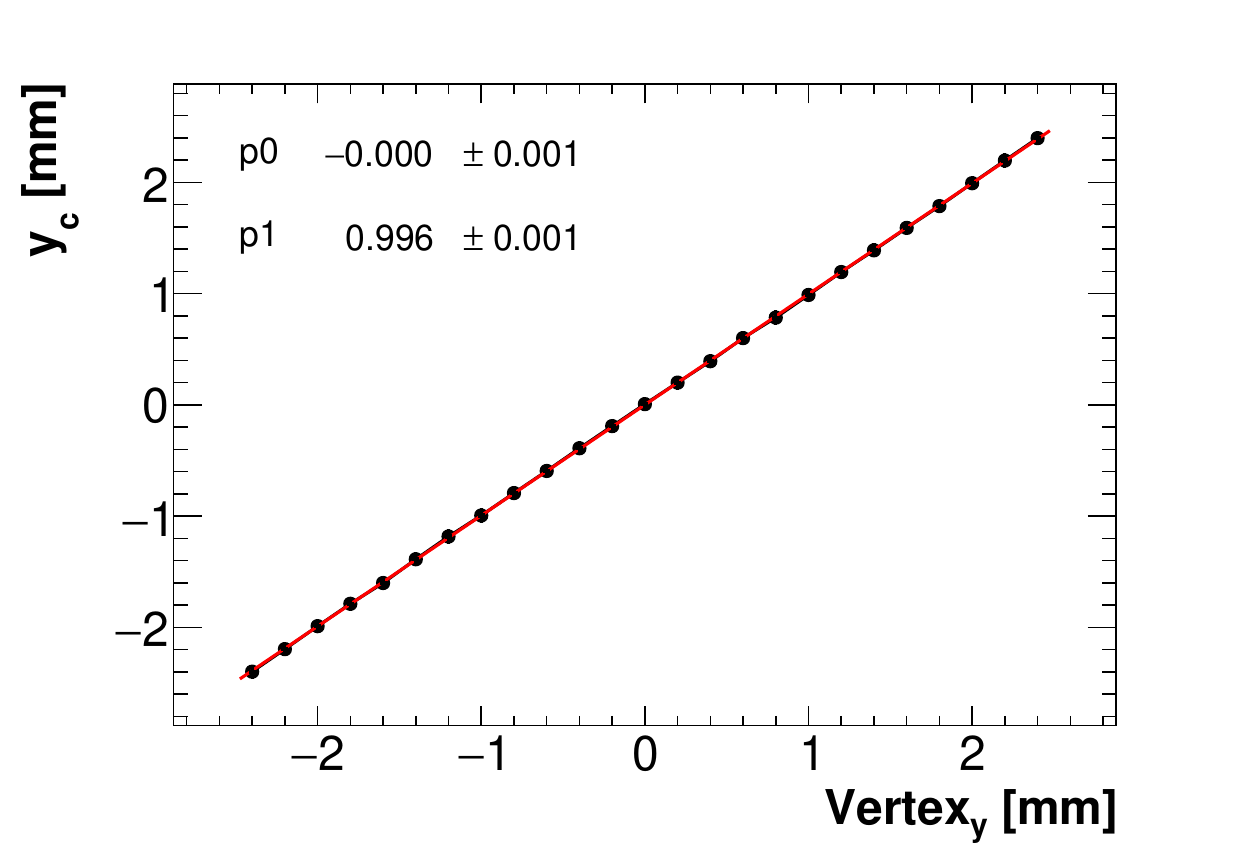}
    \caption{Linearity of the centroid reconstruction in Monte Carlo. The reconstructed centroid coordinates are plotted as a function of the generated impact position for the $x$ (left) and $y$ (right) directions. The centroid values are obtained from Gaussian fits to the reconstructed distributions in successive impact-position intervals.}
    \label{fig:centroid_mc_linearity}
\end{figure}

The reconstructed centroid positions were then compared with the average generated impact position in each interval. Figure~\ref{fig:centroid_mc_linearity} shows the resulting correlation between $x_c$ and the generated $x$ coordinate in the range $(-2.5,\,2.5)$~mm; an analogous result is obtained for the $y$ coordinate. The response is well described by a linear function over the full studied range, confirming the linearity of the centroid reconstruction.

\begin{figure}[htbp!]
    \centering
    \includegraphics[height=5.1cm]{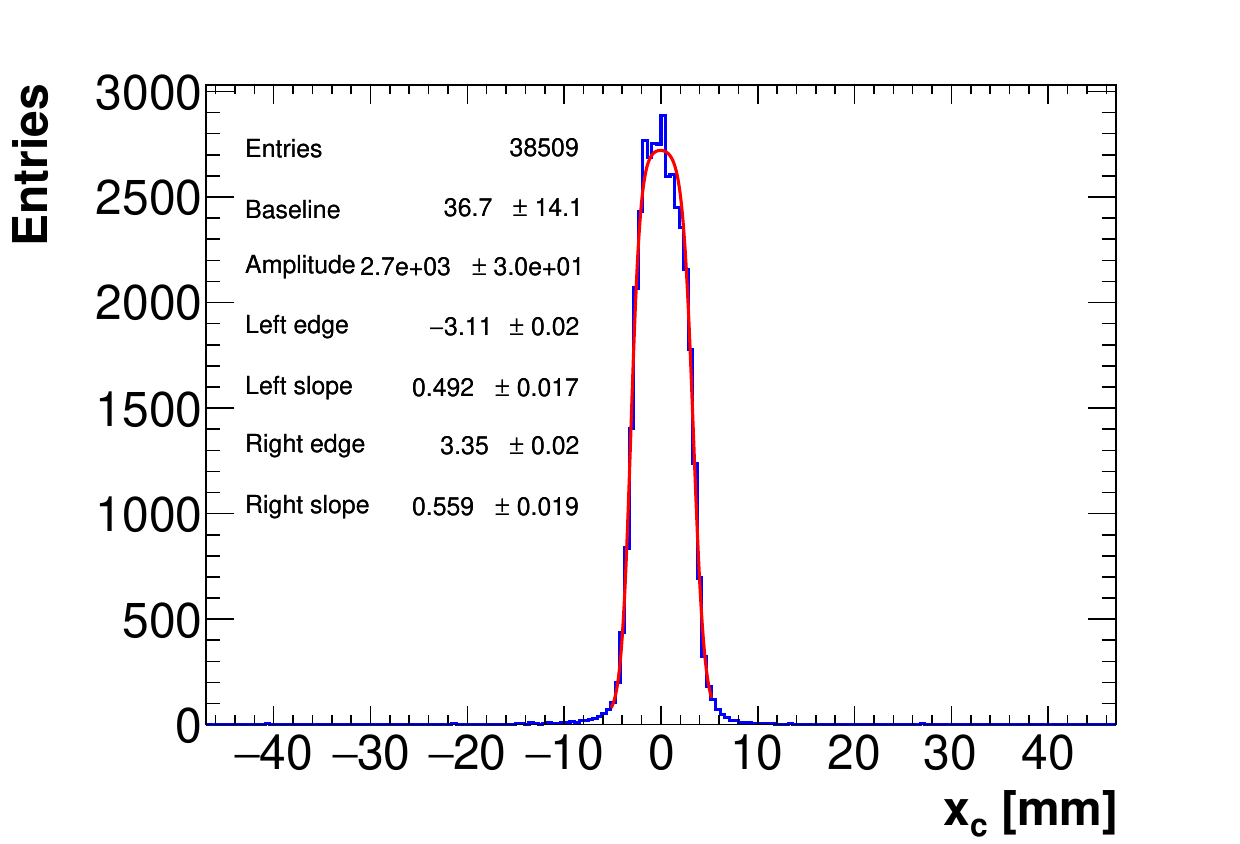}
    \includegraphics[height=5.1cm]{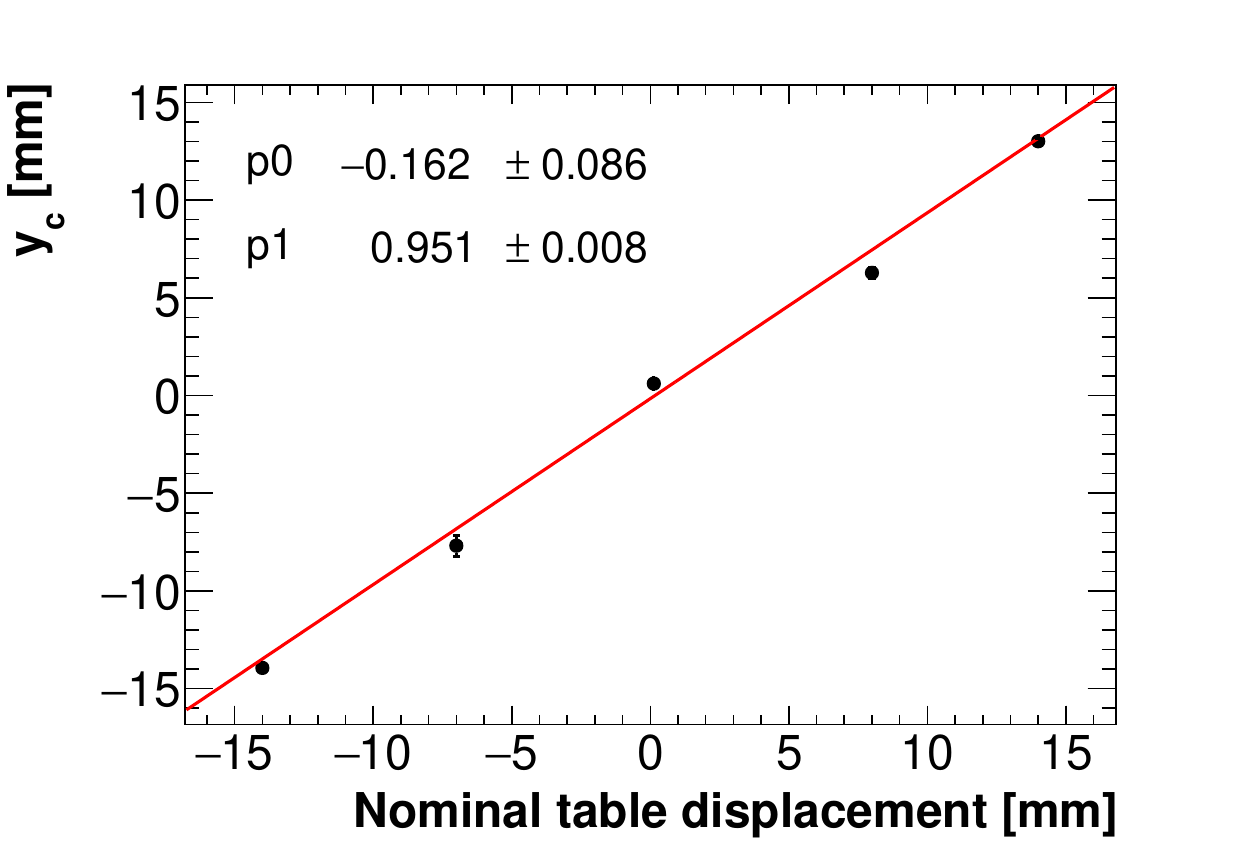}
    \caption{Left: example of horizontal centroid distribution measured during the position scan with the 74~GeV electron beam. The distribution is fitted with a double-sigmoid function to extract the centroid position and its uncertainty.
    Right: reconstructed centroid position as a function of the nominal beam position during the vertical scan of the calorimeter with 74~GeV electrons. The beam position is determined from the motorized table settings.}
    \label{fig:centroid_fit}
\end{figure}

The centroid reconstruction was subsequently validated using dedicated vertical position scans performed with a 74~GeV electron beam. The detector was translated vertically using the motorized table, changing the beam impact position on the calorimeter front face in known steps. For each table position, the reconstructed centroid distribution was projected onto the scan direction. Since these distributions exhibit a broad peak with pronounced tails, they were fitted with a double-sigmoid function, from which the centroid position and its uncertainty were extracted. Figure~\ref{fig:centroid_fit} shows examples of the resulting fits.

The reconstructed centroid positions were compared with the beam coordinates corresponding to the motorized table's nominal displacement. As shown in Fig.~\ref{fig:centroid_fit}, the detector response is well described by a linear function over the explored scan range, with a maximum absolute discrepancy of 0.45~mm at the edge of the scanned region. The same conclusions apply for both axes.

\subsection{Timing resolution}
The time resolution is evaluated by comparing the hit time of the two highest-energy cells, for each beam energy.

The time information, \textit{timestamp}, for each hit is extracted from the same template fit used for the energy reconstruction (see Section \ref{sec:Calibration and Reconstruction}). Only events with a sizable amount of collected charge in each of the considered cells are selected.

\begin{figure}[htbp!]
    \centering
    \includegraphics[width=0.53\textwidth]{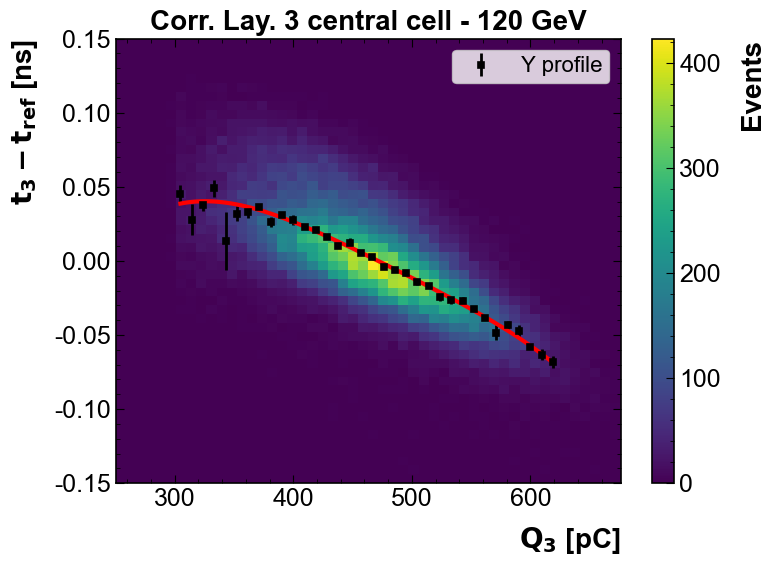}
    \hfill
    \includegraphics[width=0.44\textwidth]{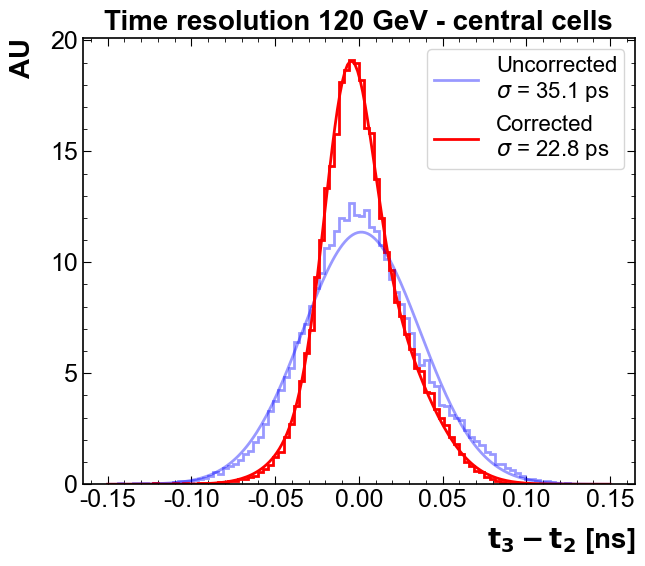}
    \caption{\textbf{Left:} example $(Q_i, \, t_i-t_{ref})$ spectrum, with the fitted correction for the deposited-energy effect overlaid. \textbf{Right:} example of $t_a - t_b$ distribution, before and after correction, where the distributions are normalized to unit integral and are both fitted with a double-Gaussian PDF.}
    \label{fig:timingfit}
\end{figure}

The timestamp must be corrected for deposited-energy dependence effects. As CRILIN has the best time resolution of all the other detectors in the setup, the correction for each CRILIN cell is calculated using an adjacent cell as time reference. An example correction is shown in Figure \ref{fig:timingfit}-left. The corrections are computed from the spectrum of $(Q_i, \, t_i-t_{ref})$, with $Q_i$ the charge of the channel to correct $i$, and $t_i-t_{ref}$ the time difference between the $i$-th channel and the reference one (shifted to have a null average). The spectrum is divided into bins of $Q_i$, and a fourth-order polynomial $f(Q_i)$ is fitted to the means of each bin. The corrected time is thus: $t_{i,corr} = t_i-f(Q_i)$. The reason $Q_i$ is used instead of the signal peak amplitude is that the latter is extracted with the same template fit as the timestamp: this would introduce extra correlations between the two variables, resulting in a nonphysical correction.

After the correction, the time resolution is evaluated from the standard deviation of the $t_a - t_b$ distribution, where $t_a,t_b$ indicate the timestamps of the most energetic cells in the layers $a$ and $b$. All time values are corrected, and the distributions are shifted to have a null average. For beam energies up to 30 GeV, the central cells of the first and second layers are used; above 30 GeV, the central cells of the second and third layers are used. An example distribution is shown in Figure \ref{fig:timingfit} right. The distributions show a residual skewness after correction. To estimate the standard deviation $\sigma_{t_a - t_b}$, the distributions are fitted with a double Gaussian PDF: the smallest $t_a - t_b$ range containing $68\%$ of the PDF's integral is taken as $2\sigma_{t_a - t_b}$. As $\sigma_{t_a - t_b}$ is the root-sum-square of the individual standard deviations of the two channels, the time resolution is $\sigma_{\Delta t} = \sigma_{t_a - t_b}/\sqrt 2$.

\begin{figure}[htbp!]
    \centering
    \includegraphics[width=0.5\textwidth]{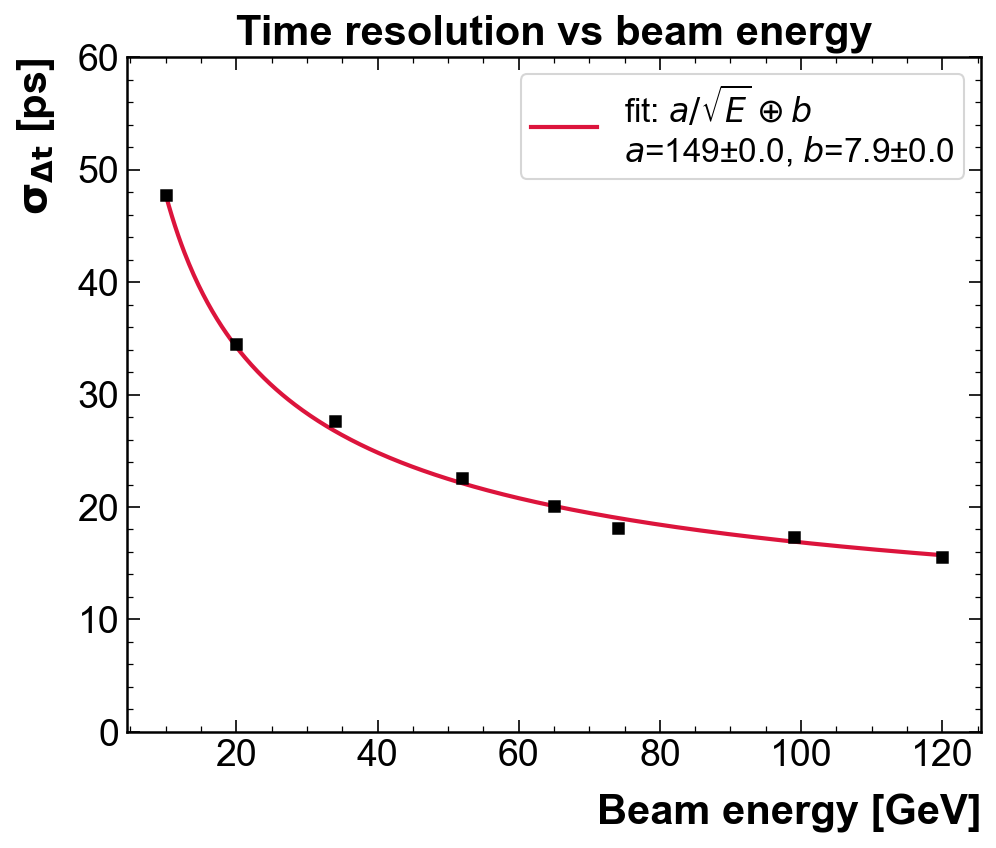}
    \caption{CRILIN time resolution as a function of beam energy, with the empirical fitted function overlaid. The error bars are smaller than the marker.}
    \label{fig:timeres0}
\end{figure}

The trend of the time resolution is evaluated as a function of the beam energy, Figure \ref{fig:timeres0}. The data points are fitted with the function
\begin{equation}
    \sigma_{\Delta t} = \frac{a}{\sqrt{E\text{ (GeV)}}} \oplus b
\end{equation}

The result shows that CRILIN achieves a sub-50 ps time resolution already for electron energies above 10 GeV, reaching sub-20 ps values above 60 GeV.

\section{Energy reconstruction}
\label{sec:energy_measurement}

\subsection{Light yield and photostatistics}
\label{sec:light_yield}

The most probable MIP response, averaged over all calibrated channels, is
\begin{equation}
    \left\langle V_{iFEE}^{\mathrm{\mu}} \right\rangle
    =
    (1.4 \pm 0.24)~\mathrm{mV}.
\end{equation}
The average ratio between the integrated charge and the peak amplitude is 
    $\left\langle Q/V \right\rangle = 1.17$. This quantity has been observed to exhibit only a weak dependence on \textit{StartCell} of a maximum of 1\% peak-to-peak.


The light yield can be estimated from the average muon response by converting the measured integrated charge into the corresponding number of photoelectrons:
\begin{equation}
    \mathrm{LY}_{\mu}
    =
    \frac{
        \left\langle V_{iFEE}^{\mathrm{\mu}} \right\rangle
        \cdot
        \left\langle Q/V \right\rangle
    }{
        E_{\mathrm{dep}}^{\mu}
        \cdot e
        \cdot G_{\mathrm{SiPM}}
    }
    =
    0.68 \pm 0.11~\mathrm{p.e./MeV},
    \label{eq:lightyield_mip}
\end{equation}
where $E_{\mathrm{dep}}^{\mu}$ is the average energy deposited by a through-going muon, $e$ is the elementary charge, and $G_{\mathrm{SiPM}} = 3.6 \times 10^{5}$ is the SiPM gain at the operating voltage $V_{\mathrm{op}}$.

An independent estimate of the light yield can be obtained from electromagnetic showers. For this purpose, the reconstructed channel amplitudes are first equalized using their respective MIP calibration coefficients and subsequently rescaled to the average MIP response. The resulting MIP-equalized energy-sum distribution is evaluated for the 99~GeV electron-beam data and normalized to the average deposited energy predicted by the Monte Carlo simulation, $E_{\mathrm{dep}}^{99~\mathrm{GeV}} \simeq 88~\mathrm{GeV}$.

\begin{figure}[htbp!]
    \centering
    \includegraphics[width=0.5\linewidth]{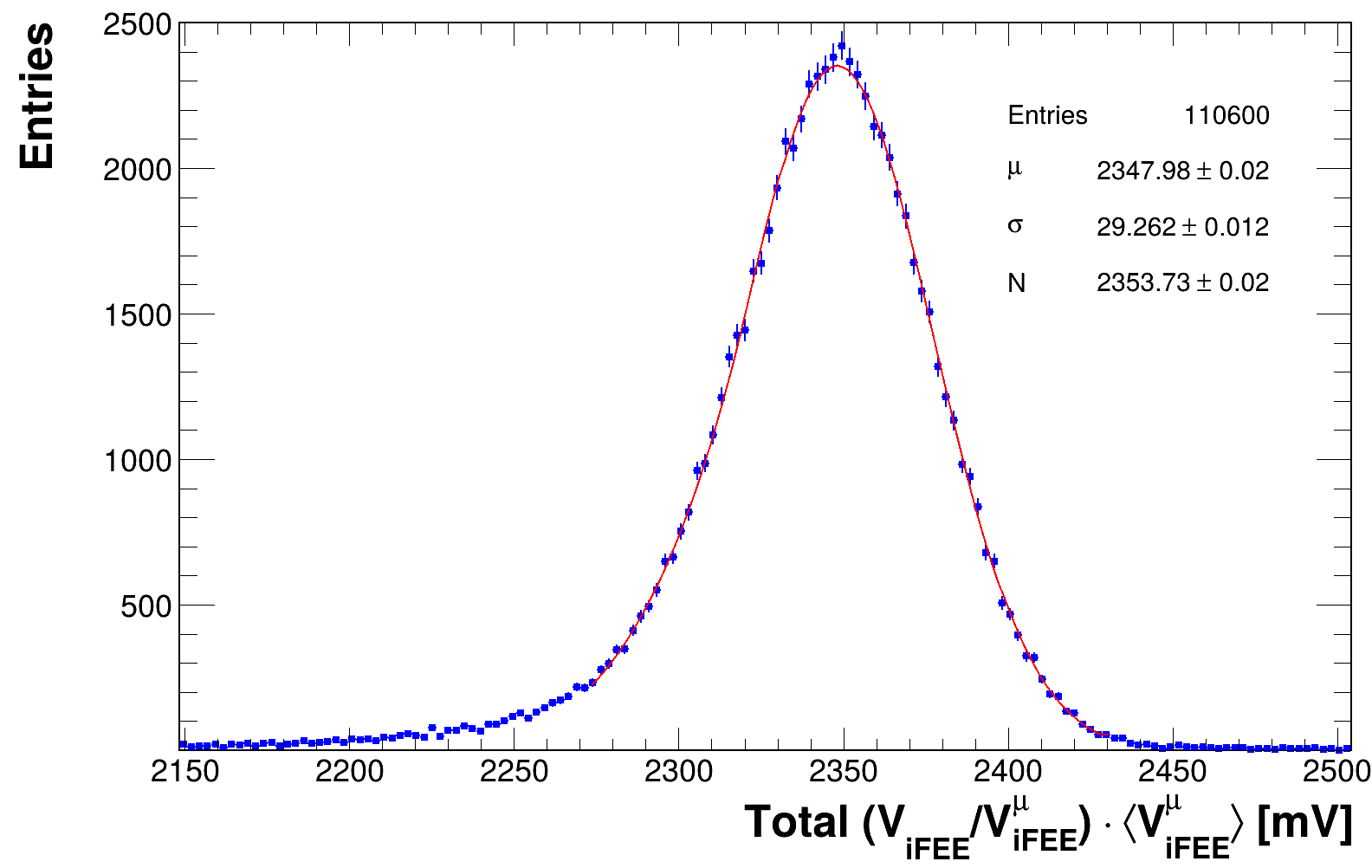}
    \caption{Distribution of the total MIP-equalized energy-sum estimator for the 99~GeV electron-beam data. The amplitude of each channel is normalized by its individual MIP calibration coefficient and rescaled to the average MIP response.}
    \label{fig:uncalib99GeV}
\end{figure}

As shown in Fig.~\ref{fig:uncalib99GeV}, the peak of the MIP-equalized energy-sum distribution is $V_{iFEE}^{99~\mathrm{GeV}} = 2353~\mathrm{mV}$, and the corresponding light yield for electromagnetic showers is
\begin{equation}
    \mathrm{LY}_{e}
    =
    \frac{
        V_{iFEE}^{99~\mathrm{GeV}}
        \cdot
        \left\langle Q/V \right\rangle
    }{
        E_{\mathrm{dep}}^{99~\mathrm{GeV}}
        \cdot e
        \cdot G_{\mathrm{SiPM}}
    }
    =
    0.54~\mathrm{p.e./MeV}.
    \label{eq:lightyield_99gev}
\end{equation}

To compare this result with the estimate obtained from MIP events, the electromagnetic-shower light yield is rescaled to the muon reference scale using the Cherenkov-response correction factor $k_{\mu/e} = 1.12$, discussed in Sec.~\ref{sec:monte_carlo}:
\begin{equation}
    \mathrm{LY}_{\mu}^{e-scaled}
    =
    k_{\mu/e} \cdot \mathrm{LY}_{e}
    =
    1.12 \cdot \mathrm{LY}_{e}
    = 0.62 ~\mathrm{p.e./MeV}.
    \label{eq:lightyield_99gev_muonscale}
\end{equation}

The light yield inferred from the 99~GeV electromagnetic-shower sample is therefore compatible, within uncertainties, with the value obtained from muon events.

\subsection{Residual corrections on the digitizers starting cell index}
After achieving a channel-by-channel calibration using MIPs, the coefficients are applied to evaluate the total energy reconstructed by the calorimeter.

\begin{figure}[htbp!]
    \centering
    \includegraphics[height=4.5cm]{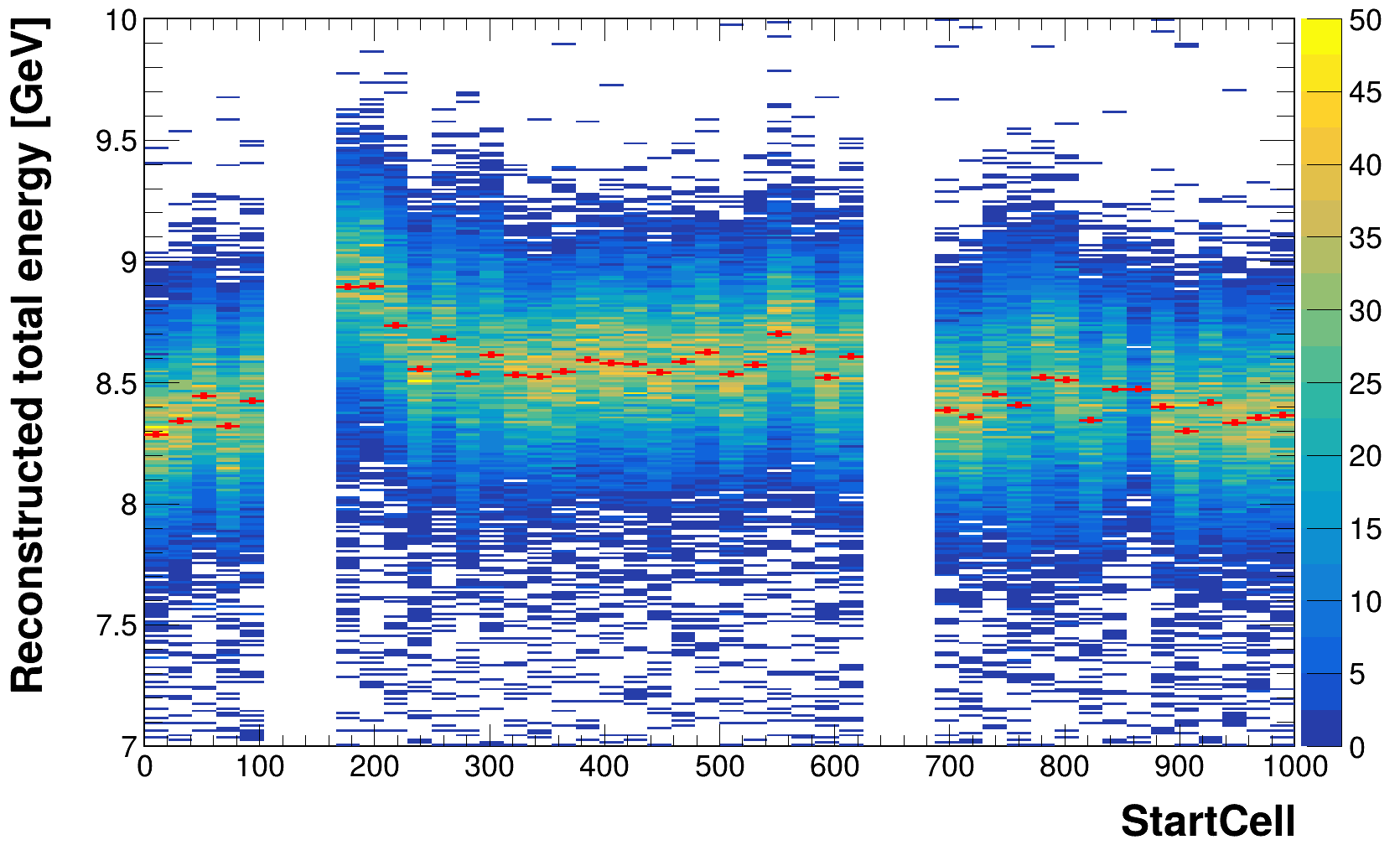}
        \includegraphics[height=4.5cm]{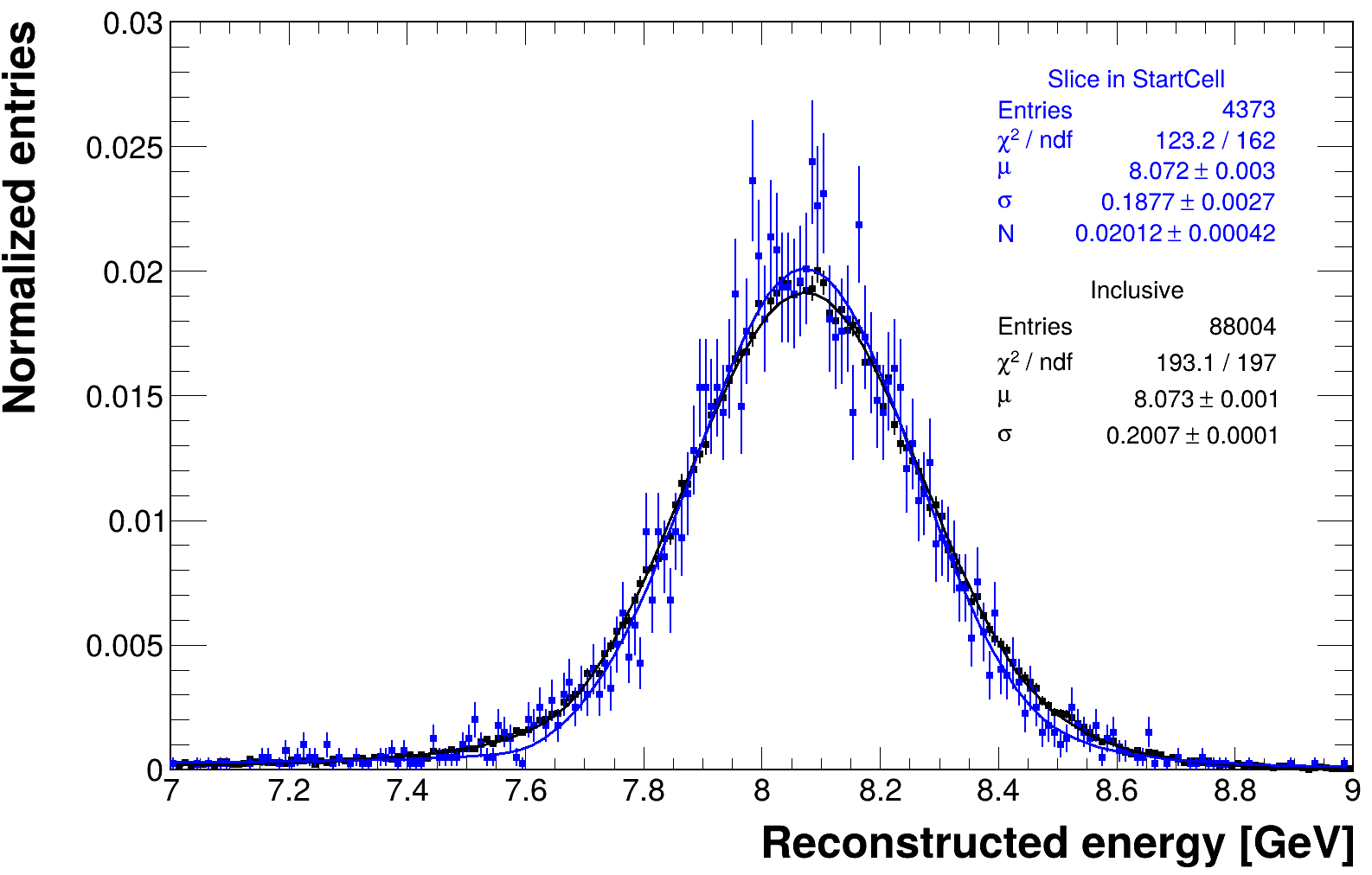}
\caption{Left: two-dimensional distribution of the total reconstructed energy, as a function of \textit{StartCell} for signal events. The red points indicate the peak positions of Gaussian fits to the projected histograms in each \textit{StartCell} bin.\\
Right: total reconstructed energy distribution for the 10~GeV run in a single \textit{StartCell} slice (blue) and in the inclusive case after residual \textit{StartCell} corrections, fitted with DCB functions.}
    \label{fig:energy_ped_corr}
\end{figure}

As shown in Fig.~\ref{fig:energy_ped_corr}-left for the 10~GeV run, the pedestal-based correction already reduces the peak-to-peak spread in the total reconstructed energy from about 25\% in the uncorrected case to approximately 6\%. However, the statistical power of the pedestal dataset available for each run is limited, because the pedestal trigger rate is only 0.5~Hz, compared with about 150~Hz during data taking. For this reason, a residual correction is applied using the peak of the total reconstructed energy in each run as a standard candle. The same fitting procedure in \textit{StartCell} slices as for pedestal events is used.

To avoid overfitting the data and artificially improving the energy resolution, a two-fold cross-validation procedure is employed. Odd and even event numbers are split into two independent subsets; the residual corrections are derived separately for each subset, and the correction obtained from odd-numbered events is applied to even-numbered events, and vice versa. This procedure is applied without any energy threshold on hits, and it has been repeated throughout the data analysis for reprocessing runs with different threshold values.

To demonstrate the absence of overfitting, the energy resolution of the 10~GeV run in a single \textit{StartCell} slice is compared in Fig.~\ref{fig:energy_ped_corr}-right with the inclusive resolution after residual \textit{StartCell} corrections. The inclusive case is actually worse by about 10\%, due to the fluctuations introduced by the correction procedure in the low-statistics \textit{StartCell} slices. All histograms are fitted with double Crystal Ball (DCB) functions~\cite{DCB}.


\subsection{Thresholds optimizations}
\label{sec:thresholds}
Since, from Section \ref{sec:noise}, the noise RMS measured in ADC counts is constant with a 15\% spread across layers, a unique threshold in ADC counts is applied to all calorimetric hits.
This threshold is optimized in the run most sensitive to noise effects, i.e., the one at 10 GeV, and for each threshold value the full \textit{StartCell} corrections, described in Section \ref{sec:Calibration and Reconstruction}, are applied.
The window between 3 and 5 ADC counts is studied, and a plateau is found between around 3.5 and 4.5 ADC counts. As a consequence, in the following, the value of 3.5 ADC counts is employed as the hit threshold.

On average, this threshold corresponds to 34, 8.5, and 5.7 MeV energy deposits, respectively, for the layers at gain 1, 4, and 6, and the channels on the digitizers at 1V, and these values are doubled for the digitizer at 2V. After threshold optimization, it is possible to evaluate the number of hits $H(E)$ and the total energy equivalent noise sum $N(E)$ as a function of beam energy $E$. In particular, $N(E)$ is evaluated as the RMS of the distribution of the per-event noise $N_{per\,ev.}$. For each event $N_{per\,ev.}$ is calculated by throwing a random vector $\vec{n} \in \mathbb{R}^{225}$, following a multivariate Gaussian distribution with the noise covariance matrix derived in Section \ref{sec:noise}, and using the following expression:
\begin{equation}
    N_{per\,ev} = \sum_{i \in \left\{ channels \right\} } n_i \Theta(E_i > T_i)
\end{equation}
where $\{channels\}$ is the set of all channels,  and $E_i, \, T_i$ are, respectively, the reconstructed energy deposit and the energy-equivalent threshold for the $i$-th channel.

\begin{figure}[htbp!]
    \centering
    \includegraphics[width=0.65\linewidth]{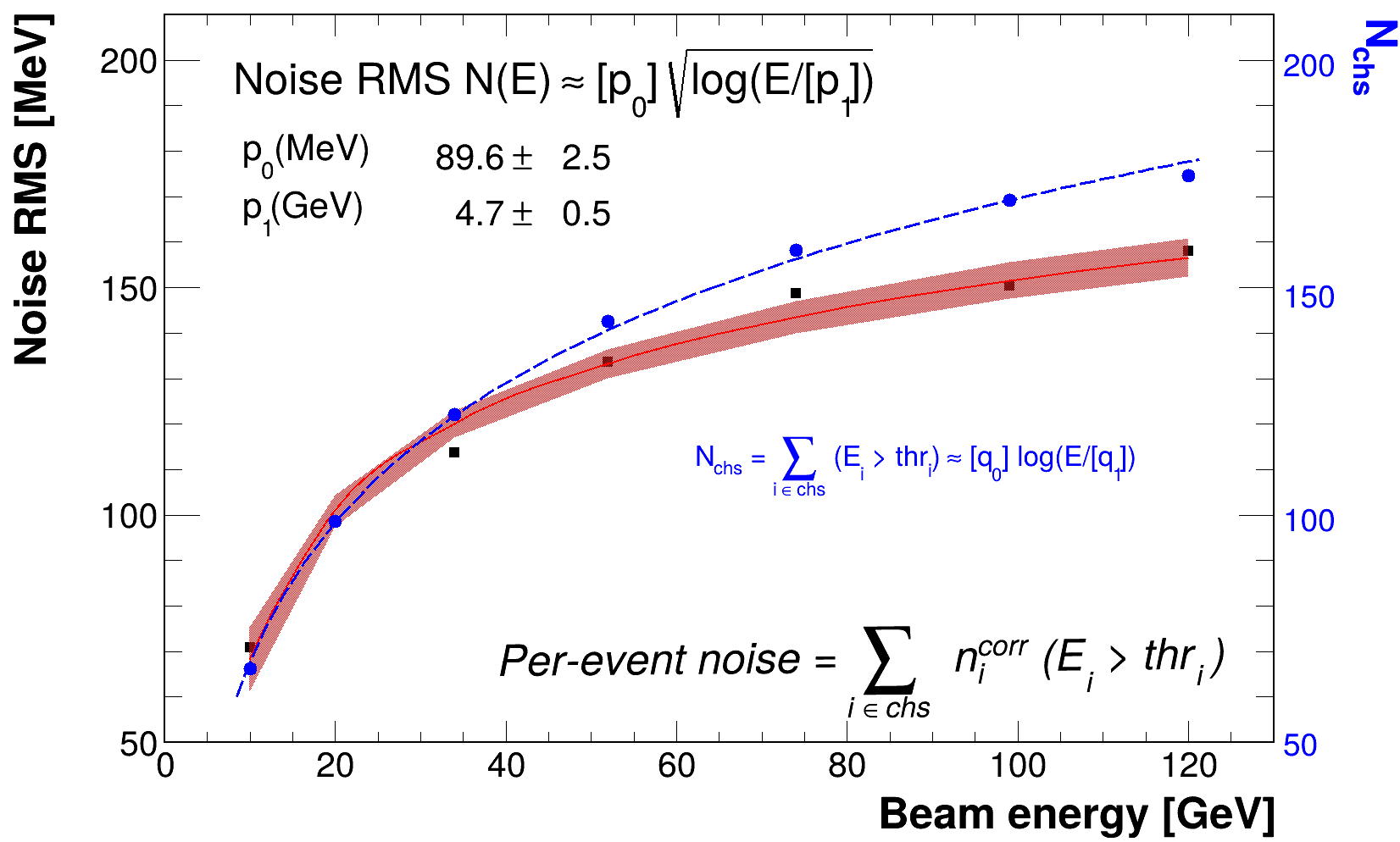}
    \caption{The functions $H(E)$ (blue) and $N(E)$ (black points) are shown, together with their fitted functions. For $N(E)$, also the 68\% confidence intervals of the fitted functions are shown in red.}
    \label{fig:n_h_e}
\end{figure}

Both $H(E)$ and $N(E)$ are reported in Figure \ref{fig:n_h_e}, where $H(E)$ is fitted with a logarithmic function, and $N(E)$ is fitted with the square root of a logarithmic function, as predicted by Rossi's theory of electromagnetic showers.

\subsection{Event-by-event corrections}

\begin{figure}[htbp!]
    \centering
    \includegraphics[width=0.99\linewidth]{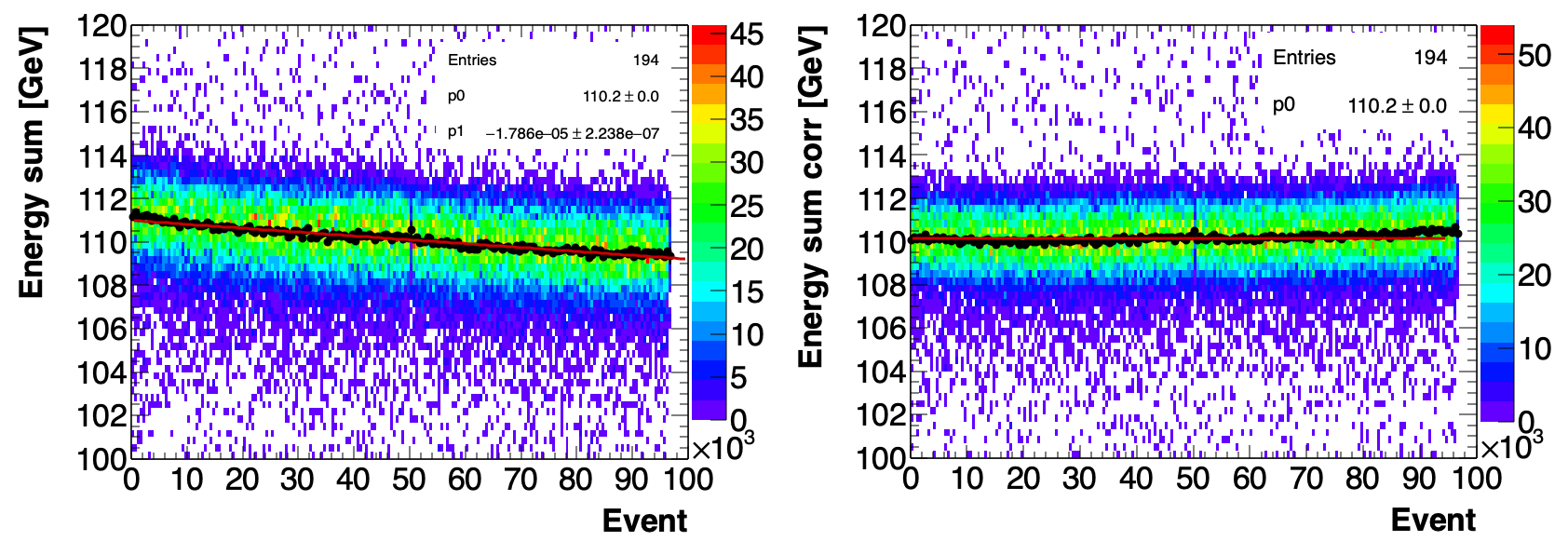}
    \caption{Time drift observed for 120 GeV before (left panel) and after (right panel) the correction. Notice that the linear fit is shifted to the center of the $X$-axis.}
    \label{fig:120drift}
\end{figure}

\paragraph{Correction of time drifts}
Since during the test beam no temperature control and monitoring was available, the only strategy available to provide a reliable measurement of the energy resolution consisted of correcting any linear drifts in the total energy values as a function of time, during each run. Some runs, especially the one at 120 GeV, as shown in Figure \ref{fig:120drift}, lasted many hours because of the low event rate due to the higher energy, and showed a decrease in the total energy of more than 1\% throughout the run.
Since no measurements of this drift are available, the maximum relative variation of the energy in each run is used as a systematic uncertainty for the energy linearity measurement.

\subsubsection{Correction as a function of the number of hits}
\begin{figure}[htbp!]
    \centering
    \includegraphics[width=0.99\linewidth]{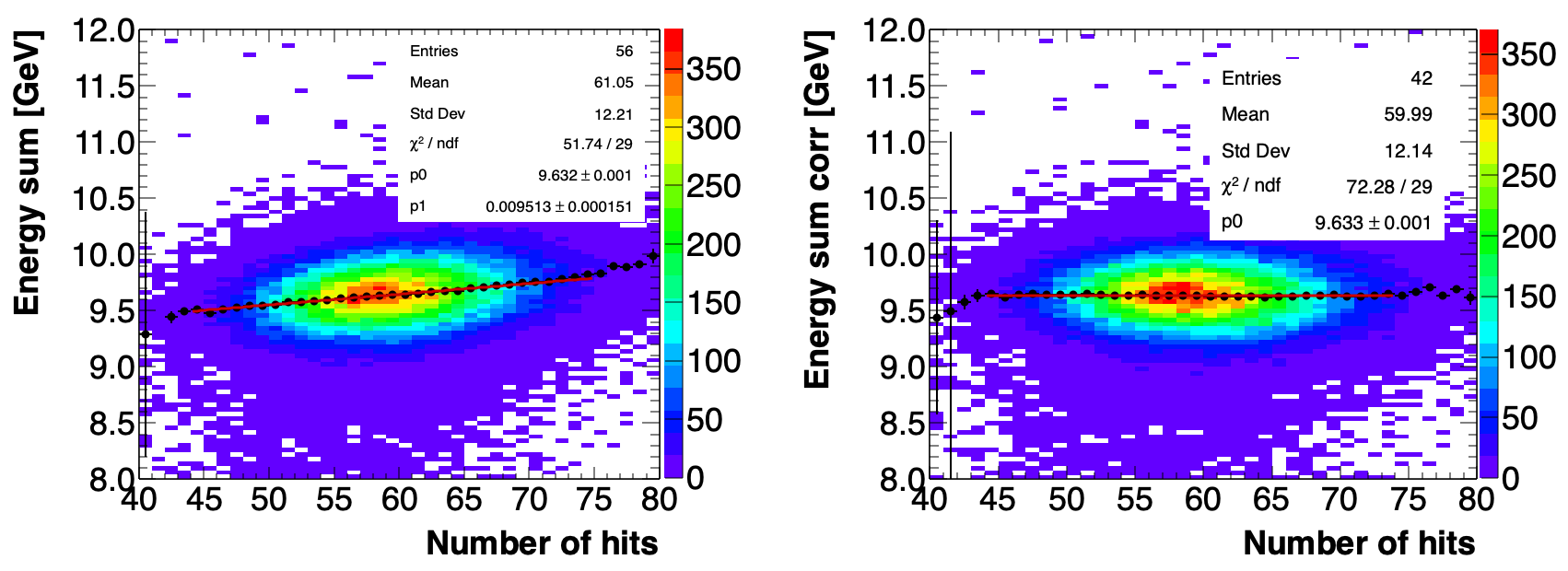}
    \caption{Energy sum as a function of the number of hits at 10 GeV before and after the linear correction. Notice that the linear fit variable is shifted to the mode of its $X$-axis to avoid variations of the mean energy sum value. }
    \label{fig:nhits_corr}
\end{figure}

Due to the presence of correlated noise, potential mis-calibrations in the channel-by-channel response, and the need to correct the effect of the containment fluctuations, a set of energy corrections based on the collections of channels above threshold (hits) would be needed. For simplicity, a one-dimensional proxy is used, i.e., the number of hits, which, as shown in Figure \ref{fig:nhits_corr} at 10 GeV, is correlated to the total energy.
Linear event-by-event corrections, as in the previous Section, are derived for each energy point, as a function of the difference between the number of hits and its average, in such a way as to have a negligible impact on the calorimeter linearity.

The correction procedure is derived using samples of fixed beam energy. In a collider environment, the incident energy is not known a priori. In practice, in a collider environment, a preliminary uncorrected energy estimate can be employed to evaluate the correction coefficients via interpolation for each event.

\subsubsection{Leakage correction with longitudinal information}

A fraction of the energy-resolution degradation is due to fluctuations in the longitudinal starting point of the shower, which can be conceptually identified with the position of the first hard bremsstrahlung or, for photons, the first pair conversion above the relevant energy threshold. The longitudinal shower start follows an exponential distribution with mean free path $X_0 \approx 0.94$~cm, which also corresponds to its standard deviation.

Fluctuations in the reconstructed total energy that are correlated with the shower-start position $z$ can be corrected if an appropriate proxy for $z$ is available. In this work, the chosen proxy is the ratio between the energy deposited in the central crystal of the second layer and the total reconstructed energy, hereafter denoted by $R_{c2}$. This correction is made possible by the longitudinal segmentation of CRILIN, which provides direct event-by-event information on the shower development and would not be available in a purely homogeneous calorimeter. For each beam energy, the mode $\hat{R}_{c2}$ of the $R_{c2}$ distribution is determined, and the total energy is then studied as a function of $(R_{c2}-\hat{R}_{c2})$ to extract the corresponding correlation coefficient.

\begin{figure}[htbp!]
    \centering
    \includegraphics[width=0.99\linewidth]{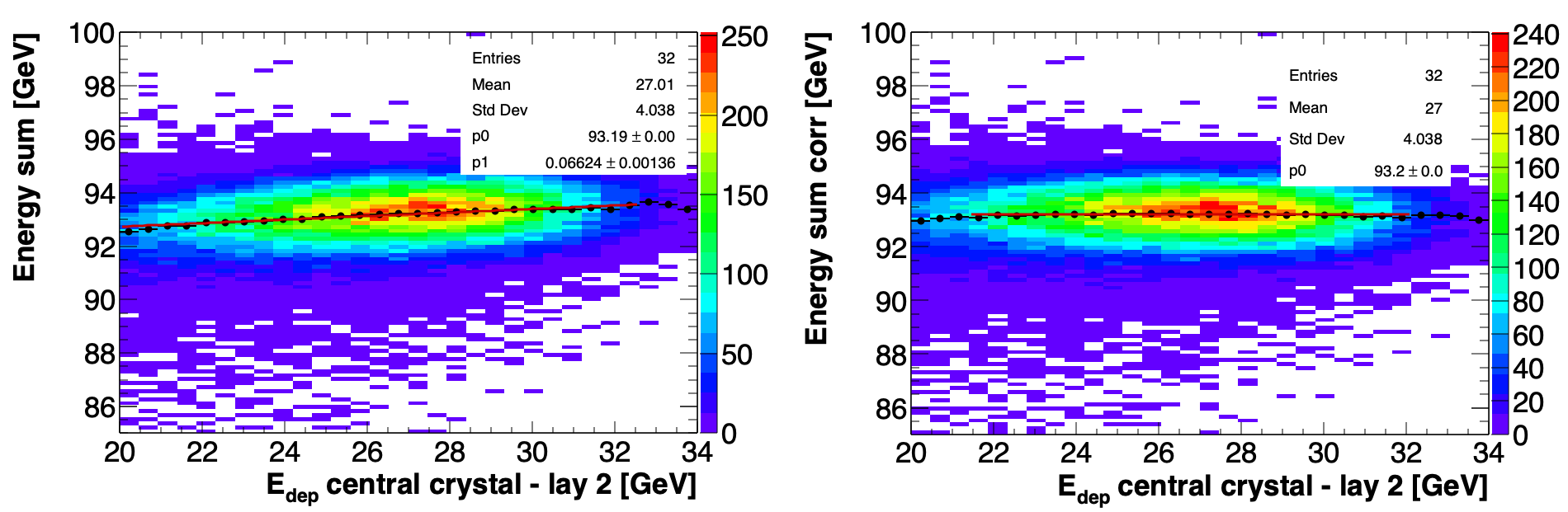}
    \caption{Energy sum as a function of the energy deposit in the central crystal of the second layer at 99 GeV before and after the linear correction. Notice that the linear fit variable is shifted to the mode of its $X$-axis to avoid variations of the mean energy sum value.}
    \label{fig:tot_rc2_corr}
\end{figure}

As shown in Fig.~\ref{fig:tot_rc2_corr} for the 99~GeV beam, the two-dimensional distribution of the total energy versus $(R_{c2}-\hat{R}_{c2})$ exhibits a clear linear dependence. The means of Gaussian functions fitted to the projections along the $Y$ axis in successive $X$ slices are then fitted with a linear function, and the resulting slope is taken as the correlation coefficient, $\rho_{R_{c2}}$.

The corrected energy is therefore obtained by adding to the uncorrected energy the term
\[
\Delta E_{R_{c2}} = \rho_{R_{c2}} \cdot (R_{c2}-\hat{R}_{c2}) .
\]

The use of the ratio between the energy deposited in the central crystal of the second layer and the total energy as a proxy for $z$ was found during a search for the best linear combination of the central-crystal energies, namely the one with the highest explained-variance ratio with respect to the total reconstructed energy, in the framework of principal component analysis (PCA) \cite{pca}, showing similar performances to the more complete PCA approach. 
It should be noted that, to avoid the potential risk of overfitting, the two corrections mentioned above are determined with 2-fold validation as already done for the StartCell-based one.
The improvements in the resolution due to these corrections are reported later on in Figure~\ref{fig:en_reso_fit}-left.
        
\subsubsection{Systematic effects due to beam position}
\begin{figure}[htbp!]
    \centering
    \includegraphics[width=0.99\linewidth]{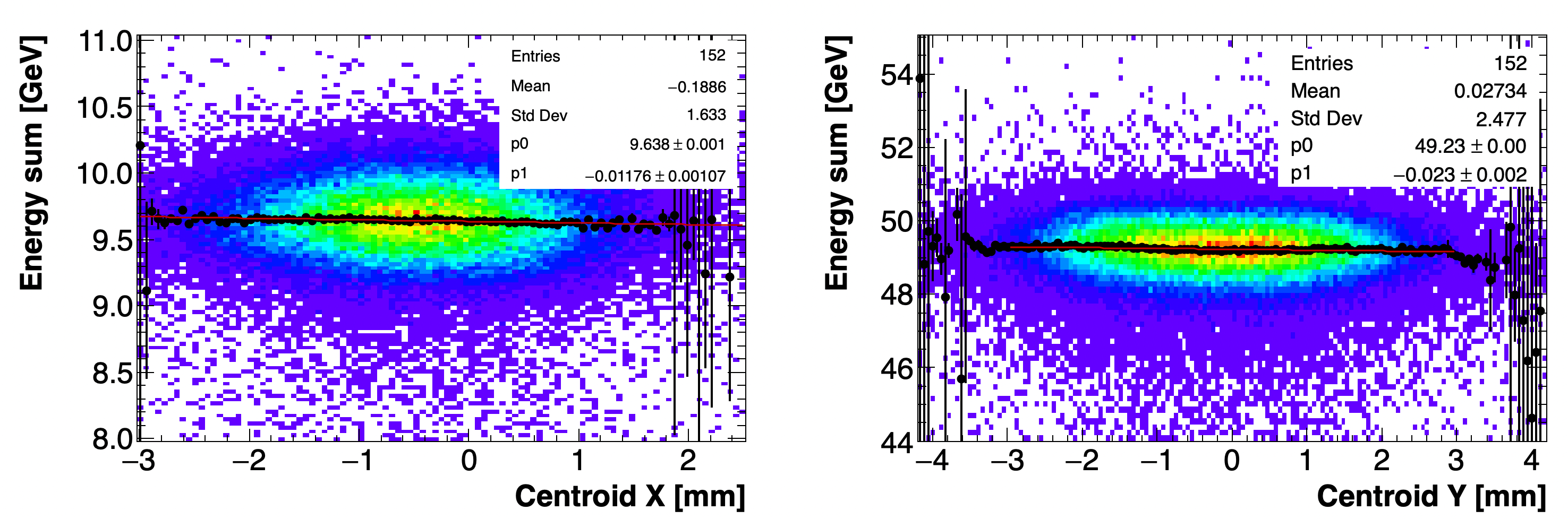}
    \caption{Energy-sum versus X–Y centroid distributions for the 10 GeV and 52 GeV runs, respectively. The relative variation of these distributions, in quadrature with the other coordinate component, is used to estimate the associated systematic uncertainty.}
    \label{fig:centroid_syst}
\end{figure}

Due to the possible imperfect calibration of all the channels, the calorimeter response is not constant with respect to the beam transverse (x, y) position. 
For this reason, a systematic uncertainty $\sigma_{xy}$ is applied to all energy measurements, depending on the run. 
Since the events were collected by triggering on $5\times5$ mm$^2$ scintillators, two systematic uncertainties $\sigma_x,\,\sigma_y$ are determined by plotting the energy as a function of the reconstructed beam position (via centroid) in X and Y separately, as shown in Figure \ref{fig:centroid_syst}, and fitting with a linear function. 
After this step, $\sigma_x$ and $\sigma_y$ are determined by dividing by $\sqrt{12}$ the peak-to-peak spread of the linear function in the centroid window corresponding to the trigger, to obtain a Gaussian error.

The final error budget $\sigma_{xy}$ is obtained by summing in quadrature $\sigma_x$ and $\sigma_y$.
This systematic uncertainty is applied as-is for energy linearity measurements, while, for energy resolution measurements, its impact is evaluated as $\sigma_{xy}^{reso} = \sqrt{\left(\sigma_{E}/E\right)^2 + \sigma_{xy}^2} - \sigma_{E}/E$. This is the only systematic uncertainty applied to the energy resolution in data, since this effect is not corrected, differently from all the others.

\subsection{Energy linearity and resolution}
\begin{figure}[htbp!]
    \centering
    \includegraphics[height=4.5cm]{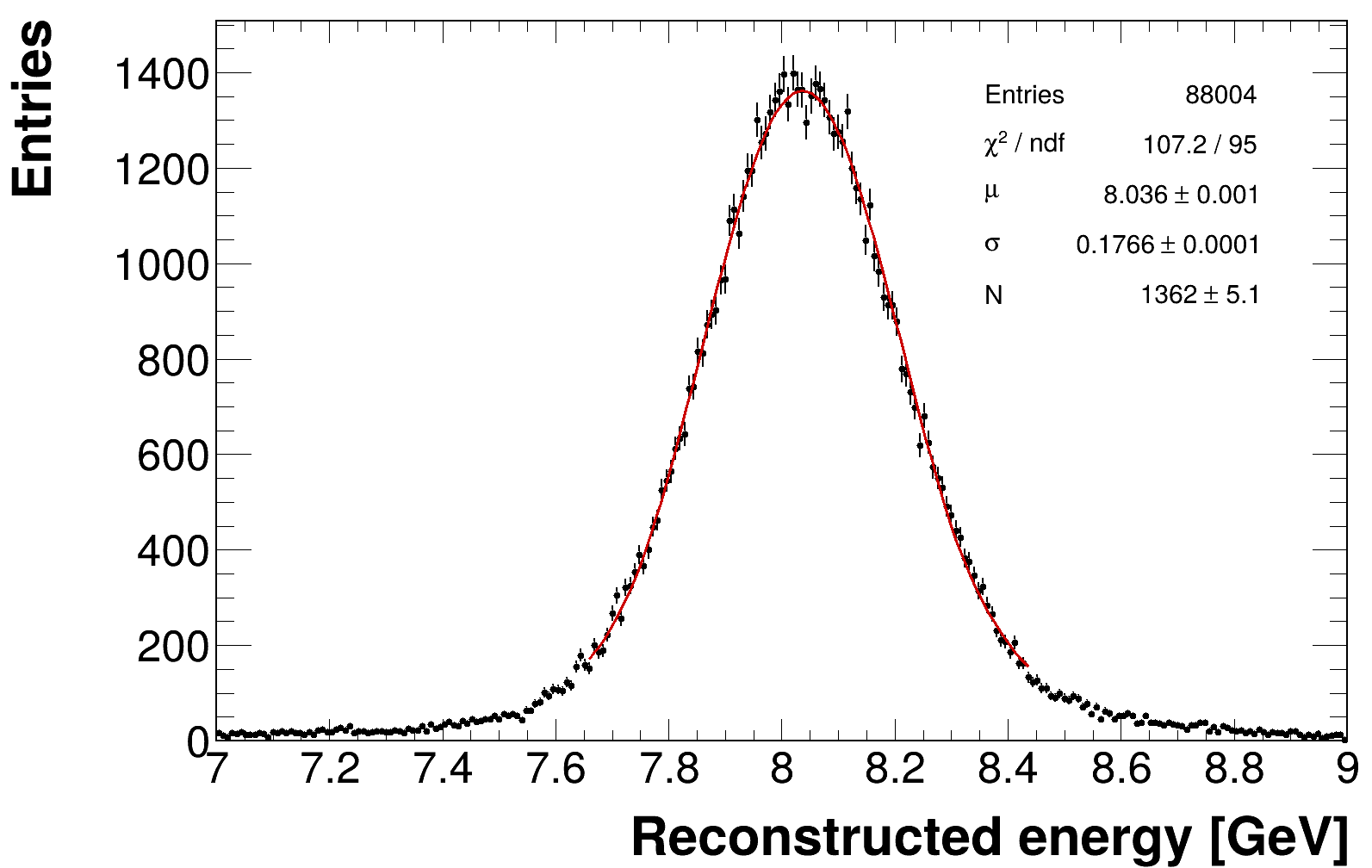}\,\,
        \includegraphics[height=4.5cm]{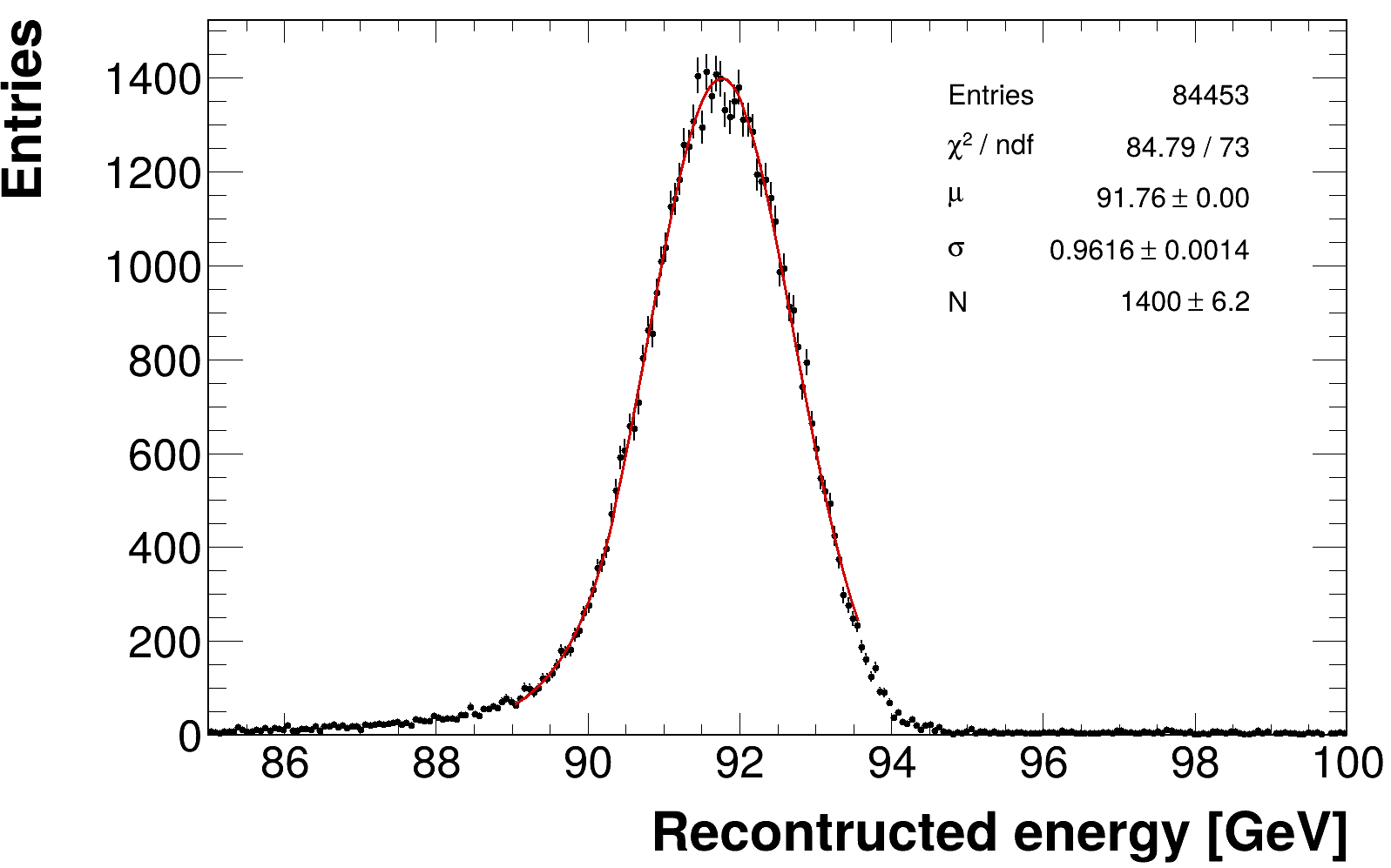}
    \caption{Histograms of energy distribution at 10 (left) and 120 (right) GeV, fitted with DCB functions.}
    \label{fig:fit_examples}
\end{figure}

All energy distributions, after all corrections, are fitted with DCB functions, and the peak $\mu$ and $\sigma$ parameters are employed to study the energy linearity and resolution.
Two example fits at 10 and 120 GeV are shown in Figure \ref{fig:fit_examples}.

\subsubsection{Linearity}
\begin{figure}[htbp!]
    \centering
    \includegraphics[width=0.75\linewidth]{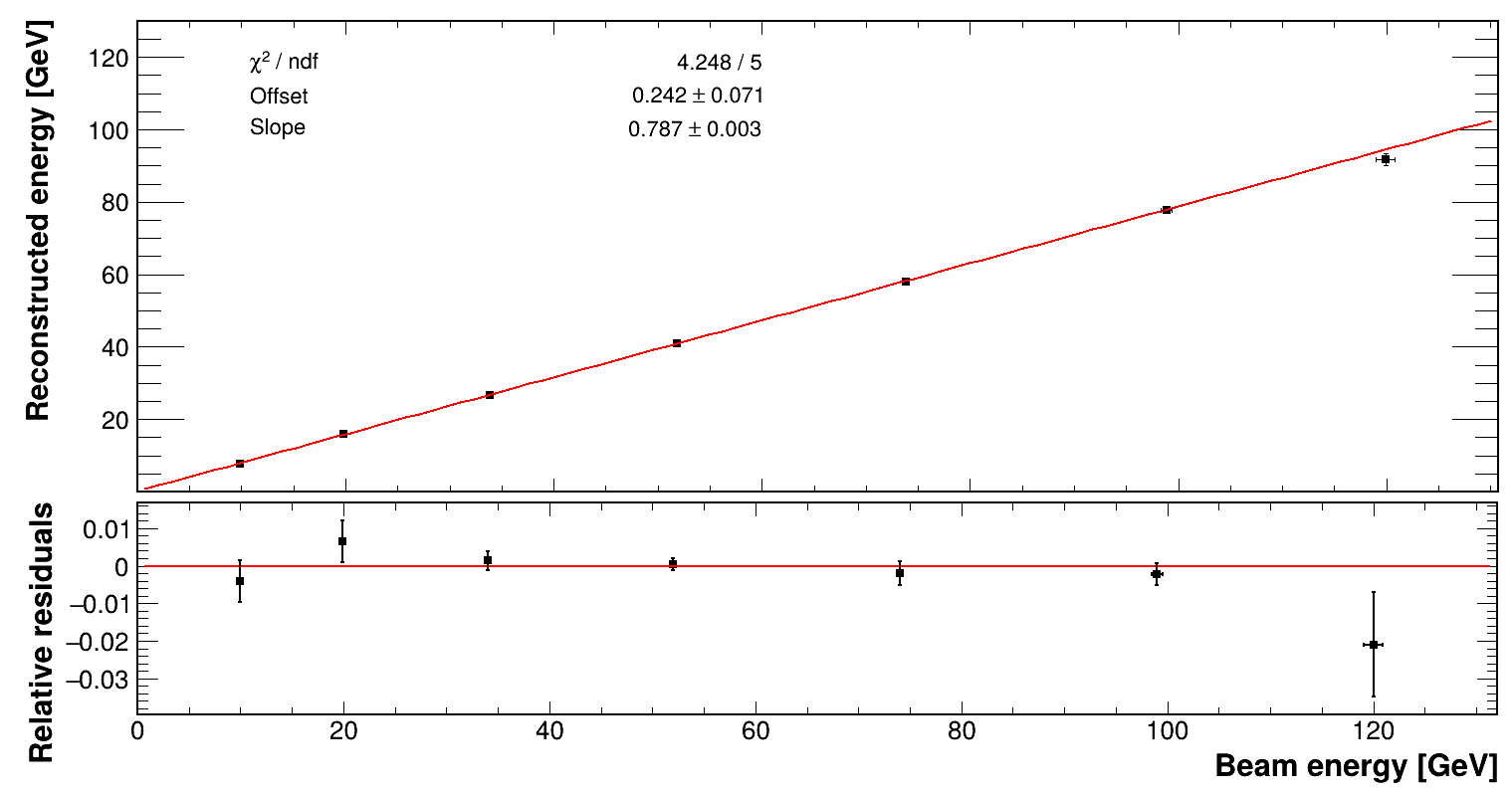}
    \caption{Top: reconstructed vs. beam energy, fitted with a linear function. Bottom: residuals with respect to the fitted linear }
    \label{fig:fit_linearity}
\end{figure}

The energy linearity is studied by means of a linear fit to the fitted peaks as a function of the beam energy, as shown in Figure \ref{fig:fit_linearity}.
The $\chi^2/\mathrm{ndf}$ value of the linear fit is $10.7/5$, corresponding to a 6\% p-value, and the relative residuals are reported in the lower pad of Figure \ref{fig:fit_linearity}, showing data/fit compatibility within 1$\sigma$ below 99 GeV (included), and a slightly larger discrepancy at 120 GeV. This is not particularly significant, since no temperature control and no HV or temperature monitor was in place during the test beam. Moreover, the runs were taken in this sequence: (99, 20, 34, 52, 74) GeV in the first 8 hours, then 120 GeV, 10 GeV, respectively, 18 and 72 hours afterwards.
Nevertheless, the maximum deviation from the fit line is around 2.5\% at 120 GeV, which is reassuring of the fact that the calorimeter contains almost all the showers, which allows proceeding further and evaluating the energy resolution.

\subsubsection{Energy resolution}
\begin{figure}[htbp!]
    \centering
        \includegraphics[height=4.8cm]{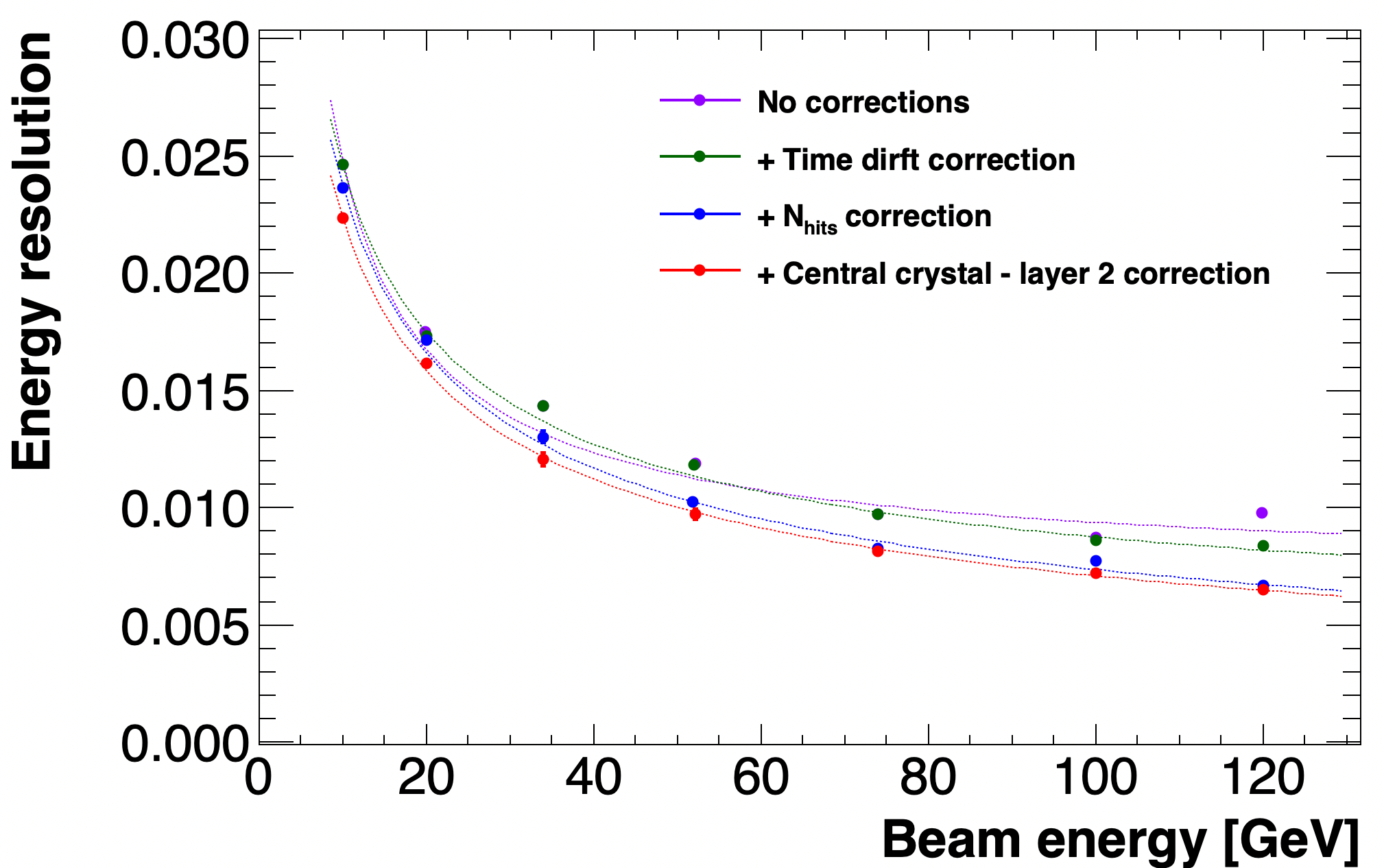}\,\,
\includegraphics[height=4.7cm]{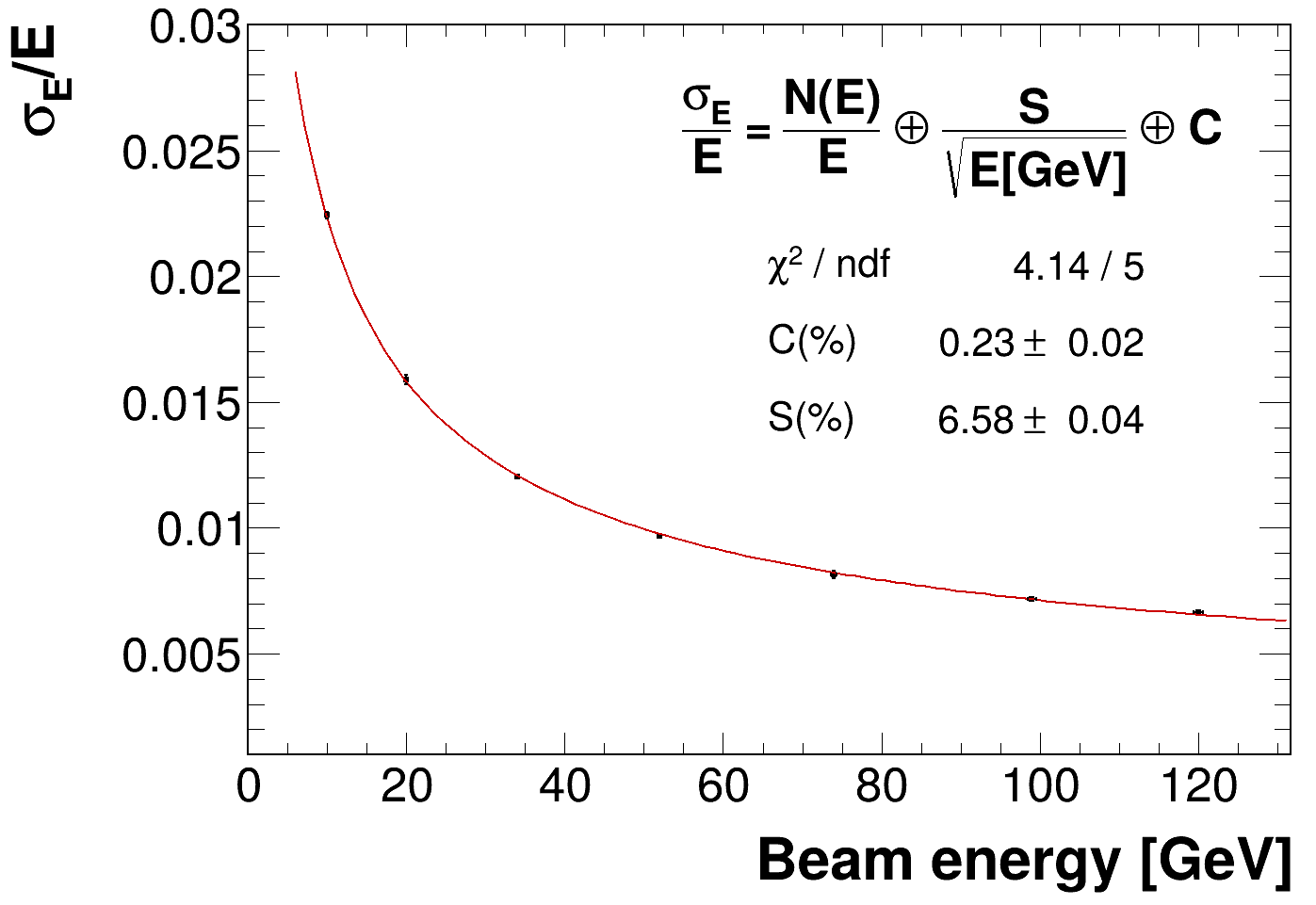}
    \caption{Left: energy resolution at all steps of the correction, with empirical functions fitted to show the trend. Right: energy resolution at the final step, fitted with the complete energy resolution model.}
    \label{fig:en_reso_fit}
\end{figure}

The ratios $\sigma/\mu$ from the fitted DCB functions for each beam energy are employed to study the energy resolution. 
The beam energy spread is subtracted in quadrature for each energy. 
The energy resolutions at each step in the flow of event-by-event corrections are shown in Figure \ref{fig:en_reso_fit}-left.

Because of the large differences (up to a factor of 3 between 10 and 120 GeV) in the number of hits as a function of the beam energy, the resolution is very difficult to describe using a standard noise term $N/E$ with a constant $N$ value.
Hence, the precise estimate of the non-constant term $N(E)$ from Section \ref{sec:thresholds} is employed, and the energy resolution is modeled as:
\begin{equation}
    \frac{\sigma_{E}}{E} = \frac{N(E)}{E} \, \oplus \frac{S}{\sqrt{E[GeV]}} \, \oplus C
\end{equation}
where $S$ and $C$ are the only parameters of interest of the fit.
As shown in Figure \ref{fig:en_reso_fit}-right, the fitted value of $S$ and $C$ are, respectively, $S = (6.58 \pm 0.04) \%$ and $C = (0.23\pm 0.02)\%$.

\subsubsection{Comparison of the energy resolution in data and simulation}
\label{sec:mccomp}
The simulation described in Section~\ref{sec:monte_carlo} is used to reproduce the energy resolution measured in the test-beam data and to assess the impact of the main instrumental effects. To this end, the resolution is evaluated for several simulation configurations, in which photoelectron-statistics fluctuations, electronic noise, and channel thresholds are introduced progressively.

\begin{figure}[htbp!]
    \centering
    \includegraphics[height=4.8cm]{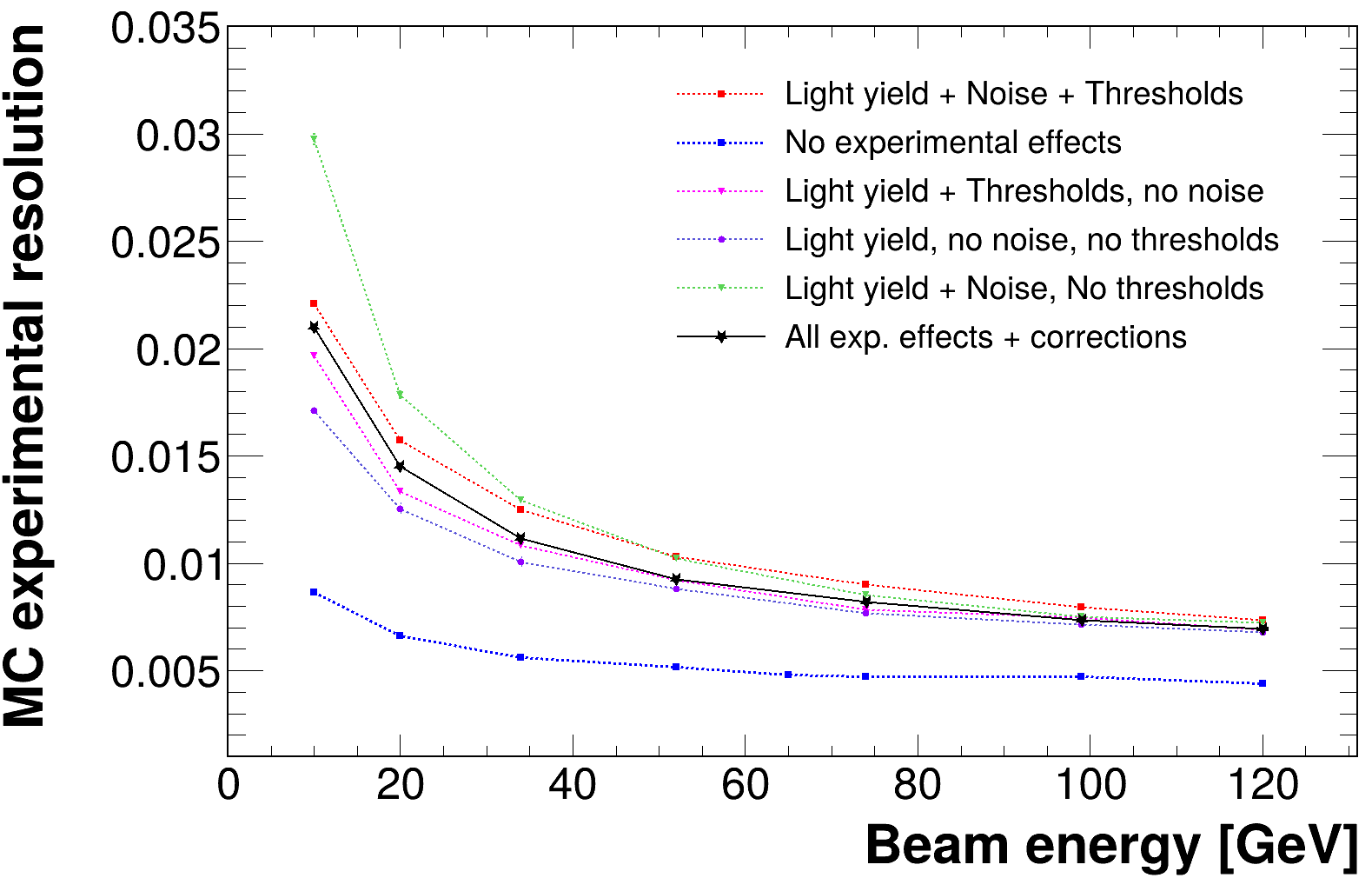}\,\,
        \includegraphics[height=4.7cm]{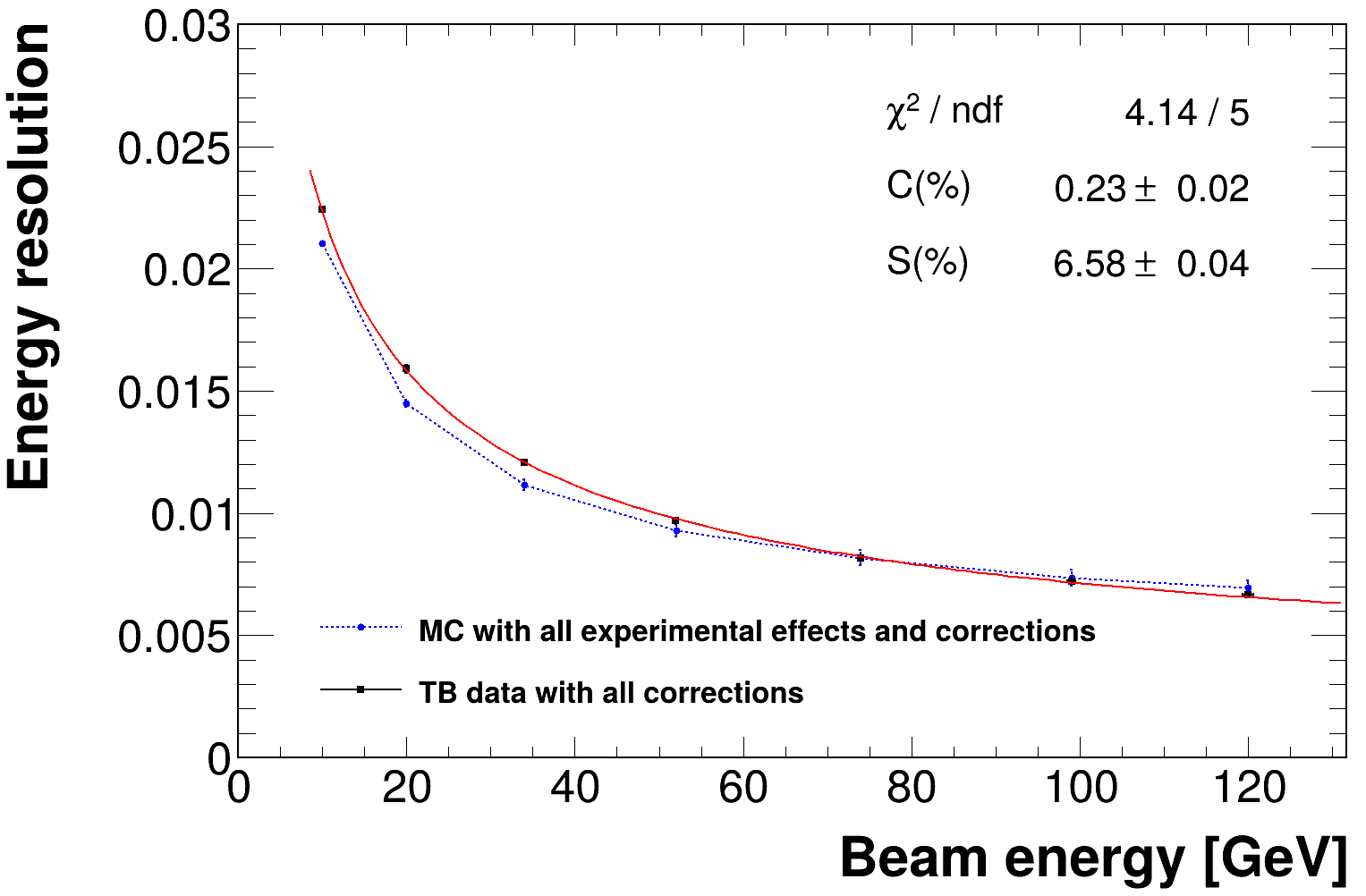}
    \caption{Left: energy resolutions obtained with different simulation configurations. Right: comparison between energy resolutions in MC and data, at the final step in the analysis, with all corrections applied.}
    \label{fig:sim_results}
\end{figure}

For each configuration, the energy reconstruction follows the same procedure adopted for the data, including the leakage correction and the double Crystal Ball fit to the reconstructed energy distributions. The resulting resolutions are summarized in Fig.~\ref{fig:sim_results}-left and Table~\ref{tab:sim_results}.

The inclusion of photoelectron-statistics fluctuations leads to a substantial degradation of the resolution. At 10~GeV, for instance, the resolution increases from 0.86\% for the ideal detector configuration to approximately 1.7\% once photo-statistics are included, showing that the finite light yield is the dominant contribution to the stochastic term. At low energies, the configuration including electronic noise but no channel thresholds yields a poorer resolution, since the thresholds suppress a significant fraction of the noise contribution.

The comparison between the test-beam data and the full simulation is shown in Fig.~\ref{fig:sim_results}-right. In the full simulation, the experimentally determined light yield, noise covariance matrix, and channel thresholds are included. Moreover, the effect due to the finite precision of the channel-by-channel calibration in the data is taken into account as an error to the energy resolution, as detailed in Section \ref{sec:mc_calib_impact}. The simulation reproduces the measured energy resolution within a maximum absolute difference of 0.15\%. This agreement supports the consistency of the detector-response model and of the implementation of the dominant instrumental effects.

\begin{table}[htbp] 
\centering 
\begin{tabular}{c|cccccc} 
\hline $E_{\rm beam}$ [GeV] & Ideal & Poisson & Poisson + Threshold & Poisson + Noise & Full Simulation\\ 
\hline 
10 & 0.87 & 1.71 & 1.97 & 2.97 & 2.21 \\
20 & 0.66 & 1.25 & 1.34 & 1.78 & 1.57 \\
34 & 0.56 & 1.01 & 1.08 & 1.29 & 1.25 \\
52 & 0.52 & 0.88 & 0.92 & 1.02 & 1.03 \\
74 & 0.47 & 0.77 & 0.78 & 0.85 & 0.90 \\
99 & 0.47 & 0.71 & 0.75 & 0.75 & 0.80 \\
120 & 0.44 & 0.68 & 0.69 & 0.72 & 0.74\\ 
\hline 
\end{tabular} 
\caption{Energy resolution obtained for the different simulation configurations. The resolution is expressed as $100\times\sigma_E/E$.}
\label{tab:sim_results} 
\end{table}


\section{Conclusions}
\label{sec:conclusions}

A large-area CRILIN prototype has been constructed and characterized in electron and muon beams at the CERN SPS. The detector consists of five longitudinal layers of $7\times7$ PbF$_2$ crystals, for a total of 225 calorimetric cells and a depth of about $22X_0$. Each crystal is read out by four UV-extended SiPMs, while the front-end electronics are placed outside the active calorimeter volume through a compact remote-routing architecture.

The channel response was reconstructed from digitized waveforms with a template-fit algorithm and equalized using through-going 150~GeV muons. The light yield was evaluated using two independent methods. The measurement based on MIP events yields $(0.68\pm0.11)$~photoelectrons/MeV. The value inferred from 99~GeV electromagnetic showers is 0.54~photoelectrons/MeV on the electromagnetic-shower energy scale; after conversion to the muon reference scale, it becomes 0.62~photoelectrons/MeV. The two determinations are compatible within their uncertainties. The RMS of the summed pedestal energy is about 208~MeV, consistent with the value of about 213~MeV obtained from the measured noise covariance matrix; this confirms the importance of inter-channel noise correlations. A common threshold of 3.5 ADC counts was selected from the 10~GeV resolution plateau.

The calorimeter response is linear within approximately 0.5\% between 10 and 100~GeV. The largest discrepancy from the linear fit is observed at 120~GeV, where the beam momentum spread is about 0.77\%, and the data were collected substantially later than the main energy scan, after unmonitored energy response drifts in time. The energy resolution is described by a stochastic term of $S=(6.58\pm0.04)\%$, and a constant term of $C (0.23\pm0.02)\%$,
where the energy-dependent noise contribution estimated from pedestal events and the beam momentum spread are taken into account. The longitudinal segmentation enables event-by-event corrections based on the number of hits and on the ratio between the energy in the central crystal of the second layer and the total energy. These corrections improve the energy resolution by correcting event-by-event fluctuations in the reconstructed shower development. 
The detector achieves a time resolution below 50~ps for electron energies above 10~GeV and below 20~ps above 60~GeV. The Geant4-based simulation, including photoelectron statistics, correlated electronic noise, and the hit thresholds used in the reconstruction, reproduces the measured energy-resolution trend. In particular, the simulation shows that the finite light yield provides the dominant experimental contribution to the stochastic term: at 10~GeV the resolution changes from 0.87\% in the ideal configuration to 1.71\% when photoelectron statistics alone are included.

These results validate the complete CRILIN detector chain, from the compact mechanics and remote readout to calibration, energy reconstruction, and precision timing. They support the further development of this fast, compact, and longitudinally segmented electromagnetic calorimeter technology for future high-energy lepton colliders. The absence of temperature control and monitoring during the test beam limits the precision of the long-term stability assessment and will be addressed in future implementations with an active cooling and monitoring system.




\section*{Acknowledgments}

This work was developed within the framework of the International Muon Collider Collaboration (IMCC), whose Physics and Detector activities aim at evaluating detector R\&D solutions for the optimization of experiments at a future multi-TeV muon collider.

This work was supported by the EU Horizon 2020 Research and Innovation Programme under Grant Agreements No.~101006726 and No.~101004761.

The authors gratefully acknowledge the support of the CERN SPS staff and, in particular, Nikolaos Charitonidis and Bastien Rae, for the help provided during the test-beam activities. They also thank the LNF Research Division, the ENEA NUC-IRAD-GAM Laboratory (Casaccia R.C.), the Servizio Progettazione e Costruzioni Meccaniche, and the Electronics and Automation Service of the Laboratori Nazionali di Frascati for their technical and logistical support. 

The authors wish to express their gratitude to Tommaso Spadaro for useful discussions on the Geant4 physics lists, and Mattia Campana for precious logistical support and for the design of the PHP online monitor interface.

A special acknowledgment is due to Antonio Soave and to Artel S.r.l. for their essential contribution to the realisation of the PCBs and of the rigid-flex circuits, as well as for helping accelerate the production process despite the delays caused by the international supply-chain situation.

\appendix
\section*{Appendix}

\section{Kapton-based interconnect and remote routing}
\label{app:kapton_routing}
The electrical connection between the SiPM readout boards and the remote front-end electronics is implemented through a dedicated rigid-flex printed circuit. This solution was developed to preserve the compactness and longitudinal segmentation of the calorimeter while avoiding active components, discrete signal cables, and additional services within the active detector volume.

\begin{figure}[htbp!]
    \centering
    \includegraphics[width=0.95\textwidth]{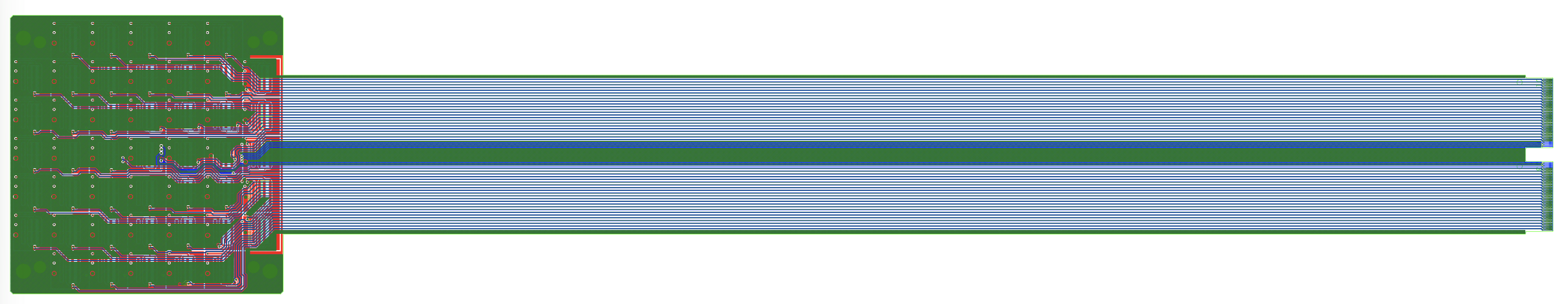}
    \\
    \includegraphics[width=0.95\textwidth]{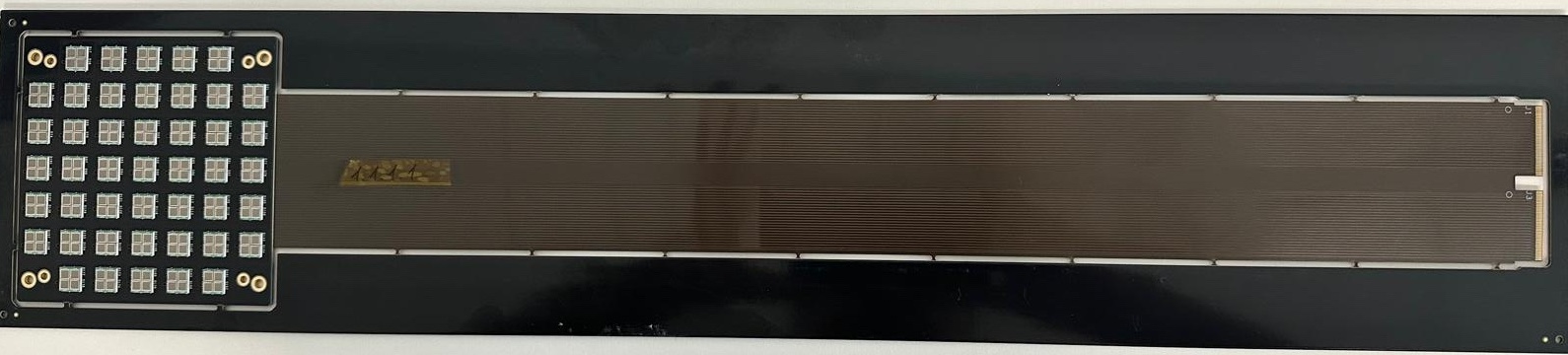}
    \caption{Rigid-flex interconnect between a $7\times7$ SiPM layer and the rear transition board. Top: routing layout of the Kapton-based flexible section. Bottom: assembled SiPM layer with the rigid-flex interconnect.}
    \label{fig:kapton_routing}
\end{figure}

Each calorimeter layer is instrumented with a $7\times7$ SiPM matrix mounted on a rigid printed-circuit board placed directly behind the corresponding PbF$_2$ crystal matrix, see Fig.~\ref{fig:kapton_routing}. The rigid section of the interconnect supports the SiPM matrix and provides the interface to the flexible routing section. Analog signals and bias voltages are carried from the SiPM board to a rear transition board, which provides the connection to the remote front-end electronics.

The flexible section is based on Kapton and has an overall thickness of approximately $300~\mu\mathrm{m}$. It contains embedded $50~\Omega$ stripline transmission lines for the analog signal paths. Each signal line is enclosed by dedicated upper and lower ground conductors, forming an individually shielded stripline. This layout provides controlled impedance, limits electromagnetic interference and inter-channel cross-talk, and preserves the fast SiPM pulse shape during transmission to the rear transition region. Dedicated shielded traces are used in parallel to distribute the SiPM bias voltages from the transition board to the photosensors.

The routing layout preserves the channel ordering between the $7\times7$ SiPM matrix, the transition board, and the remote front-end electronics. The flexible sections from the five calorimeter layers are accommodated within the external mechanical envelope of the detector: their combined thickness remains below the approximately $2~\mathrm{mm}$ thickness of the external aluminum shell. The flexible circuits also accommodate the relative displacement between individual longitudinal modules and the rear transition region, avoiding mechanical stress on the electrical connections during assembly, operation, or maintenance.

The transition boards are housed in the rear basket of the calorimeter mechanics. From this point, analog signals and bias voltages are transmitted over approximately $3.2~\mathrm{m}$ to the front-end crate through coaxial cables. No active amplification is present in the SiPM region. This choice reduces local power dissipation and radiation exposure, while retaining a controlled-impedance signal path to the front-end boards. The architecture also simplifies the installation, testing, and replacement of individual calorimeter layers.

\section{Comparison of Geant4 physics lists}
\label{app:physics_list}
\begin{figure}[H]
    \centering
    \includegraphics[height=4.7cm]{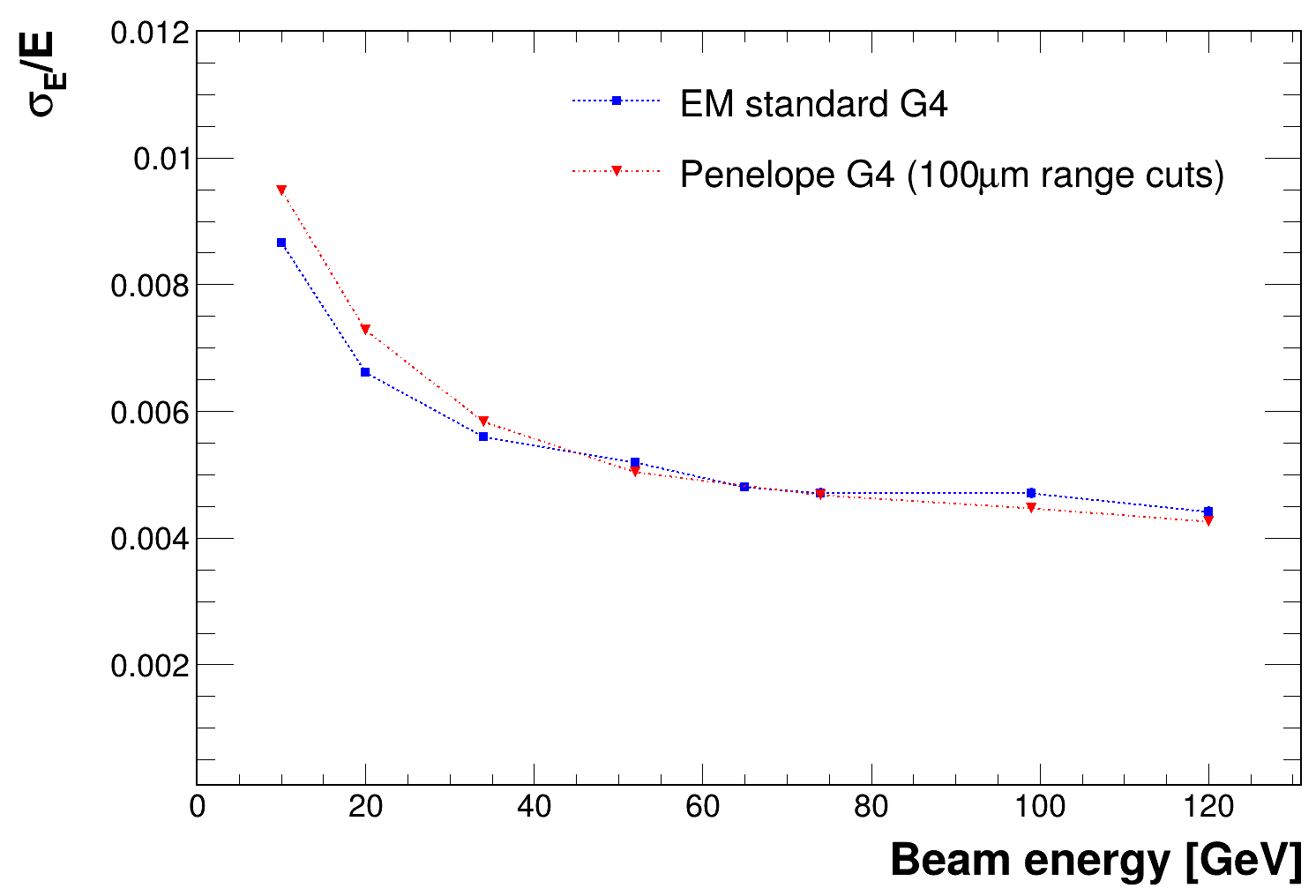}\,\,
        \includegraphics[height=4.7cm]{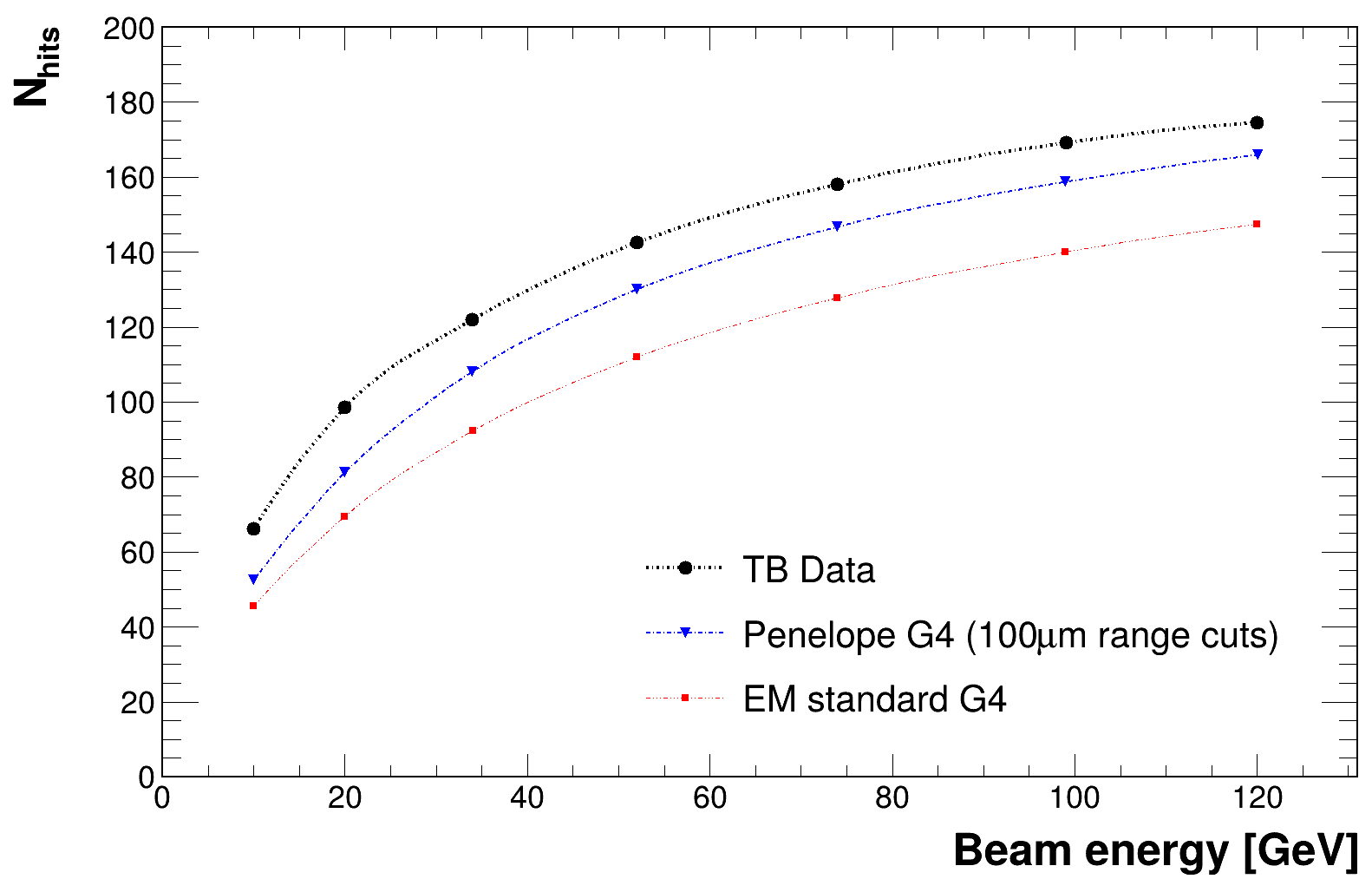}
        \caption{Left: comparison of the energy resolution without experimental effects between PENELOPE-based and standard EM Geant4 physics lists. Right: comparison of the number of hits between the two physics lists and data.}
        \label{fig:comp_liste}
\end{figure}

A comparison has been performed between the PENELOPE-based physics lists, which are more accurate below 1 GeV, and the standard electromagnetic (EM) physics lists in Geant4.
For a high-granularity ECAL like CRILIN, with hit thresholds less than 35 MeV and a Cherenkov response, the accuracy of the physics description on the tens and hundreds of MeV energy scale could be very important. For the PENELOPE-based simulation, a 100 $\mu$m range in the steps has been set in Geant4, motivated by the fact that the Cherenkov threshold in PbF$_{2}$ for electrons is around 100 keV in kinetic energy, which is lost by ionization in the MIP case in around 100 $\mu$m. As a cross-check, a dedicated simulation of 10~GeV electrons was performed using a reduced production cut corresponding to a range of 50~\(\mu\)m. No significant differences were observed with respect to the results presented in this Appendix. The choice of a 100~\(\mu\)m range is therefore adopted as a compromise between computational cost and the accuracy of the electromagnetic shower modeling.
The simulation pipeline described in Sec. \ref{sec:monte_carlo} with the standard Geant4 EM physics list has been performed, but the event-by-event corrections have not been applied. The comparison, as shown in Figure \ref{fig:comp_liste}, is therefore based on the energy resolution without instrumental effects, i.e., with no light yield photo-statistics, noise, and threshold application.
The energy resolutions obtained with the standard Geant4 EM physics list and with the PENELOPE model are compared in the left panel. While the two configurations provide consistent results over most of the explored energy range, the PENELOPE physics list predicts a slightly improved energy resolution at low electron beam energies.
The number of hits for both physics lists is also compared in the right panel with the one found in data, showing that PENELOPE, despite not being compatible, with differences in the range of 10-15 channels, achieves a better agreement with data, with respect to the standard Geant4 EM physics list.

\section{Precision of the calibration procedure}
\label{sec:Calibration precision}
The precision of the channel-by-channel equalization was investigated using the two dedicated overnight MIP runs employed for the relative calibration. Each run contains approximately $10^{6}$ triggered events collected over about 6~hours under stable beam conditions with 150 GeV muons. 

For each calorimeter channel, the selected through-going MIP events were ordered by acquisition time and divided into consecutive slices of 135000 events each. This choice provides sufficient statistics to perform a reliable Langauss fit even for the peripheral channels while preserving enough temporal sampling points to study the response evolution throughout the run.

The same fitting procedure described in Sec.~\ref{subsec:mip_calibration} was applied to each slice, yielding a sequence of MPV measurements as a function of the cumulative number of acquired events. Channels with at least four slices yielding acceptable Langauss fits ($\chi^2/\mathrm{NDF}<1.7$) were retained for the stability analysis. The corresponding MPV values were then fitted with a linear function to quantify possible long-term drifts of the detector response.

\begin{figure}[htbp!]
    \centering
    \includegraphics[width=0.49\textwidth]{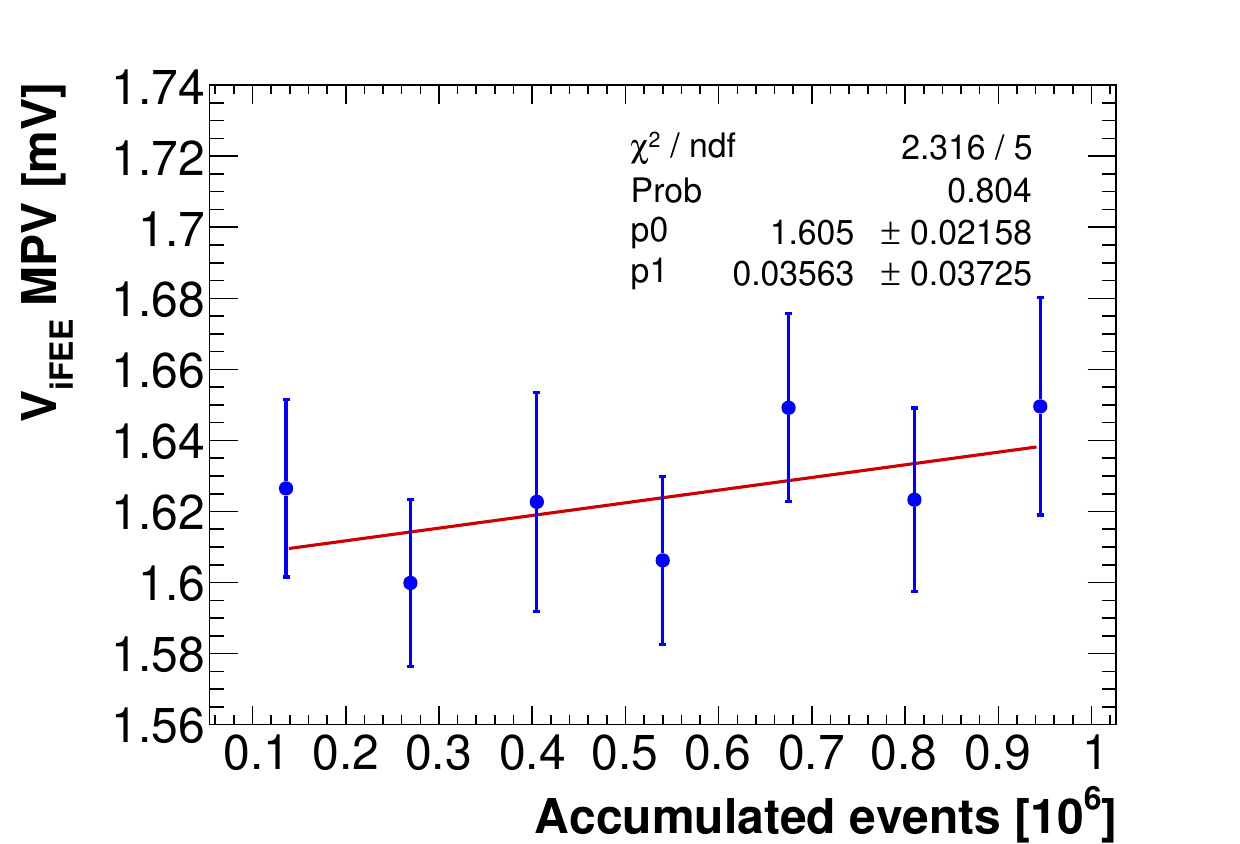}
    \includegraphics[width=0.49\textwidth]{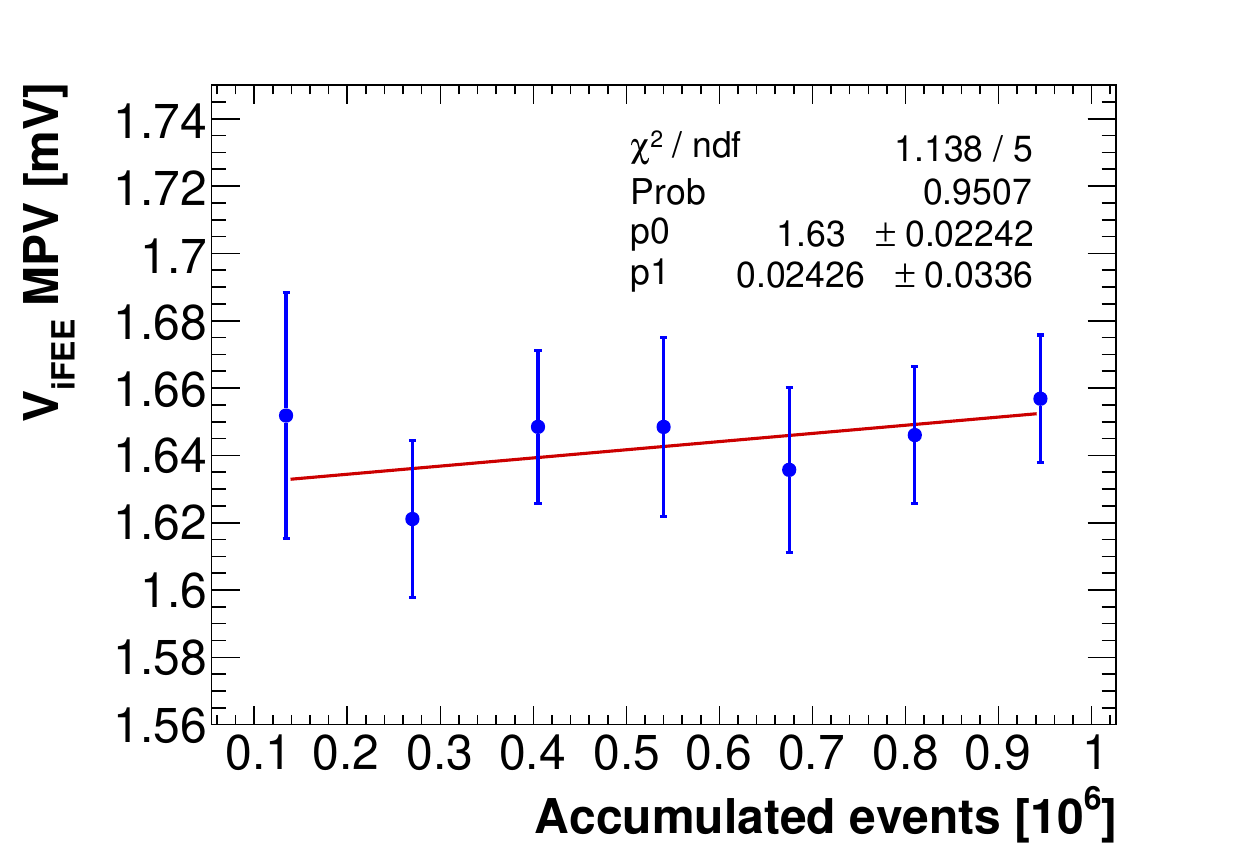}
    \caption{Evolution of the fitted MIP MPV as a function of the cumulative number of acquired events for two representative calorimeter channels (left: channel 60, right: channel 216). The solid lines show the corresponding linear fits.}
    \label{fig:stability_examples}
\end{figure}

Representative examples of the MPV evolution for different calorimeter channels are shown in Fig.~\ref{fig:stability_examples}. While most channels exhibit a nearly constant response throughout the run, a small fraction shows a measurable monotonic variation. 

\begin{figure}[htbp!]
    \centering
    \includegraphics[height=5cm]{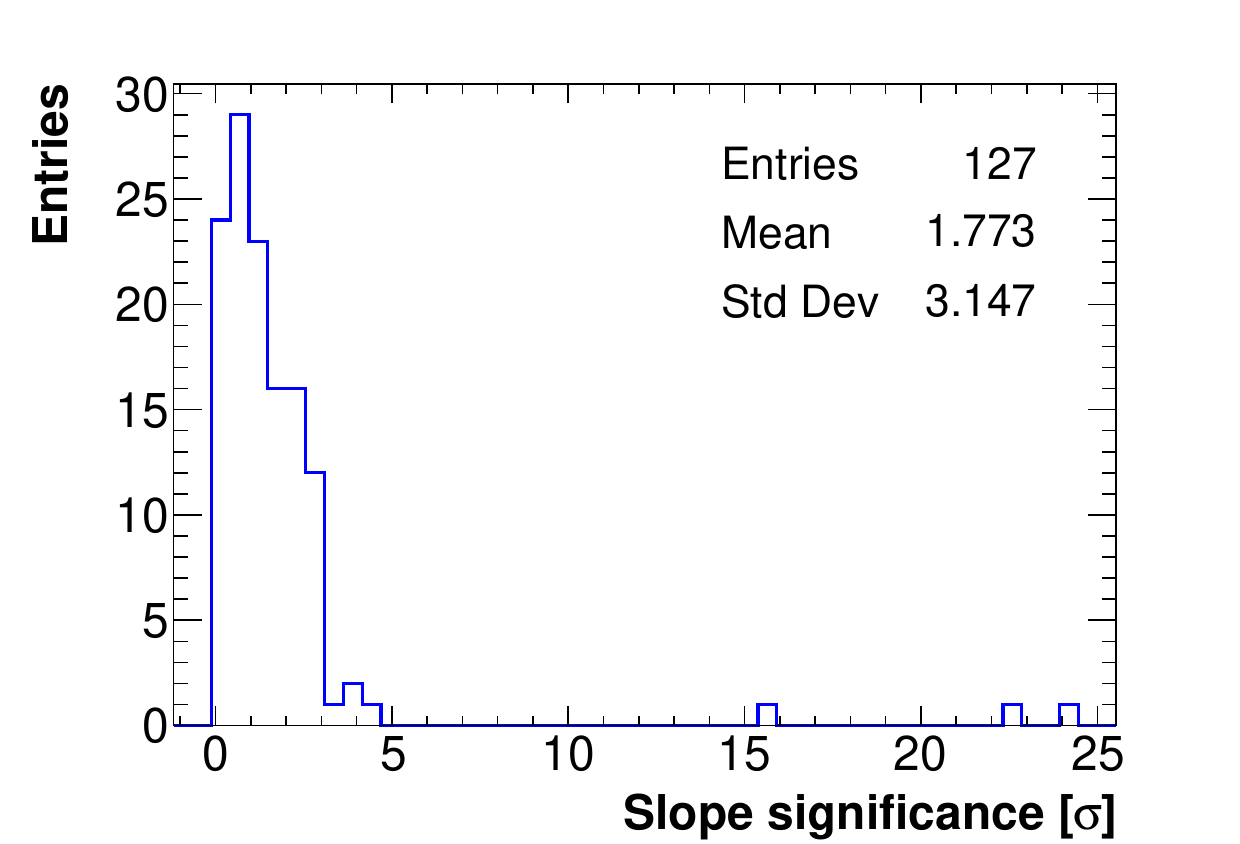}
        \includegraphics[height=5cm]{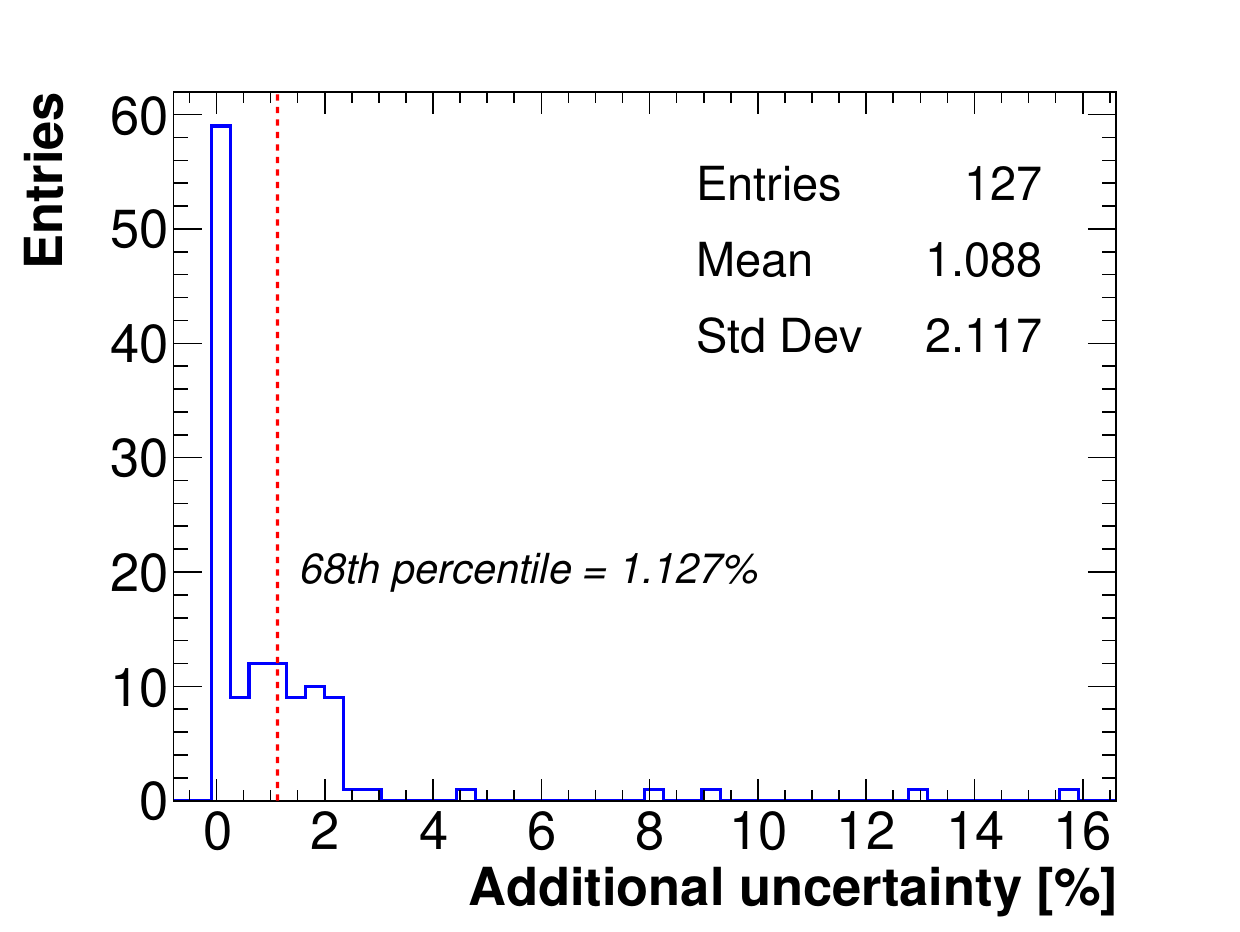}
\caption{Left: distribution of the significance of the fitted linear slopes, expressed as the ratio between the fitted slope and its uncertainty. Right: distribution of the additional uncertainty associated with the stability of the MIP response, obtained from the residual fluctuations around the fitted linear trends.}
    \label{fig:slope_significance}
\end{figure}

The significance of the observed drift was quantified from the ratio between the fitted slope and its uncertainty. The resulting distribution is shown in Fig.~\ref{fig:slope_significance}. Approximately three-quarters of the calorimeter channels have slopes compatible with zero within $2\sigma$, indicating that no significant time dependence is observed for the majority of the detector. 

For each channel, the spread of the equalization coefficient was extracted by requiring the reduced $\chi^2$ of a constant fit to be unity after introducing an additional uncorrelated uncertainty added in quadrature to the statistical errors,
\begin{equation}
\sigma_{\mathrm{syst}}=\sqrt{\chi^2/\mathrm{NDF}-1}\,\langle\sigma_{\mathrm{stat}}\rangle,
\label{eq:sigma_syst}
\end{equation}
where $\langle\sigma_{\mathrm{stat}}\rangle$ denotes the average statistical uncertainty on the MPV measurements.

The distribution of the extracted equalization spreads is shown in Fig.~\ref{fig:slope_significance}. Most channels exhibit a spread close to 1\%. The first bin at $\sigma_{\mathrm{syst}}=0$ corresponds to channels with $\chi^2/\mathrm{NDF}\le1$, for which the observed fluctuations are fully consistent with the statistical uncertainties and no additional uncertainty contribution is required. A limited number of outliers with larger values are observed. The 68th percentile of the distribution is found at $\sigma_{\mathrm{syst}}=1.127\%$, which is adopted as an upper limit at the 68\% confidence level for the uncertainty of the equalization coefficients.

\label{sec:stability}

\section{Impact on the energy resolution of the finite calibration precision}
\label{sec:mc_calib_impact}
\begin{figure}
    \centering
    \includegraphics[height=4.5cm]{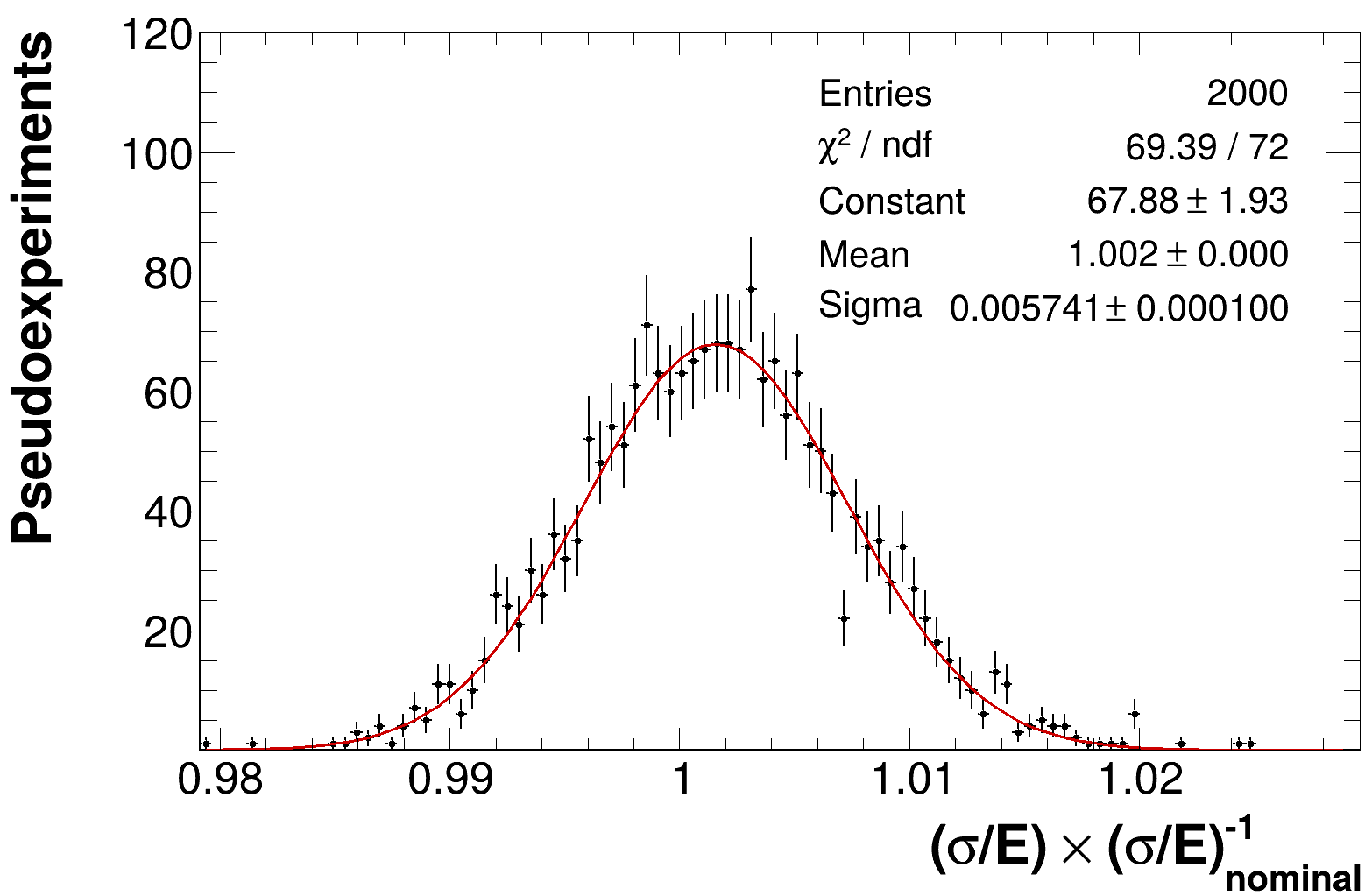}\,\,
    \includegraphics[height=4.5cm]{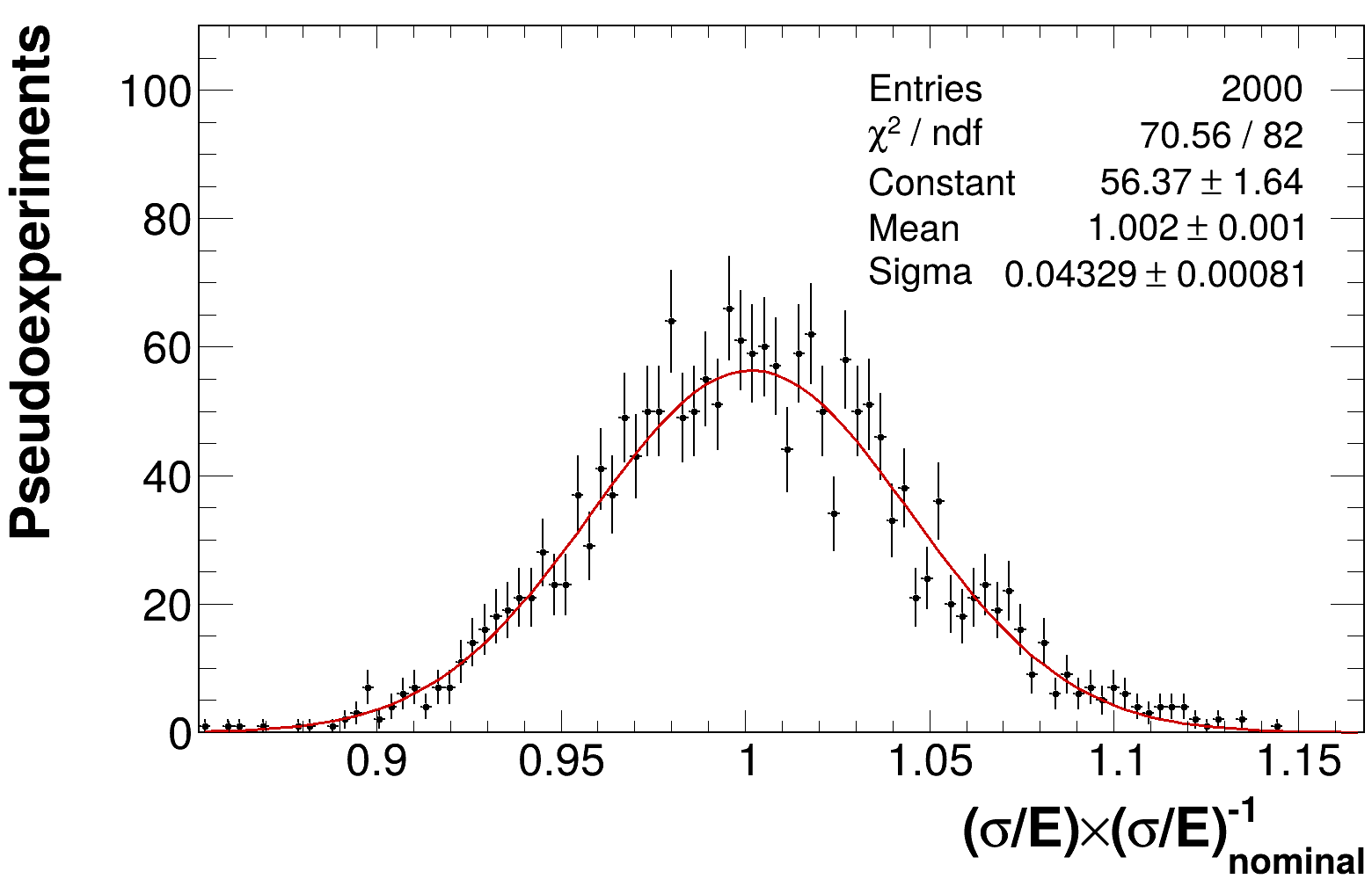}
    \caption{Distributions of $r = (\sigma/E) \cdot (\sigma/E)^{-1}_{nominal}$ from 2000 pseudo-experiments at 10 (left) and 120 GeV (right)}.
    \label{fig:toys}
\end{figure}

\begin{figure}
    \centering
    \includegraphics[width=0.7\linewidth]{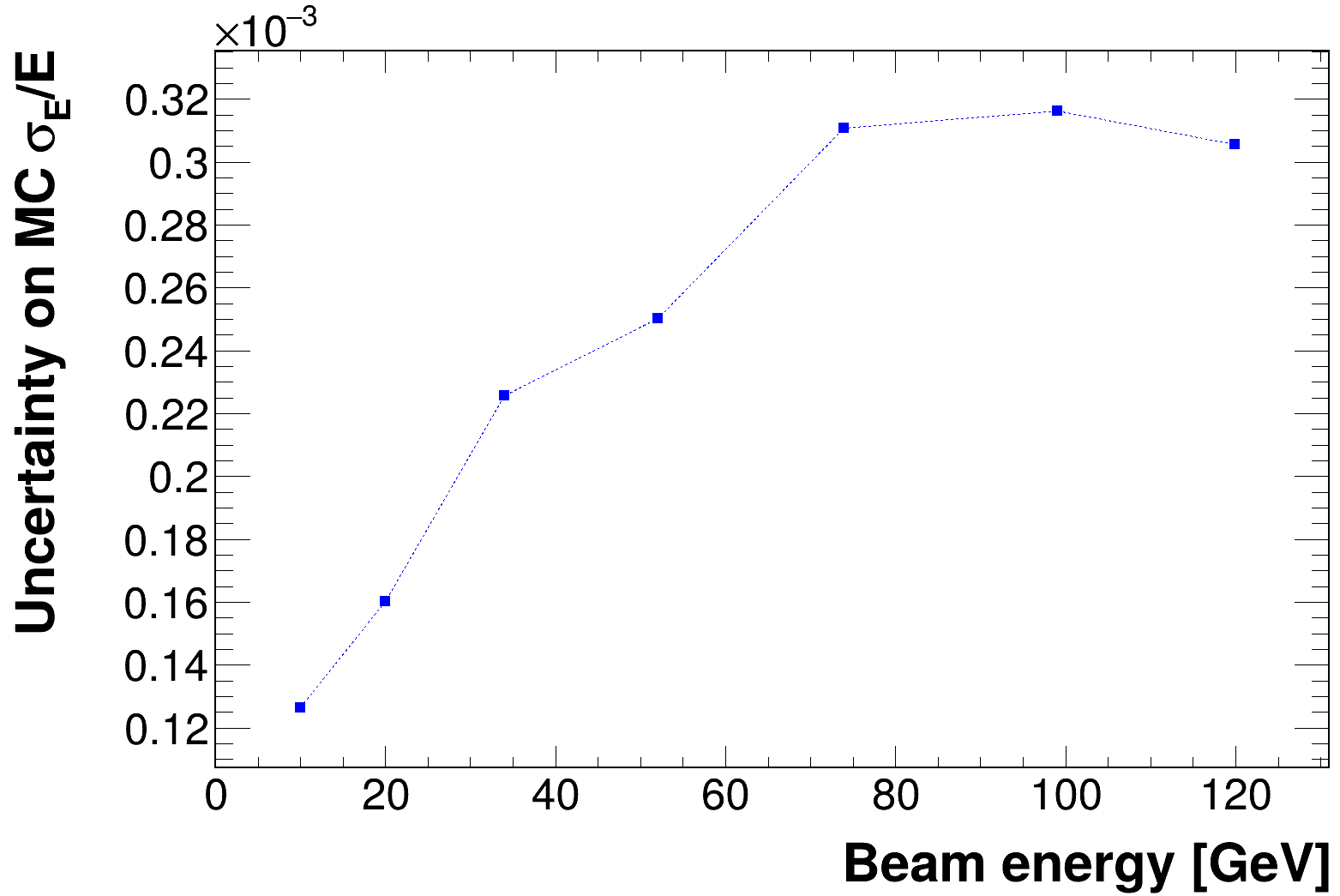}
    \caption{Uncertainties on MC energy resolution due to the finite precision of the channel-by-channel calibration, as a function of beam energy.}
    \label{fig:mc_calib_spread}
\end{figure}

To quantify the impact on energy resolution of the finite channel-by-channel calibration precision (1.127\%) derived in Section \ref{sec:stability}, Monte Carlo events are used
For this study, with all experimental effects and corrections applied, as detailed in Section \ref{sec:mccomp}.
A series of 2000 pseudo-experiments was performed for each energy point, and for each one a random equalization vector was generated using a Gaussian distribution for each channel, with mean equal to 1 and standard deviation equal to 1.127\%.
For each pseudo-experiment, a Gaussian fit in the [-2, 2] $\sigma$ range was performed, and the resulting $\sigma/E$ value was compared to the $\sigma/E$ value fitted in the nominal case, and the ratio $r = (\sigma/E) \cdot (\sigma/E)^{-1}_{nominal}$ was evaluated.
The distribution of $r$ was obtained from the pseudo-experiments for each energy point, and it is shown in Figure \ref{fig:toys} for beam energies of 10 and 120 GeV, where, respectively, standard deviations in $r$ of about 0.57\% and 4.3\% were found.
Since the effects due to the finite precision of the channel-by-channel equalization are not implemented in the MC simulation, a relative uncertainty is added to the MC energy resolution corresponding to the standard deviations in $r$, for each energy point, and these uncertainties are applied to the right plot in Figure \ref{fig:sim_results}. The absolute uncertainties on MC energy resolution, i.e., standard deviations on $r$ multiplied by the MC energy resolutions, are reported as a function of beam energy in Figure \ref{fig:mc_calib_spread}.

\clearpage

\bibliographystyle{unsrtnat}
\bibliography{biblio}

 \end{document}